\documentclass{IEEEoj}
\usepackage{cite}
\usepackage{amsmath,amssymb,amsfonts}
\usepackage{algorithmic}
\usepackage{graphicx,color}
\usepackage{textcomp}
\def\BibTeX{{\rm B\kern-.05em{\sc i\kern-.025em b}\kern-.08em
    T\kern-.1667em\lower.7ex\hbox{E}\kern-.125emX}}
\AtBeginDocument{\definecolor{ojcolor}{cmyk}{0.93,0.59,0.15,0.02}}
\def\OJlogo{\vspace{-14pt}\includegraphics[height=20pt]{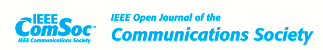}}
\usepackage{balance}
\usepackage{tikz}
\usepackage{multirow}
\usetikzlibrary{arrows, calc, decorations.markings, positioning}
\usepackage{longtable}
\usepackage{tabularray}
\usepackage{subfloat}
\usepackage{subcaption}
\usepackage{algorithm}
\usepackage{algorithmic}
\usepackage{bm}
\usepackage{nomencl}
\usepackage{etoolbox}
\usepackage{framed}
\newcounter{changebar}

\makeatletter
\newcommand{\cboff}{\let\end@float\ltx@end@float}
\newcommand{\cbon}{\let\end@float\cb@end@float}
\makeatother
\usepackage{mdframed}
\newmdenv[
leftmargin = -2pt,
innerleftmargin =1ex,
innertopmargin = 0pt,
innerbottommargin = 0pt,
innerrightmargin = 0pt,
rightmargin = 0pt,
linecolor=blue,
linewidth = 1pt,  
topline = false,
rightline = false,
bottomline = false
]{added}

\makenomenclature
\renewcommand\nomgroup[1]{%
	\item[\bfseries
	\ifstrequal{#1}{A}{List of Abbreviations}{%
		\ifstrequal{#1}{V}{List of Key Variables}{%
			\ifstrequal{#1}{O}{Other symbols}{}}}%
	]}
\makeatletter
\newenvironment{timeline}[6]{%
	\newcommand{\startyear}{#1}
	\newcommand{\tlendyear}{#2}
	
	\newcommand{\yearcolumnwidth}{#3}
	\newcommand{\rulecolumnwidth}{#4}
	\newcommand{\entrycolumnwidth}{#5}
	\newcommand{\timelineheight}{#6}
	
	\newcommand{\templength}{}
	
	\newcommand{\entrycounter}{0}
	
	\long\def\ifnodedefined##1##2##3{%
		\@ifundefined{pgf@sh@ns@##1}{##3}{##2}%
	}
	
	\newcommand{\ifnodeundefined}[2]{%
		\ifnodedefined{##1}{}{##2}
	}
	
	\newcommand{\drawtimeline}{%
		\draw[timelinerule] (\yearcolumnwidth+5pt, 0pt) -- (\yearcolumnwidth+5pt, -\timelineheight);
		\draw (\yearcolumnwidth+0pt, -10pt) -- (\yearcolumnwidth+10pt, -10pt);
		\draw (\yearcolumnwidth+0pt, -\timelineheight+15pt) -- (\yearcolumnwidth+10pt, -\timelineheight+15pt);
		
		\pgfmathsetlengthmacro{\templength}{neg(add(multiply(subtract(\startyear, \startyear), divide(subtract(\timelineheight, 25), subtract(\tlendyear, \startyear))), 10))}
		\node[year] (year-\startyear) at (\yearcolumnwidth, \templength) {\startyear};
		
		\pgfmathsetlengthmacro{\templength}{neg(add(multiply(subtract(\tlendyear, \startyear), divide(subtract(\timelineheight, 25), subtract(\tlendyear, \startyear))), 10))}
		\node[year] (year-\tlendyear) at (\yearcolumnwidth, \templength) {\tlendyear};
	}
	
	\newcommand{\entry}[2]{%
		\pgfmathtruncatemacro{\lastentrycount}{\entrycounter}
		\pgfmathtruncatemacro{\entrycounter}{\entrycounter + 1}
		
		\ifdim \lastentrycount pt > 0 pt%
		\node[entry] (entry-\entrycounter) [below of=entry-\lastentrycount] {##2};
		\else%
		\pgfmathsetlengthmacro{\templength}{neg(add(multiply(subtract(\startyear, \startyear), divide(subtract(\timelineheight, 25), subtract(\tlendyear, \startyear))), 10))}
		\node[entry] (entry-\entrycounter) at (\yearcolumnwidth+\rulecolumnwidth+10pt, \templength) {##2};
		\fi
		
		\ifnodeundefined{year-##1}{%
			\pgfmathsetlengthmacro{\templength}{neg(add(multiply(subtract(##1, \startyear), divide(subtract(\timelineheight, 25), subtract(\tlendyear, \startyear))), 10))}
			\draw (\yearcolumnwidth+2.5pt, \templength) -- (\yearcolumnwidth+7.5pt, \templength);
			\node[year] (year-##1) at (\yearcolumnwidth, \templength) {##1};
		}
		
		\draw ($(year-##1.east)+(2.5pt, 0pt)$) -- ($(year-##1.east)+(7.5pt, 0pt)$) -- ($(entry-\entrycounter.west)-(5pt,0)$) -- (entry-\entrycounter.west);
	}
	
	\newcommand{\plainentry}[2]{% plainentry won't print date in the timeline
		\pgfmathtruncatemacro{\lastentrycount}{\entrycounter}
		\pgfmathtruncatemacro{\entrycounter}{\entrycounter + 1}
		
		\ifdim \lastentrycount pt > 0 pt%
		\node[entry] (entry-\entrycounter) [below of=entry-\lastentrycount] {##2};
		\else%
		\pgfmathsetlengthmacro{\templength}{neg(add(multiply(subtract(\startyear, \startyear), divide(subtract(\timelineheight, 25), subtract(\tlendyear, \startyear))), 10))}
		\node[entry] (entry-\entrycounter) at (\yearcolumnwidth+\rulecolumnwidth+10pt, \templength) {##2};
		\fi
		
		\ifnodeundefined{invisible-year-##1}{%
			\pgfmathsetlengthmacro{\templength}{neg(add(multiply(subtract(##1, \startyear), divide(subtract(\timelineheight, 25), subtract(\tlendyear, \startyear))), 10))}
			\draw (\yearcolumnwidth+2.5pt, \templength) -- (\yearcolumnwidth+7.5pt, \templength);
			\node[year] (invisible-year-##1) at (\yearcolumnwidth, \templength) {};
		}
		
		\draw ($(invisible-year-##1.east)+(2.5pt, 0pt)$) -- ($(invisible-year-##1.east)+(7.5pt, 0pt)$) -- ($(entry-\entrycounter.west)-(5pt,0)$) -- (entry-\entrycounter.west);
	}
	
	\begin{tikzpicture}
		\tikzstyle{entry} = [%
		align=left,%
		text width=\entrycolumnwidth,%
		node distance=10mm,%
		anchor=west]
		\tikzstyle{year} = [anchor=east]
		\tikzstyle{timelinerule} = [%
		draw,%
		decoration={markings, mark=at position 1 with {\arrow[scale=1.5]{latex'}}},%
		postaction={decorate},%
		shorten >=0.4pt]
		
		\drawtimeline
	}
	{
	\end{tikzpicture}
	\let\startyear\@undefined
	\let\tlendyear\@undefined
	\let\yearcolumnwidth\@undefined
	\let\rulecolumnwidth\@undefined
	\let\entrycolumnwidth\@undefined
	\let\timelineheight\@undefined
	\let\entrycounter\@undefined
	\let\ifnodedefined\@undefined
	\let\ifnodeundefined\@undefined
	\let\drawtimeline\@undefined
	\let\entry\@undefined
}
\makeatother

\begin{document}
\receiveddate{XX Month, XXXX}
\reviseddate{XX Month, XXXX}
\accepteddate{XX Month, XXXX}
\publisheddate{XX Month, XXXX}
\currentdate{XX Month, XXXX}
\doiinfo{OJCOMS.2022.1234567}

\title{The Road to Near-Capacity CV-QKD Reconciliation: An FEC-Agnostic Design}

\author{XIN LIU (Graduate Student Member, IEEE), CHAO XU (Senior Member, IEEE), YASIR NOORI (Senior Member, IEEE), SOON XIN NG (Senior Member, IEEE), LAJOS HANZO (LIFE Fellow, IEEE)}
%\author{XIN LIU\authorrefmark{1}(Graduate Student Member, IEEE), CHAO XU\authorrefmark{1} (Senior Member, IEEE), YASIR NOORI\authorrefmark{1}(Senior Member, IEEE), SOON XIN NG\authorrefmark{1}(Senior Member, IEEE), LAJOS HANZO\authorrefmark{1}(LIFE Fellow, IEEE)}
\affil{School of Electronics and Computer Science, University of Southampton, SO17 1BJ Southamoton, UK}
%\affil{?School oof Electronics and Computer Science, University of Southampton, SO17 1BJ Southamoton, UK}
\corresp{CORRESPONDING AUTHOR: L. Hanzo (e-mail: lh@ecs.soton.ac.uk.)}
\authornote{~L. Hanzo would like to acknowledge the financial support of the Engineering and Physical Sciences Research Council projects EP/W016605/1, EP/X01228X/1, EP/Y026721/1 and EP/W032635/1 as well as of the European Research Council's Advanced Fellow Grant QuantCom (Grant No. 789028).}
\markboth{The Road to Near-Capacity CV-QKD Reconciliation: An FEC-Agnostic Design}{Xin Liu \textit{et al.}}

\begin{abstract}
{\color{black}{New near-capacity continuous-variable quantum key distribution (CV-QKD) reconciliation schemes are proposed, where both the authenticated classical channel (ClC) and the quantum channel (QuC) for QKD are protected by separate forward error correction (FEC) coding schemes. More explicitly, all of the syndrome-based QKD reconciliation schemes found in literature rely on syndrome-based codes, such as low-density parity-check (LDPC) codes. Hence at the current state-of-the-art the channel codes that cannot use syndrome decoding such as the family of convolutional codes (CCs) and polar codes cannot be directly applied. Moreover, the ClC used for syndrome transmission in these schemes is generally assumed to be idealistically error-free, where the realistic additive white Gaussian noise (AWGN) and Rayleigh fading of the ClC have not been taken into account.
		To circumvent this limitation, a new codeword-based - rather than syndrome-based - QKD reconciliation scheme is proposed, where Alice sends an FEC-protected codeword to Bob through a ClC, while Bob sends a separate FEC protected codeword to Alice through a QuC. Upon decoding the codeword received from the other side, the final key is obtained by applying a simple modulo-2 operation to the local codeword and the decoded  remote codeword. 
		As a result, \textbf{first of all}, the proposed codeword-based QKD reconciliation system ensures protection of both the QuC and of the ClC. \textbf{Secondly}, the proposed system has a similar complexity at both sides, where both Alice and Bob have an FEC encoder and an FEC decoder. \textbf{Thirdly}, the proposed system makes QKD reconciliation compatible with a wide range of FEC schemes, including polar codes, CCs and irregular convolutional codes (IRCCs), where a near-capacity performance can be achieved for both the QuC and for the ClC.
		Our simulation results demonstrate that thanks to the proposed regime, the performance improvements of the QuC and of the ClC benefit each other, hence leading to an improved secret key rate (SKR) that inches closer to both the Pirandola-Laurenza-Ottaviani-Banchi (PLOB) bound and to the maximum achieveable rate bound.}}
\end{abstract}

\begin{IEEEkeywords}
	Continuous variable quantum key distribution (CV-QKD), multidimensional reconciliation, low-density parity check (LDPC) codes, irregular convolutional codes (IRCC), secret key rate (SKR), near-capacity codes.
\end{IEEEkeywords}

%\IEEEspecialpapernotice{(Invited Paper)}

\maketitle

%\title{}
%\author{Xin Liu, Chao Xu, Yasir Noori, Soon Xin Ng, Lajos Hanzo
%	\thanks{Manuscript created October, 2020; This work was developed by the IEEE Publication Technology Department. This work is distributed under the \LaTeX \ Project Public License (LP) ( http://www.latex-project.org/ ) version 1.3. A copy of the LP, version 1.3, is included in the base \LaTeX \ documentation of all distributions of \LaTeX \ released 2003/12/01 or later. The opinions expressed here are entirely that of the author. No warranty is expressed or imied. User assumes all risk.}
%}
\nomenclature[A]{5G}{Fifth Generation}
\nomenclature[A]{6G}{Sixth Generation}
\nomenclature[A]{AES}{Adcanced Encryption Standard}
\nomenclature[A]{AWGN}{Additive White Gaussian Noise}
\nomenclature[A]{B5G}{Beyond 5G}
\nomenclature[A]{B92}{Bennett-92}
\nomenclature[A]{BB84}{Bennett-Brassard-1984}
\nomenclature[A]{BBM92}{Bennett-Brassard-Mermin-1992}
\nomenclature[A]{BCH}{Bose-Chaudhuri-Hocquenghem}
\nomenclature[A]{BER}{Bit Error Rate}
\nomenclature[A]{BF}{Bit-Flipping}
\nomenclature[A]{BI-AWGN}{Binary-Input Additive White Gaussian Noise}
\nomenclature[A]{BLER}{Block Error Rate}
\nomenclature[A]{BP}{Belief Propagation}
\nomenclature[A]{BPSK}{Binary Phase-Shift Keying}
\nomenclature[A]{BSC}{Binary Symmetric Channel}
\nomenclature[A]{CC}{Convolutional Code}
\nomenclature[A]{ClC}{Classical Channel}
\nomenclature[A]{CM}{Covariance Matrix}
\nomenclature[A]{CN}{Check Node}
\nomenclature[A]{CV-QKD}{Continuous Variable Quantum Key distribution}
\nomenclature[A]{DES}{Data Encryption Standard}
\nomenclature[A]{DH}{Diffie-Hellman}
\nomenclature[A]{DR}{Direct Reconciliation}
\nomenclature[A]{DV-QKD}{Discrete Variable Quantum Key distribution}
\nomenclature[A]{E91}{Ekert-91}
\nomenclature[A]{ECDH}{Elliptic-Curve Diffie-Hellman}
\nomenclature[A]{ECDSA}{Elliptic Curve Digital Signature Algorithm}
\nomenclature[A]{EPR}{Einstein-Podolsky-Rosen}
\nomenclature[A]{EXIT}{Extrinsic Information Transfer}
\nomenclature[A]{FEC}{Forward Error Correction}
\nomenclature[A]{GG02}{Grosshans-Grangier-2002}
\nomenclature[A]{IRCC}{Irregular Convolutional Code}
\nomenclature[A]{LDPC}{Low Density Parity-check}
\nomenclature[A]{LG09}{Leverrier-Grangier-2009}
\nomenclature[A]{LLR}{Log-likelihood Ratio}
\nomenclature[A]{MI}{Mutual Information}
\nomenclature[A]{MIMO}{Multiple-Input Multiple-Output}
\nomenclature[A]{NG}{Next-generation}
\nomenclature[A]{OFDM}{Orthogonal Frequency Division Multiplexing}
\nomenclature[A]{OTP}{One-Time Pad}
\nomenclature[A]{PCM}{Parity-Check Matrix}
\nomenclature[A]{PLOB}{Pirandola-Laurenza-Ottaviani-Banchi}
\nomenclature[A]{PM}{Phase-matching}
\nomenclature[A]{QKD}{Quantum Key Distribution}
\nomenclature[A]{QRNG}{Quantum Random Number Generator}
\nomenclature[A]{QuC}{Quantum channel}
\nomenclature[A]{RR}{Reverse Reconciliation}
\nomenclature[A]{RSA}{Rivest-Shamir-Adleman}
\nomenclature[A]{SARG04}{Scarani-Aci{\'e}n-Ribordy-Gisin-2004}
\nomenclature[A]{SHA}{Secure Hash Algorithm}
\nomenclature[A]{SKR}{Secret Key Rate}
\nomenclature[A]{SNR}{Signal-to-Noise Ratio}
\nomenclature[A]{SPA}{Sum-Product Algorithm}
\nomenclature[A]{TF}{Twin-Field}
\nomenclature[A]{THz}{Terahertz }
\nomenclature[A]{URC}{Unitary Rate Code}
\nomenclature[A]{VN}{Variable Node}
\nomenclature[A]{QK}{Quantum key}
\nomenclature[A]{CK}{Classical key}
\nomenclature[A]{D2D}{Device-to-device}
\nomenclature[A]{UAV}{Unmanned Aerial Vehicle}
\nomenclature[V]{$\hat{X}_{A}$}{the quadrature component transmitted by Alice}
\nomenclature[V]{$\hat{X}_{B}$}{the quadrature component transmitted by Bob}
\nomenclature[V]{$\hat{X}_{E}$}{the excess noise quadrature component introduced by Eve}
\nomenclature[V]{$V_s$}{the variance of Gaussian signals transmitted over QuC}
\nomenclature[V]{$\mathbf{x}$}{the rest of raw data of Alice}
\nomenclature[V]{$\mathbf{x}^\prime$}{the normalized version of $\mathbf{x}$}
\nomenclature[V]{$\mathbf{y}$}{the rest of raw data of Bob}
\nomenclature[V]{$\mathbf{y}^\prime$}{the normalized version of $\mathbf{y}$}
\nomenclature[V]{$\mathbf{b}$}{the random bit stream}
\nomenclature[V]{$\mathbf{b}^\prime$}{the random bit stream after interleaving}
\nomenclature[V]{$\hat{\mathbf{b}}$}{the decoded bit stream of $\mathbf{b}$}
\nomenclature[V]{$\mathbf{u}$}{the spherical codes of $\mathbf{b^\prime}$}
\nomenclature[V]{$\widetilde{\mathbf{u}}$}{the noisy version of $\mathbf{u}$}
\nomenclature[V]{$\mathbf{s}$}{the syndrome side information}
\nomenclature[V]{$\mathbf{M}(\mathbf{y^\prime}, \mathbf{u})$}{the mapping function sent from Bob to Alice}
\nomenclature[V]{$N$}{the codeword length of LDPC codes}
\nomenclature[V]{$K$}{the information length of LDPC codes}
\nomenclature[V]{$K_f$}{the SKR}
\nomenclature[V]{$P_B$}{the BLER in the reconciliation}
\nomenclature[V]{$\beta$}{the reconciliation efficiency}
\nomenclature[V]{$I_{A,B}$}{the mutual information between Alice and Bob}
\nomenclature[V]{$\chi_{BE}$}{the Holevo information between Bob and Eve}
\nomenclature[V]{$\xi_{\text{ch}}$}{the excess noise}
\nomenclature[V]{$\alpha$}{the attenuation of a single-mode optical fibre}
\nomenclature[V]{$v_{el}$}{the electronic noise}
\nomenclature[V]{$\eta$}{the homodyne detector efficiency}
	%	$\hat{X}_{B}$&the quadrature component received by Bob\\
	%	$\hat{X}_{E}$& the excess noise quadrature component introduced by Eve\\
	%	$V_s$& the variance of Gaussian signals transmitted over QuC\\
	%	$\mathbf{x}$& the rest of raw data of Alice\\
	%	$\mathbf{x}^\prime$& the normalized version of $\mathbf{x}$\\
	%	$\mathbf{y}$& the rest of raw data of Bob\\
	%	$\mathbf{y}^\prime$& the normalized version of $\mathbf{y}$\\
	%	$\mathbf{b}$& the random bit stream\\
	%	$\mathbf{b}^\prime$& the random bit stream after interleaving\\
	%	$\widetilde{\mathbf{b}}$& the decoded bit stream\\
	%	$\mathbf{u}$& the spherical codes of $\mathbf{b^\prime}$\\
	%	$\mathbf{v}$& the noisy version of $\mathbf{u}$\\
	%	$\mathbf{s}$& the syndrome side information\\
	%	$\mathbf{M}(\mathbf{y^\prime}, \mathbf{u})$ & the mapping function sent from Bob to Alice\\
	%	$N$& the codeword length of LDPC codes\\
	%	$K$ & the information length of LDPC codes\\
	%	$K_f$& the SKR\\
	%	$P_B$ & the BLER in the reconciliation\\
	%	$\beta$& the reconciliation efficiency\\
	%	$I_{A,B}$ &the mutual information between Alice and Bob\\
	%	$\chi_{BE}$ & the Holevo information between Bob and Eve\\
	%	$\xi_{\text{ch}}$& the excess noise\\
	%	$\alpha$& the attenuation of a single-mode optical fibre\\
	%	$v_{el}$& the electronic noise\\
	%	$\eta$& the homodyne detector efficiency\\
	%\end{longtblr}
	%%\end{table}
\printnomenclature
\section{Introduction}\label{intro}
%\subsection{Background}
Given the increasing penetration of commercial fifth generation (5G) services, since 2020 researchers have embarked on the exploration of future wireless systems such as beyond 5G (B5G) and sixth generation (6G) communication.
In this context, quantum science has the promise of supporting a range of appealing application scenarios\cite{hanzo2012,Botsinis2019,Botsinis2013,Maslov2018}.
More explicitly, on one hand, quantum computing provides revolutionary acceleration in the information processing speed, which is envisioned to substantially improve the computing efficiency in B5G applications and to facilitate powerful new solutions for optimizing next-generation (NG) systems \cite{Gui2020}.
However, the commercialization of quantum computing may also impose a threat to the conventional cryptosystems \cite{Porambage2021,coppersmith1994data,nechvatal2001report,rivest1978method,hellman1976new,miller1986use,koblitz1987elliptic,mani2021error,mollin2000introduction}.
%To elaborate further, the conventional cryptography encompasses three categories, which are symmetric cryptography, asymmetric cryptography, and secure hash algorithms (SHA).
%First of all, the symmetric cryptographies include the data encryption standard (DES) \cite{coppersmith1994data} and the  advanced encryption standard (AES) \cite{nechvatal2001report}, where a single public key is used for both the encryption and decryption process.
%Secondly, the asymmetric cryptography algorithms, where a public and a private key are used for encryption and decryption respectively,   include the Rivest-Shamir-Adleman (RSA) \cite{rivest1978method}, Diffie-Hellman (DH) \cite{hellman1976new}, elliptic curve digital signature algorithm (ECDSA), and the elliptic-curve Diffie-Hellman (ECDH)  \cite{miller1986use,koblitz1987elliptic} technique, which are based on high-complexity mathematical problems such as  the factorization of large prime numbers, discrete logarithm and Elliptic curves, respectively.
%Thirdly, the SHA-2 and SHA-3  systems \cite{mani2021error} have the advantage of providing a mechanism to ensure the integrity of a file \cite{mollin2000introduction}.
These classical cryptography algorithms can provide computational security, which is practically unbreakable within a relatively short period of time when using state-of-the-art computational sources.
However, conventional cryptography may be endangered by the progress in advanced quantum computing techniques.
More explicitly, Shor's powerful algorithm that is capable of efficiently factorizing large prime numbers and of solving elliptic curve problems can impose a serious threat on the classic asymmetric cryptography \cite{shor1994algorithms}.
Similarly, Grover's search algorithm will also make symmetric cryptography insecure \cite{grover1996fastsearch,grover1997quantum}.
Hence, a quantum-safe cryptosystem is needed to tackle this threat.
Against this backdrop, quantum key distribution (QKD) as one of the promising technologies can play an important role in providing  sufficiently secure and reliable data transmission for next-generation communication systems \cite{Wang2020,Hosseinidehaj2019,Cao2022,wang2019inter,kish2020feasibility,you2021towards,Fujiwara20222QKDnetwork,Stanco2022ArchitectureQKD}.
More explicitly, a QKD scheme instructs both the transmitter Alice and the receiver Bob to encrypt their confidential messages with the reconciled keys generated at both sides.
This so-called one-time pad  (OTP) system has been proven by Shannon to be information-theoretically secure \cite{shannon1949communication}.
Furthermore, the QKD-based cryptosystem possesses the capability of eavesdropping detection based on the no-cloning theorem and Heisenberg's uncertainty principle.
%Therefore, the QKD-based cryptosystem can guarantee  a secure key generation process with the presence of eavesdropping.
\begin{figure}[tbp]
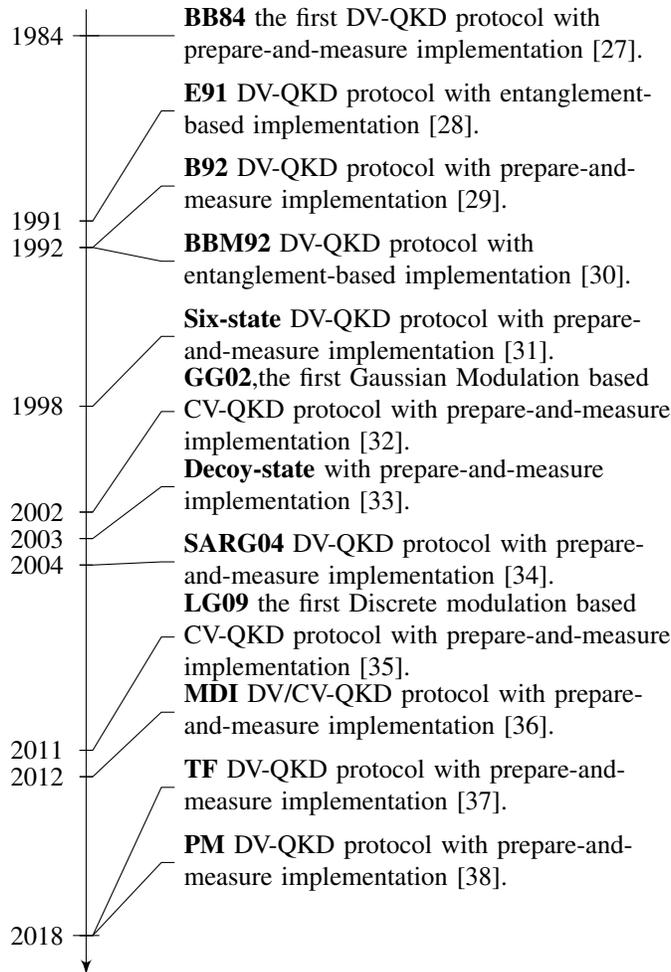

	\centering
	\begin{timeline}{1984}{2018}{1cm}{1cm}{6.5cm}{0.55\textheight}
		\entry{1984}{\textbf{BB84} the first DV-QKD protocol with prepare-and-measure implementation  \cite{BB84}.}
		\entry{1991}{\textbf{E91} DV-QKD protocol with entanglement-based implementation \cite{ekert1991quantum}.}
		\entry{1992}{\textbf{B92} DV-QKD protocol with prepare-and-measure implementation \cite{bennett1992quantum}.}
		\entry{1992}{\textbf{BBM92} DV-QKD protocol with entanglement-based implementation \cite{BBM92}.}
		\entry{1998}{\textbf{Six-state} DV-QKD protocol  with prepare-and-measure implementation \cite{bruss1998optimal}.}
		\entry{2002}{\textbf{GG02},the first Gaussian Modulation based CV-QKD protocol  with prepare-and-measure implementation \cite{grosshans2002continuous}.}
		\entry{2003}{\textbf{Decoy-state}  with prepare-and-measure implementation \cite{hwang2003quantum}.}
		\entry{2004}{\textbf{SARG04} DV-QKD protocol  with prepare-and-measure implementation \cite{scarani2004quantum}.}
		\entry{2011}{\textbf{LG09} the first Discrete modulation based CV-QKD protocol with prepare-and-measure implementation  \cite{LG09}.}
		\entry{2012}{\textbf{MDI} DV/CV-QKD protocol  with prepare-and-measure implementation \cite{lo2012measurement}.}
		\entry{2018}{\textbf{TF} DV-QKD protocol  with prepare-and-measure implementation  \cite{lucamarini2018overcoming}.}
		\entry{2018}{\textbf{PM} DV-QKD protocol   with prepare-and-measure implementation \cite{MaPhysRevX2018}.}
	\end{timeline}
	\caption{State-of-the-art QKD protocols}
	\label{fig:development_milestone}
\end{figure}

The earliest QKD protocol can be traced back to 1984, which is the Bennett-Brassard-1984 (BB84) protocol \cite{BB84}.
Since then, a variety of QKD protocols have been proposed, which can be divided into two types, i.e. discrete variable QKD (DV-QKD) and continuous variable QKD (CV-QKD).
The state-of-the-art of DV-QKD and CV-QKD schemes is summarized  at a glance in Fig.~\ref{fig:development_milestone}.
%Fig.~\ref{fig:development_milestone} represents some classic QKD protocols that encompass protocols both for CV and DV.
More specifically, the landmark BB84 protocol  \cite{BB84} has spawned the family of DV-QKD exemplified by the Ekert-91 (E91) \cite{ekert1991quantum}, Bennett-Brassard-Mermin-1992 (BBM92) \cite{BBM92}, Bennett-92 (B92) \cite{bennett1992quantum}, six-state \cite{bruss1998optimal}, decoy-state \cite{hwang2003quantum}, Scarani-Aci{\'e}n-Ribordy-Gisin-2004 (SARG04) \cite{scarani2004quantum}, Twin-field (TF) \cite{lucamarini2018overcoming}, and phase-matching (PM) \cite{MaPhysRevX2018} protocols. 
Furthermore, the first  CV-QKD protocol was the Gaussian modulation assisted Grosshans-Grangier-2002 (GG02) protocol \cite{grosshans2002continuous}, which was followed by the discrete modulation based CV-QKD Leverrier-Grangier-2009 (LG09) \cite{LG09} protocol.
A comprehensive overview of QKD protocols can be found in \cite{pirandola2020advances,xu2020secure,lo2014secure,diamanti2015distributing,Ge2023DDIQKD}.

%The differences of them are summarised in Table \ref{Table:Two kinds of QKD}.
Table \ref{Table:Two kinds of QKD} offers a comparison between the two types of QKD.
First of all, for light sources, typically the single photon or the attenuated laser source is utilized in DV-QKD, whilst the coherent state or squeezed state solution is used for CV-QKD.
Secondly, the {DV-QKD} modulates or maps information onto the discrete degrees of freedom of a single photon, such as its polarization or phase. 
By contrast, the CV-QKD information is modulated or mapped onto the quadrature components of electromagnetic fields\cite{Hosseinidehaj2019}.
%\footnote{This is a description from quantum aspect. More specifically, one form of a continuous variable quantum system is represented by $M$ bosonic modes, corresponding to $M$ quantized radiation modes of the electromagnetic field, i.e., $M$ quantum harmonic oscillators \cite{Hosseinidehaj2019,Weedbrook2012}. The concept of $M$ modes here is a mathematical and conceptual framework that simplifies the description and analysis of electromagnetic field interactions, quantization of energy, and practical applications in experiments.}
Finally, single-photon detection is required for DV-QKD, which is expensive to implement and yet has a low key rate.
By contrast, for CV-QKD either homodyne or heterodyne detection is utilized, which has convenient compatibility with the operational network infrastructure \cite{Weedbrook2012,Hosseinidehaj2019}.
\begin{table}[tb]
	\centering
	\caption{Comparisons between two types of QKD}
	{ 	\begin{tabular}{|l|r|r|}
			%\toprule
			\hline
			\quad &{DV-QKD} & {CV-QKD}\\
			\hline
			\multirow{2}{*}{Light source}& Single photon or &  Coherent state or\\
			& attenuated laser&squeezed state\\ \hline
			\multirow{2}{*}{Modulation} &Polarization or&Quadrature components of \\
			&phase& electromagnetic fields\\\hline
			\multirow{2}{*}	{Detection} &Single-photon  & Homodyne or\\ 
			&detection&Heterodyne detection\\
			%\bottomrule
			\hline
	\end{tabular}}
	\label{Table:Two kinds of QKD}
\end{table}

Recently, some authors have studied the feasibility of  {CV-QKD} for NG wireless communication systems operating at microwave and terahertz (THz) frequencies \cite{Ottaviani2020, He2020, Liu2018, Weedbrook2010}.
More explicitly, multiple-input  multiple-output (MIMO) and orthogonal frequency division multiplexing (OFDM) air interface techniques have been utilized for increasing the limited secure transmission distance caused by the high path loss of the THz band \cite{Gabay2006,kundu2021mimo,kundu2022channel,kundu2023mimo,zhang2023millimetre,Zhao2019,Zhao2018,Zhang2018}. 
Furthermore, some recent achievements in THz hardware implementations such as detectors, power-efficient sources and antennas \cite{lin2020heterodyne,ikamas2021homodyne,cattaneo2021superconducting,rain2021wave}, can facilitate the practical implementation of CV-QKD in NG communication systems. 
Therefore, this paper mainly focuses on the study of CV-QKD.

As an important step of classical post-processing in QKD, reconciliation plays an pivotal role in ensuring that both the transmitter  and the receiver  rely on the same bit stream  and use it as the reconciled key. 
More explicitly, the reconciliation process is based on error correction used for mitigating the deleterious effects of noise  and interference imposed by Eve \cite{Laudenbach2018}. For instance, a simple Hamming code was utilized in the reconciliation step to correct the bit errors in the raw key string shared by the satellite and the ground station for the experimental satellite-to-ground QKD system used in the \textit{Micius} experiment \cite{liao2017satellite}. 
Inspired by this development, some more advanced forward error correction (FEC) codes have also been investigated, such as  low-density parity-check (LDPC) codes \cite{Zhang2020,guo2020free,shirvanimoghaddam2016design,mani2021error,milicevic2017key,milicevic2018quasi,gumucs2021novel}, polar codes \cite{wen2021rotation,tang2023polar,Zhao2018,kim2017reconciliation}, rateless codes \cite{zhou2019continuous,Asfaw2019}, and their diverse variants.
As a further advance, instead of using a fixed FEC code rate, adaptive-rate  reconciliation schemes were proposed in \cite{zhang2021rate,Zhou2021,Zhang2021polarRecon}, where the secret key rate (SKR) and the secure transmission distance were optimized for different signal-to-noise ratios (SNRs). Moreover, a Raptor-like LDPC code was harnessed for QKD in \cite{Zhou2021}, where the rate-compatible nature of the raptor code  was exploited for  reducing the cost of constructing new matrices for low-rate LDPC codes harnessed at low SNRs. {\color{black}{In contrast to the conventional CV-QKD reconciliation, where a so-called single decoding attempt based algorithm was used, a multiple decoding attempt based method was adopted in \cite{gumucs2021novel} to improve the SKR performance.}} Furthermore, a large block length based LDPC coded scheme was analyzed in \cite{ai2022optimised}, where a near-capacity performance was achieved for transmission over the quantum channel (QuC).

A list of LDPC coded QKD reconciliation schemes is seen at a glance in Table II. In a nutshell, there are two main types of reconciliation methods, namely the multidimensional \cite{leverrier2008multidimensional,PhysRevApplied.19.044023} and the slice-based reconciliation method \cite{lodewyck2007quantum,bloch2005efficient}.
The former achieves better performance in the low-SNR region, which is suitable for longe-range CV-QKD transmission, while the latter in the high-SNR domain, which is suitable for short-distance CV-QKD systems\footnote{As for the multidimensional reconciliation, it attains higher reconciliation efficiency than slice based reconciliation due to the fact that there is no quantization process, which can cause performance degradation, and also that the capacity of the virtual established channel gets closer to the capacity of additive white Gaussian noise (AWGN) channel at a low SNR \cite{jouguet2011long}. However, its throughput is limited to 1 bit, hence making it more suitable for long-range CV-QKD transmission system. By contrast, the slice based reconciliation, especially the multilevel coding and multistage decoding aided slice based reconciliation, has the capability of extracting more than 1 bit of information per channel use (bpcu), especially for  higher SNRs. This is achieved at the cost of poor quantization performance in the low SNR region, making it more suitable for a short range CV-QKD transmission system.}.
The soft-decision LDPC decoding adopted for QKD in  \cite{mani2021error,Mani21} outperforms the hard-decision decoding algorithm of \cite{bloch2005efficient}, but at the cost of a higher complexity. \textbf{However}, a major issue  is that all the existing studies assume that the classical channel (ClC) used for syndrome transmission is error-free. In practice, the ClC is contaminated both by fading and noise, hence error correction is required for both the QuC and the classical syndrome-feedback channel. Consequently, for the multidimensional reconciliation scheme, the receiver has to perform two separate FEC decoding actions, namely one for the QuC and one for the ClC\footnote{Note that the QuC and ClC of CV-QKD will be detailed in Section \ref{system descriptions}-\ref{equivlanet_setup}.}.
This creates an imbalance in terms of the reconciliation complexity, heavily burdening one side.
Furthermore, the classic syndrome-based QKD reconciliation system is limited to syndrome-based codes such as LDPC codes, while the family of convolutional codes (CCs) that are often included in communication standards \cite{xu2019near,xu2019sixty} have not been used in the open literature. Against this background, the novel contributions of this work are as follows:

\begin{itemize}
	\item Firstly, the block error rate (BLER) performance is analyzed in the context of  syndrome-based reconciliation systems, where the ClC is initially assumed to be error-free, and both the bit-flipping (BF) and belief propagation (BP) based decoding algorithms are harnessed.
	%Comparison has been made to demonstrate that the BLER and BER performance of such  syndrome-based system with BP decoding outperforms that of the syndrome-based system with BF decoding, which would impose a huge effect on the final secret key rate (SKR). Hence, the performance analysis thereafter are mainly conducted with the aid of BP decoding due to its superior decoding performance.
	More explicitly, we revise Gallager's sum-product algorithm (SPA) for LDPC codes using BP, where both the codeword transmitted through the QuC and the side information conveying the syndrome through the authenticated ClC can  be accepted as the input of the modified SPA. Our performance results confirm that the revised BP decoder substantially outperforms the conventional BF decoder in terms of the secret key rate (SKR) of the QKD system. 
	\item Secondly, for the first time in the literature, the effect of a realistic  imperfect ClC is characterized for syndrome transmission from Bob to Alice, where reverse reconciliation (RR) is considered and the effects of both fading as well as of noise are taken into account. We demonstrate that the QKD system requires error correction for both the quantum and ClC. Consequently, the receiver has to perform FEC decoding of the potentially corrupted encoded syndrome for transmission over the ClC, and FEC decoding of the corrupted reference key sent from Bob over the ClC, making the decoding complexity unbalanced that burdens the receiver side. This calls for clean-slate considerations for a new QKD system design.
	\item{\color{black}{Thirdly, a new bit-difference based CV-QKD reconciliation scheme is proposed, where Bob transmits the key through the QuC to Alice, and Alice carries out decoding with the aid of the bit-difference side information sent by Bob through the ClC to Alice. The bit-difference side information is constituted by the vector of bit differences between the key and a legitimate LDPC codeword. This regime allows us to use any  arbitrary FEC codes. Our performance results confirm that for a specific FEC this new system has the same performance as the conventional syndrome-based CV-QKD \cite{Laudenbach2018}, but again, it is compatible with any FEC schemes, including polar codes, CCs and irregular convolutional codes (IRCCs).}
	}
	\item {\color{black}{Since the bit-difference vector based CV-QKD system still requires Alice to perform FEC decoding for both the QuC and ClC, a new codeword-based QKD reconciliation system is proposed. In this system, Alice sends a FEC-protected classical key (CK) to Bob through the ClC, while Bob sends a separate FEC protected quantum key (QK) to Alice through the QuC\footnote{{\color{black}{Note that in our proposed codeword-based reconciliation system the QK is defined as the specific part of the key that is transmitted through the QuC, while the CK is defined as the remaining part of the key that is transmitted through the ClC. This is different from the terminology of key used in Systems A-C, where the key is only transmitted through the QuC with the aid of some side information.}}}. Upon a FEC decoding performed at both sides, the final key to be used for the message encryption is  the modulo-2 sum of the CK and QK\footnote{{\color{black}{We note that in the conventional syndrome-based QKD \cite{Laudenbach2018}, even if Eve infers the syndrome from the ClC, she still cannot extract the QK from the QuC. Similarly, in the proposed system, even if Eve obtains the CK that is suitable for any FEC codes, she still cannot acquire the QK from the QuC. The QKD's Heisenberg's uncertainty principle remains valid for the quantum transmission. As a benefit, the SKR will be improved by using our powerful IRCC FEC schemes for both the ClC and the QuC despite considering realistic imperfect channels.}}}. As a result, for the first time in the open literature, our proposed codeword-based CV-QKD system achieves the following novelties. \textbf{Firstly}, the proposed scheme ensures protection of both the QuC and the ClC by FECs. \textbf{Secondly}, the system conceived has a symmetric complexity, where both Alice and Bob have an FEC encoder and an FEC decoder. \textbf{Thirdly}, the proposed QKD reconciliation scheme is compatible with a wide range of FEC schemes, including polar codes, CCs and IRCCs, where a near-capacity performance can be achieved for both the QuC and for the ClC.}}
	
	\item Our performance results demonstrate that with the aid of IRCCs, near-capacity performance can be achieved for both the quantum and the ClC, which leads to an improved SKR that inches closer to both the Pirandola-Laurenza-Ottaviani-Banchi (PLOB) bound \cite{Pirandola2017} and the maximum achieveable rate bound \cite{hanzo2009near}. Therefore, the proposed codeword-based QKD reconciliation system facilitates flexible FEC deployment and it is capable of increasing the secure transmission distance.
\end{itemize}
%\subsection{Gap analysis}
%The corresponding gap analysis table is listed in Table~\ref{table:gap analysis}.
\begin{table*}[htbp]
	\small 
	\centering
	\caption{Novel contributions of this work in comparison to the state-of-the-art schemes.}
	\begin{tabular}{|m{0.22\textwidth}|c|c|c|c|c|c|c|c|c|c|c|c|c|c|}
		\hline
		Contributions              & \textbf{This work} &\cite{leverrier2008multidimensional} &\cite{mani2021error} &\cite{Mani21}&\cite{Zhang2020}&\cite{guo2020free}&\cite{shirvanimoghaddam2016design}&\cite{milicevic2017key}&\cite{milicevic2018quasi}&{\color{black}{ \cite{gumucs2021novel}}}& \cite{ai2018quantum}&\cite{Zhou2021} &\cite{ai2022optimised}&\cite{bloch2005efficient} \\\hline \hline
		DV-QKD(BSC)        &  & & & &&&&&&&$\checkmark$& & &\\\hline
		CV-QKD(AWGN)          &  $\checkmark$ &  $\checkmark$& $\checkmark$ &$\checkmark$&$\checkmark$ &$\checkmark$&$\checkmark$&$\checkmark$&$\checkmark$&$\checkmark$& &$\checkmark$&$\checkmark$& \\\hline
		Hard-decoding           &  $\checkmark$ &&&&&& &&&& &&&$\checkmark$\\\hline
		Soft-decoding incorporates syndromes          &  $\checkmark$ && $\checkmark$ &$\checkmark$ &$\checkmark$&$\checkmark$&$\checkmark$&$\checkmark$&$\checkmark$&$\checkmark$&$\checkmark$ &$\checkmark$&$\checkmark$&\\\hline
		%Imperfect ClC &  $\checkmark$& & & & \\\hline
		AWGN/Rayleigh for the classical authenticated channel &  $\checkmark$&& & &&& &&&&&& &\\\hline
		%Throughput v.s. performance trade-off                 &  $\checkmark$(to do) &  $\checkmark$&$\checkmark$ &$\checkmark$& \\\hline
		%Polar code based QKD scheme                &  $\checkmark$(almost finished) & & &\\\hline
		%Comparison with polar code  based QKD scheme                &  $\times$ & & &\\\hline
		%Comparison with CCs based QKD scheme    &  $\times$& & & \\\hline
		%Convolutional codes based QKD scheme    &  $\checkmark$(to do)& & & \\\hline
		%This paper          & $\checkmark$&$\checkmark$ & $\checkmark$&$\checkmark$ \\\hline
		Balanced decoding complexity for Alice and Bob  & $\checkmark$ && & &&&&& &&&& &\\\hline
		Compatible application with LDPC, CC, polar codes, IRCC & $\checkmark$ & &&&& & &&&&&& &\\\hline
		Near-capacity for both classical \& quantum part & $\checkmark$&& &&& & &&&& &&&\\\hline
	\end{tabular}
	\label{table:gap analysis}
	%\vspace{-0.5cm}
\end{table*}
The rest of this paper is structured as follows. Section \ref{system descriptions} describes one of the  classic CV-QKD protocols \cite{Laudenbach2018,mani2021error}, relying on a commonly utilized reconciliation scheme. 
Furthermore, some  LDPC basics are introduced together with the modified BP\footnote{The modified BP decoding algorithm is  different from the original BP decoding algorithm, because the check node update BP contains a sign flipping term that depends on the syndrome  information.
	The revised Gallager SPA is summarized in Algorithm \ref{alg:alg1}.} decoding algorithm used in the reconciliation schemes. Following this, different system designs are proposed and compared in Section \ref{system comparison}.
{\color{black}{The corresponding security analysis in terms of SKR is conducted in Section \ref{SKR}.}}~Then, Section \ref{simulation} presents the BLER and BER performance of different systems, where the performance of the proposed FEC aided CV-QKD is analyzed. Finally, Section \ref{conclusion}  provides our main conclusions and future research ideas. {\color{black}{The structure of this paper is shown in Fig.~\ref{outline}.}}
{\color{blue}{
\begin{figure}[t]
	\centering
	\includegraphics[width=3.3in]{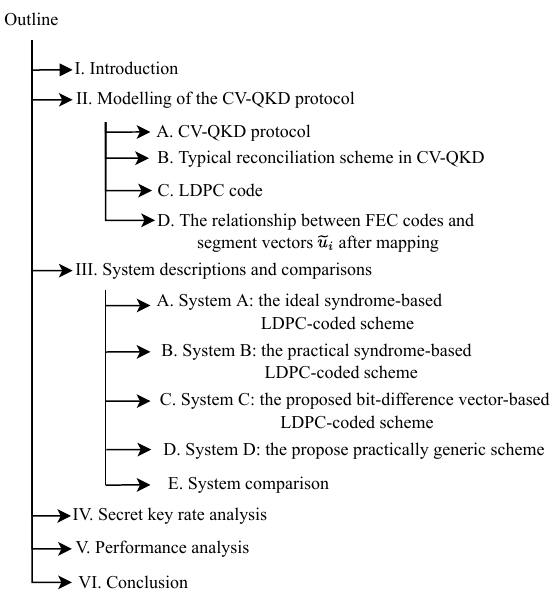}
	\caption{{\color{black}{Structure of this paper.}}}
	\label{outline}
\end{figure}}}

\textit{Notations}: In this paper, bold uppercase and lowercase represent matrices and vector, respectively; $\left\|\cdot\right\|$ denotes the Frobenius norm, and $\left(\cdot\right)^T$ denotes the transpose operation.
%A list of abbreviations is offered in Table \ref{table:List_of_Abbreviations} and a list of variables is offered in Table \ref{table:List_of_Variables}.
A list of abbreviations and a list of variables are offered in the beginning of our paper.

\section{Modelling of the CV-QKD protocol}\label{system descriptions}
In this section, a general CV-QKD scheme\footnote{In this paper, the Gaussian modulated coherent state based CV-QKD protocol is considered.} is modelled, which contains both the quantum transmission and classical post-processing. Following this, the important post-processing step of multidimensional reconciliation  is detailed. Finally, the modified BP decoding is conceived for the reconciliation scheme. 
\subsection{CV-QKD protocol}
The basic QKD protocol is shown in Fig.~\ref{reconciliation_protocol_in_QKD_whole}(a), which has a quantum processing part and a classical post-processing part. As for the quantum processing part, Alice prepares Gaussian-modulated coherent states for transmission to Bob.
After receiving the signal, Bob makes a measurement relying  either on homodyne or on heterodyne detection.
This is followed by  the classical post-processing part.
Explicitly, the signal $\mathbf{y}$ of Fig.~\ref{reconciliation_protocol_in_QKD_whole}(a) is a sifted and potentially channel-infested version of $\mathbf{x}$, which suffers from the hostile action of the QuC.
Observe in the figure that the post-processing part contains four steps, namely the  sifting, parameter estimation, reconciliation, and privacy amplification.
\subsubsection{Quantum transmission part} Firstly, Alice generates a pair of independent Gaussian distributed random variables, denoted as  ${q}_{A}, {p}_{A} \sim \mathcal{N}\left({0}, V_{s} \right)$, where $V_s$ is the variance of the initial Gaussian signal.
Then she uses the random variables ${q}_{A}, {p}_{A} $ to generate a coherent state $\left|\alpha\right\rangle $ associated with
$\alpha=q_{A}+j p_{A}$ for transmission.
As for Eve, we consider an optimal eavesdropping attack, namely the so-called Gaussian collective attack that can be implemented by the Gaussian entangling cloner attack, where Eve has full control over the channel \cite{grosshans2003virtual}.
Generally, Eve prepares the ancilla modes, which are two-mode squeezed states also known as Einstein-Podolsky-Rosen (EPR) states, with variance $W$.
The modes of the EPR states can be described by the operators $\hat{E}$ and $\hat{E}^{\prime \prime}$, where Eve keeps one of the modes such as $\hat{E}^{\prime \prime}$ and injects the other mode $\hat{E}$ into the channel. After the interaction with Alice's state Eve gets the output result $\hat{E}^{\prime}$.
Eve then collectively detects both modes of $\hat{E}^{\prime}$ and $\hat{E}^{\prime \prime}$, gathered from each  run of the protocol, in a final coherent measurement.
%Therefore, based on the description of entangling cloner attack model, the input-output relation of the two-port ($2\times2$) beamspliter can be presented as \cite{kundu2021mimo}
%\begin{equation} \label{input-output-relation}
%	\left[\begin{array}{l}
	%		\hat{a}_{\mathrm{out}, 1} \\
	%		\hat{a}_{\mathrm{out}, 2}
	%	\end{array}\right]=\left[\begin{array}{cc}
	%		\sqrt{\eta} & \sqrt{1-\eta} \\
	%		-\sqrt{1-\eta} & \sqrt{\eta}
	%	\end{array}\right]\left[\begin{array}{l}
	%		\hat{a}_{\mathrm{in}, 1} \\
	%		\hat{a}_{\mathrm{in}, 2}
	%	\end{array}\right]
%\end{equation}
%Next it is assumed that there is a {SISO} transmission system. Let's view the first output of the beamsplitter as what Bob receives and the two inputs of the beamsplitter as what Alice and Eve sends, so based on (\ref{input-output-relation}) the relation of Bob and Alice as well as Eve can be denoted as
Based on this, the output mode at Bob's side can be expressed as
\begin{equation}\label{input_output_relation_SISO}
	\hat{{a}}_{{B}}=\sqrt{T}\hat{{a}}_{{A}}+\sqrt{1-T}\hat{{a}}_{{E}},
\end{equation}
where $\sqrt{T}$ represents the transmission coefficient of the link between Alice and Bob, $\hat{{a}}_{{A}}$ and $\hat{{a}}_{{E}}$ respectively represent the transmitted mode of Alice associated with the coherent state $\left|\alpha\right\rangle $ and the injected Gaussian mode of Eve, and $\sqrt{1-T}\hat{a}_{{E}}$ can be considered as a noise term.

For each of the received modes, Bob applies homodyne measurement to one of the randomly chosen quadratures, i.e. the $Q$ or the $P$ quadrature.
After the measurement, the input-output relationship  between Alice and Bob is given by
\begin{equation}
	\hat{X}_{B}=\sqrt{T} \hat{X}_{A}+\sqrt{1-T} \hat{X}_{E},
\end{equation}
and the input-output relationship of Eve's ancilla mode is

\begin{equation}
	\hat{X}_{E^{\prime}}=-\sqrt{1-T} \hat{X}_{A}+\sqrt{T} \hat{X}_{E},
\end{equation}
where $ \hat X_{B} $ is the received quadrature component, which is measured at Bob, $ \hat X_{A} $ is the  quadrature component transmitted by Alice, $ \hat X_{E} $ is the excess noise quadrature component introduced by Eve, and $ \hat X_{E'} $ is the ancilla quadrature component stored in Eve's quantum memory.
Note that the variable $ \hat X $ corresponds to one of the
two quadrature components $ \{\hat{q}, \hat{p}\} $, so that we have $\hat{X} \in \left\{\hat{q}, \hat{p}\right\}$, which is held for $ \hat X_{A}, \hat X_{B} $, $ \hat X_{E} $ and $ \hat X_{E'} $.
The variance of Alice's transmitted mode is $V_{A}=V_{s}+V_{0}$, where $ V_s $ is the variance of the initial Gaussian signal and
$ V_0 $ is the variance of the vacuum state, and $V_E=W$
is the variance of the excess noise injected by Eve.
The variance of the vacuum state can be expressed as
\begin{equation}\label{vacuum_state_thermal_noise}
	V_{0}=2 \bar{n}+1,
\end{equation}
where $\bar{n}=\left[\exp \left(h f_{c} / K_{B} T_{e}\right)-1\right]^{-1}$ while $h$ is Planck's constant,
$k_B$ is Boltzmann's constant and $T_e$ is the environmental
temperature in Kelvin.
%Finally, the variance of the $ i $-th received mode at Bob is given by
%\begin{equation}
%	V\left(\hat{X}_{B}\right)=T V_{a}+\left(1-T\right) W.
%\end{equation}
\subsubsection{Classical post-processing part} 
\begin{figure*}[t]
	\centering
	\includegraphics[width=1.0\linewidth]{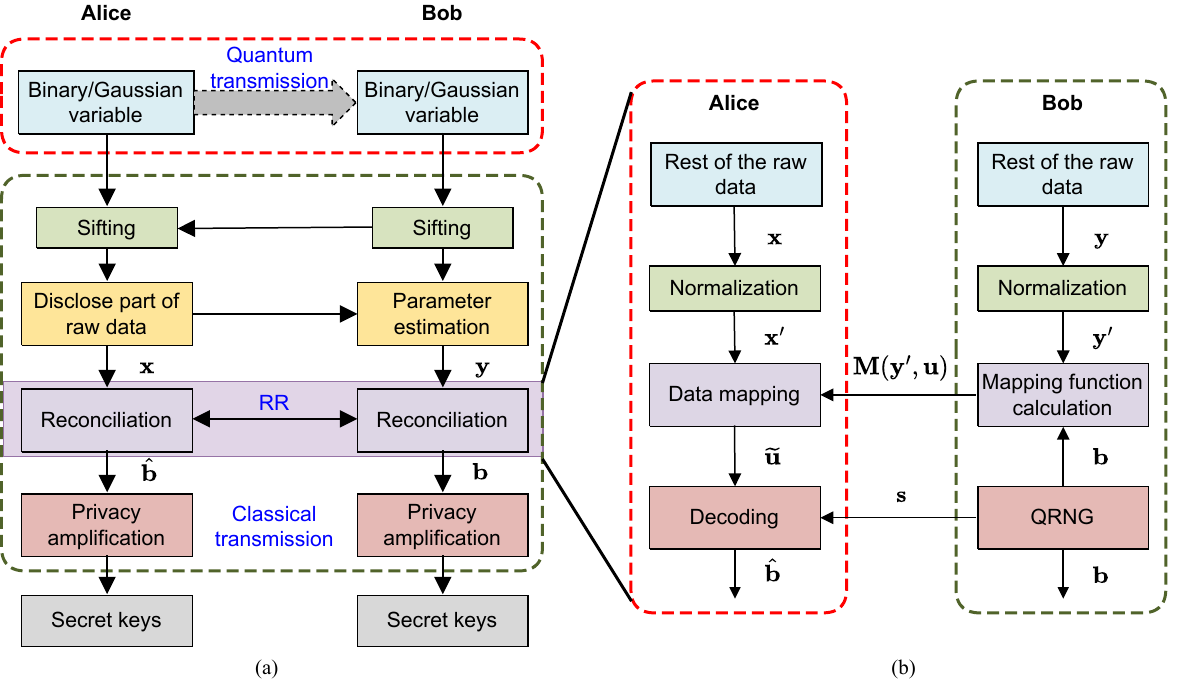}
	\caption{(a) Schematic diagram of a QKD protocol. Note that a binary variable is utilized in {DV-QKD} systems, whilst a Gaussian variable is utilized in {CV-QKD} systems. Moreover, as for the quantum transmission shown here, it contains the process of converting the binary/Gaussian variable to quantum states and that of converting the quantum states to binary/Gaussian variable, which is the quantum measurement. (b) Schematic of the multidimensional RR scheme  in QKD, where $\mathbf{x}$ and $\mathbf{y}$ are two correlated Gaussian sequences, while $\mathbf{x^\prime}$  and $\mathbf{y^\prime}$ represent
		their normalized counterparts; $\mathbf{M}(\mathbf{y^\prime}, \mathbf{u})$ represents the mapping
		function sent from Bob to Alice; $\mathbf{b}$ denotes the initial sequence
		generated by QRNG; $\mathbf{u}=\left(\frac{(-1)^{\mathbf{b}^\prime(1)}}{\sqrt{8}}, \frac{(-1)^{\mathbf{b}^\prime(2)}}{\sqrt{8}}, \ldots, \frac{(-1)^{\mathbf{b}^\prime(8)}}{\sqrt{8}}\right)$ denotes the spherical codes of $\mathbf{b}^\prime$, which is the interleaved bit stream of $\mathbf{b}$; $\widetilde{\mathbf{u}}$
		is the sequence before decoding and  $\hat{\mathbf{b}}$ is the decoding result
		that is equal to $\mathbf{b}$ when the decoding is successful; $\mathbf{s}$ denotes
		the additional side information, which is normally the syndrome calculated based on Bob's bit stream. Note that the dimensionality D is set 8. }
	\label{reconciliation_protocol_in_QKD_whole}
\end{figure*}

\begin{itemize}
	\item \textit{Sifting}: In the sifting step of Fig.~\ref{reconciliation_protocol_in_QKD_whole}(a), both Alice and Bob retain the data associated with those specific states, whose preparation and measurement basis happen to be the same, given that both their bases are randomly chosen.
	More explicitly, in the BB84 DV-QKD example, Bob randomly chooses one of two legitimate polarization bases to measure his  data received from Alice.
	Then they both publicly  communicate with each other to agree about the particular bit-indices, where the measurement basis of Bob is the same as the preparation basis of Alice.
	\item \textit{Parameters estimation}: In this step of Fig.~\ref{reconciliation_protocol_in_QKD_whole}(a), Alice and Bob will reveal and compare a random subset of the data, which allows them to estimate some parameters, such as the transmissivity (pathloss coefficient), excess noise, and the SNR of the channel. Then the mutual information (MI) between them is calculated to judge whether this channel is secure enough for supporting their communication.
	%The secrecy capacity or secret key rate (SKR) denoted as $R$ is the difference between the MI of Alice and Bob $I_{AB}$ and that of Bob and Eve $I_{BE}$, given reverse reconciliation (RR) is taken into account. here since RR scheme is more suitable to have a longer secure transmission with low SNR regions \cite{leverrier2008multidimensional,weedbrook2012continuous}.
	If the MI between Alice and Bob is higher than Eve's information concerning the key, the channel is deemed to be secure enough for supporting secret keys transmission, otherwise, the transmission aborts and a new random process is initiated.
	\item \textit{Reconciliation}: The reconciliation step of Fig.~\ref{reconciliation_protocol_in_QKD_whole}(a) relies on error correction.
	There are two styles of reconciliation, namely direct reconciliation (DR) and RR.
	As for DR, Bob corrects his data according to Alice's data, while Alice's data remains unmodified.
	By contrast, in RR, Alice corrects her data according to Bob's data and Bob's data remains unmodified.
	Usually, RR is preferred since it can provide longer secure transmission distance than that of DR.
	More explicitly, in DR, the channel's transmission coefficient must be above 0.5 to provide a non-zero SKR, while there is no such limitation in the RR case\cite{Weedbrook2010,Weedbrook2012b}.
	\item \textit{Privacy amplification}:
	%A practical revers reconciliation scheme will be elaborated later on.
	Finally, the last step of Fig.~\ref{reconciliation_protocol_in_QKD_whole}(a) is privacy amplification harnessed for reducing Eve's probability of successfully guessing (a part of) the keys, since Eve has a certain amount of information concerning the key. A hashing function may be used for privacy amplification.
	For example, a universal hashing function can be used to complete the privacy amplification via turning the reconciled key stream into a shorter-length final key stream. As for the amount by which the reconciled key is shortened, this depends on how much information Eve has gained about the key.
\end{itemize}
%\begin{figure}[t]
%	\centering
%	\includegraphics[width=3.3in]{figures/protocol_framework13.pdf}
%	\caption{Schematic diagram of a QKD protocol. Note that a binary variable is utilized in {DV-QKD} systems, whilst a Gaussian variable is utilized in {CV-QKD} systems. Moreover, as for the quantum transmission shown here, it contains the process of converting the binary/Gaussian variable to quantum states and that of converting the quantum states to binary/Gaussian variable, which is the quantum measurement. }
%	\label{reconciliation_protocol_in_QKD}
%\end{figure}
\subsection{Typical reconciliation scheme in CV-QKD}\label{equivlanet_setup}
\begin{figure}[!htbp]
	\centering
	\subfloat[1-dimensional representation]{\includegraphics[width=2.3in]{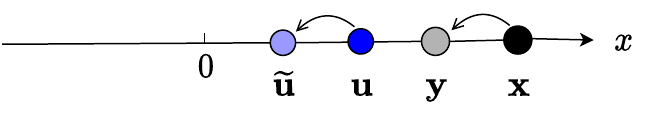}%
		\label{fig:1dimensional_representation}
	}
	\hfil
	%\vspace{0.2cm} 
	\subfloat[2-dimensional representation]{\includegraphics[width=2.3in]{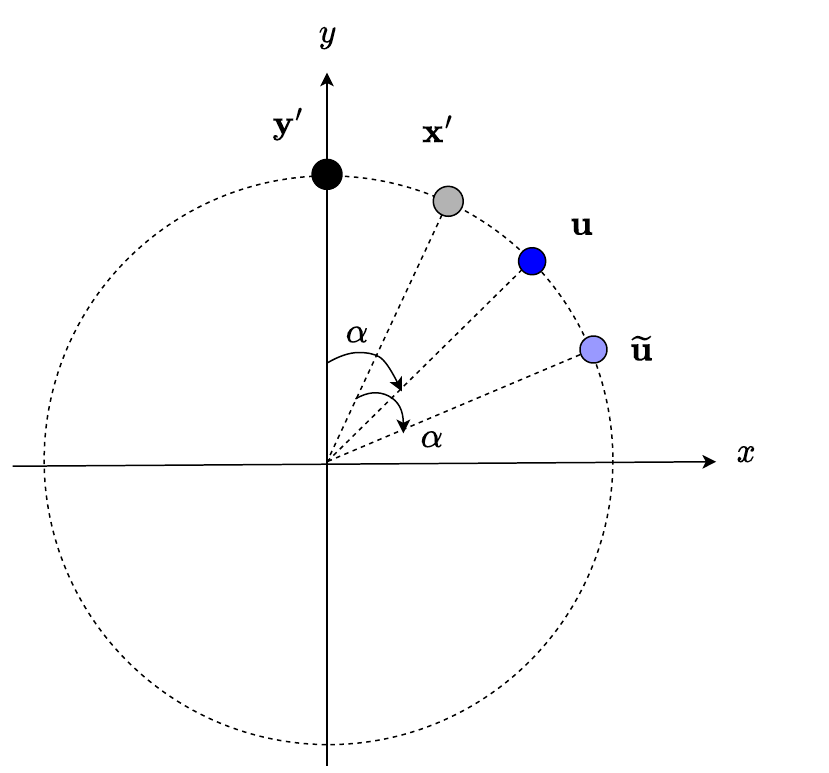}%
		\label{fig:2dimensional_representation}
	}\hfil
	\subfloat[3-dimensional representation]{\includegraphics[width=2.3in]{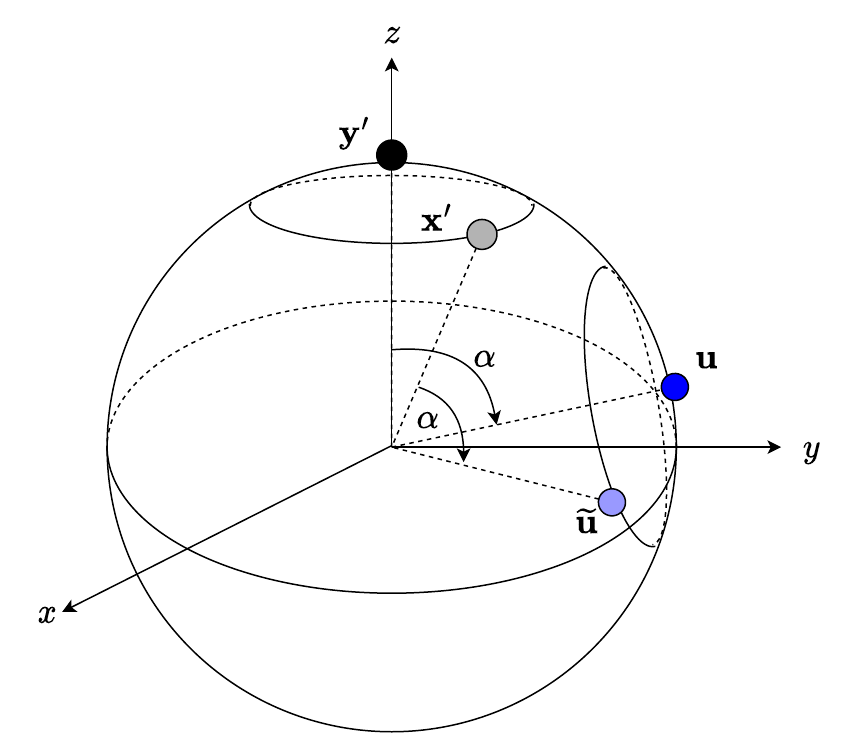}%
		\label{fig:3dimensional_representation}
	}
	\caption{The representation of the  noise conversion process for $D=1, 2$ and $3$ based on \cite{leverrier2008multidimensional}.}
	\label{fig:multidimensional_representation}
\end{figure}

Again, a multidimensional reconciliation method is considered, since it exhibits better performance in the lower SNR region, which may translate into a longer secure transmission distance \cite{leverrier2008multidimensional}.
The multidimensional reverse reconciliation process is shown in Fig. \ref{reconciliation_protocol_in_QKD_whole}(b).
%\begin{figure}[t]
%	\centering
%	\includegraphics[width=3.3in]{Figures/protocol_framework24.pdf}
%	\caption{Schematic of the multidimensional RR scheme  in QKD, where $\mathbf{x}$ and $\mathbf{y}$ are two correlated Gaussian sequences, while $\mathbf{x^\prime}$  and $\mathbf{y^\prime}$ represent
%		their normalized counterparts; $\mathbf{M}(\mathbf{y^\prime}, \mathbf{u})$ represents the mapping
%		function sent from Bob to Alice; $\mathbf{b}$ denotes the initial sequence
%		generated by QRNG; $\mathbf{u}=\left(\frac{(-1)^{\mathbf{b}^\prime(1)}}{\sqrt{8}}, \frac{(-1)^{\mathbf{b}^\prime(2)}}{\sqrt{8}}, \ldots, \frac{(-1)^{\mathbf{b}^\prime(8)}}{\sqrt{8}}\right)$ denotes the spherical codes of $\mathbf{b}^\prime$, which is the interleaved bit stream of $\mathbf{b}$; $\widetilde{\mathbf{u}}$
%		is the sequence before decoding and  $\widetilde{\mathbf{b}}$ is the decoding result
%		that is equal to $\mathbf{b}$ when the decoding is successful; $\mathbf{s}$ denotes
%		the additional side information, which is normally the syndrome calculated based on Bob's bit stream. Note that the dimensionality D is set 8.
%		% QRNG, quantum random number generator.
%	}
%	\label{reconciliation_protocol_in_QKD2}
%\end{figure}
After the disclosure of the raw data to be used for parameter estimation, as seen in Fig.~\ref{reconciliation_protocol_in_QKD_whole}(a), the rest of their raw data $x:=\hat X_{A}^{\prime}$ and $y:=\hat X_{B}^{\prime}$ is constituted by a pair of correlated Gaussian distributed sequences, where $x \sim {\cal N}\left( {0,{\sigma_x ^2}} \right)$, and $y=x+n, n \sim {\cal N}\left( {0,{\sigma_n ^2}} \right)$.
Then both Alice and Bob choose $D$  for representing the number of dimensions in the multidimensional reconciliation, which defines how the sequence of transmit data is partitioned into shorter segments.
It was shown in \cite{leverrier2008multidimensional} that the mapping function used in the multidimensional reconciliation process only exists in $\mathbb{R},\mathbb{R}^2,\mathbb{R}^4,\mathbb{R}^8$ dimensions, which corresponds to $D=1,2,4,8$, due to its algebraic structure as proven by Theorem 2 in \cite{leverrier2008multidimensional}.
Moreover, it was demonstrated in \cite{mani2021error,leverrier2008multidimensional,zhou2019continuous,milicevic2018quasi} that an eight-dimensional reconciliation scheme ($D=8$) outperformed  the schemes associated with $D=1,2,4$ in terms of the BLER performance attained.
Therefore, usually the eight-dimensional $(D=8)$ reconciliation scheme is  adopted for practical CV-QKD systems \cite{leverrier2008multidimensional, zhou2019continuous, mani2021error}.
The main steps of multidimensional RR can be described as follows.
\begin{enumerate}
	\item Firstly, the rest of the raw data of Alice and Bob, can be viewed as a pair of sequences, denoted as $\mathbf{x}$ and $\mathbf{y}$. The length of the two sequences is set to the FEC codeword length $N$. Then they are partitioned into $I=N/8$ number of shorter segments, denoted as $\mathbf{x}=\left[\mathbf{x}_1;\mathbf{x}_2;...;\mathbf{x}_I\right]$ and $\mathbf{y}=\left[\mathbf{y}_1;\mathbf{y}_2;...;\mathbf{y}_I\right]$, where $\mathbf{x}_{i}$, $\mathbf{y}_{i}, i=1,2,...,I$, are $8 \times 1$ column vectors.  
	\item Both Alice and Bob will normalize each 8-element segment of $\mathbf{x}$ and $\mathbf{y}$ in order to get a uniformly distributed 8-element vector, which is reminiscent of producing equi-probable $2^8$-ary symbols.
	To elaborate on the resultant eight-dimensional reconciliation scheme, the normalized data in the form of the vectors $\mathbf{x}_{i}^\prime$ and $\mathbf{y}_{i}^\prime$ can be obtained by $\mathbf{x}_{i}^\prime  = \frac{\mathbf{x}_{i}}{{\left\|\mathbf{ x}_i \right\|}}$ and
	$\mathbf{y}_i^\prime  = \frac{\mathbf{y}_i}{{\left\| \mathbf{y}_i \right\|}}$, where we have $\left\| \mathbf{x}_i \right\| = \sqrt {\left\langle {\mathbf{x}_i,\mathbf{x}_i} \right\rangle }  = \sqrt {\sum\nolimits_{d = 1}^8 {\mathbf{x}_i\left(d\right)^2} } $
	and $\left\| \mathbf{y}_{i} \right\| = \sqrt {\left\langle {\mathbf{y}_{i},\mathbf{y}_{i}} \right\rangle }  = \sqrt {\sum\nolimits_{d = 1}^8 {\mathbf{y}_i\left(d\right)^2} } $.
	Hence, both the normalized vectors $\mathbf{x}_{i}^\prime$ and  $\mathbf{y}_{i}^\prime$ are uniformly distributed on the surface of the 8-dimensional unit-radius sphere. Therefore, spherical codes \cite{leverrier2008multidimensional}, where all  codewords lie on a sphere centered on 0 can play the same role for CV-QKD as binary codes for DV-QKD.
	\item Bob randomly generates a binary stream $\mathbf{b}$ using a quantum random number generator (QRNG)\footnote{Note that the QRNG  generates classical random numbers.},  whose length  is the same as the FEC codeword length $N$.  Then, the random bit sequence   $\mathbf{b}$ will be interleaved into $\mathbf{b}^\prime$ and the resultant sequence $\mathbf{b}^\prime$ is partitioned into $\mathbf{b}^\prime=\left[\mathbf{b}_1^\prime;\mathbf{b}_2^\prime;...;\mathbf{b}_I^\prime\right]$, where $\mathbf{b}_{i}^\prime$ is an $8$-element binary column vector. Then each segment $\mathbf{b}_{i}^\prime, i=1,2,...,I$, will be mapped to the 8-dimensional unit-radius sphere of $\mathbf{u}_{i}=\left(\frac{(-1)^{\mathbf{b}_{i}^\prime(1)}}{\sqrt{8}}, \frac{(-1)^{\mathbf{b}_{i}^\prime(2)}}{\sqrt{8}}, \ldots, \frac{(-1)^{\mathbf{b}_{i}^\prime(8)}}{\sqrt{8}}\right)$. %$\mathbf{b}_{i}=\left( {{b_1},{b_2},....,{b_8}} \right)$ \cite{Collantes_QRNG_2017}, then maps it onto the unit 7-sphere $\mathbf{u}=\left(\frac{(-1)^{b_{1}}}{\sqrt{8}}, \frac{(-1)^{b_{2}}}{\sqrt{8}}, \ldots, \frac{(-1)^{b_{8}}}{\sqrt{8}}\right)$.
	\item Bob  calculates the mapping function for each segment based on the vectors $\mathbf{u}_{i}$ and $\mathbf{y}_{i}^\prime$. This mapping function is used to map $\mathbf{y}_{i}^\prime$ to $\mathbf{u}_{i}$ so as to find the relationship between the normalized Gaussian vector $\mathbf{y}_{i}^\prime$ and the modulated stream $\mathbf{u}_{i}$, which is represented by a phase rotation between $\mathbf{y}_{i}^\prime$ and $\mathbf{u}_{i}$ in the case of $D=2$, as can seen in Fig.~\ref{fig:multidimensional_representation}.
	More details about how the mapping function works for our scheme can be seen in our following discourse.
	On the other hand, Bob also has to calculate the side information represented by the syndrome $\mathbf{s}$ used for assisting the decoding process. This side-information decoding is slightly different for different reconciliation schemes. To elaborate further, initially we assume that {LDPC} codes are adopted in the reconciliation scheme considered in this paper. However, it is not necessary to encode $\mathbf{b}$ using {LDPC} codes, where the side information $\mathbf{s}$ could be the  syndrome calculated from $\mathbf{b}$. But again, the side information is not necessarily constituted by the syndromes in other application scenarios. {\color{black}{For example, frozen bits are used as side information in polar code-based reconciliation schemes \cite{Zhao2018,Zhang2021polarRecon}}}. Then Bob publicly transmits both the mapping function $\mathbf{M}_{i}\left(\mathbf{y}_{i}^\prime,\mathbf{u}_{i}\right)$ and the syndrome $\mathbf{s}$ to Alice through the classical communication channel. The details of the mapping function calculation can be found in \cite{leverrier2008multidimensional} and are also shown in Appendix~\ref{Mapping}.
	\item Alice then applies the same mapping function to her normalized segment $\mathbf{x}_{i}^\prime$ in order to map the Gaussian variables to $\widetilde{\mathbf{u}}_{i}=\mathbf{M}_{i}\left(\mathbf{y}_{i}^\prime, \mathbf{u}_{i}\right) \mathbf{x}_{i}^\prime$, which is actually the noisy version of $\mathbf{u}_{i}$. Hence, the difference between the variable $\mathbf{u}_{i}$ and its noisy version $\widetilde{\mathbf{u}}_{i}$ can reflect the quality of the QuC. Hence it may be exploited for eavesdropping detection.
	\item After the mapping operation harnessed for each segment at Alice's side,  she then concatenates all the segments into a sequence $\widetilde{\mathbf{u}}=\left[\widetilde{\mathbf{u}}_1;\widetilde{\mathbf{u}}_2;...;\widetilde{\mathbf{u}}_I\right]$ having the length of $N$.
	Furthermore, the sequence  $\widetilde{\mathbf{u}}$ is turned into $\widetilde{\mathbf{u}}^\prime$ after deinterleaving. She finally carries out the decoding of $\widetilde{\mathbf{u}}^\prime$ with the aid of the syndrome $\mathbf{s}$ calculated by Bob and obtains the secret key $\hat{\mathbf{b}}$.
\end{enumerate}

\begin{figure}[!htb]
	\centering
	\includegraphics[width=\linewidth]{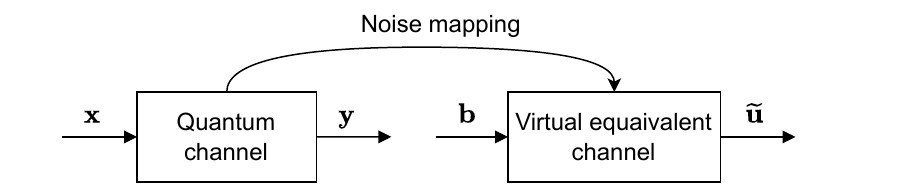}
	\caption{The relationship between the QuC and the virtual equivalent channel.}
	\label{fig:virutal_quantum_channel}
\end{figure}

In summary, the core idea of multidimensional reconciliation is to convert the noise in the QuC to the ClC via using the specific mapping functions $\mathbf{M}_{i}\left(\mathbf{y}_i^\prime,\mathbf{u}_i\right)$.
As a consequence, the noisy version  $\widetilde{\mathbf{u}}$ of $\mathbf{b}$ is obtained, hence the family of commonly used FEC schemes can be applied to CV-QKD.
More specifically, Fig.~\ref{fig:multidimensional_representation} demonstrates this conversion process from three different dimensionalities\footnote{As stated that the dimensionality of multidimensional reconciliation can be chosen to be 1, 2, 4 and 8. Here for convenience we exemplify this process via using visible 1, 2 and 3 dimensionalities.}, that are $D$=1, 2, 3, respectively.
In Fig.~\ref{fig:1dimensional_representation}, the noisy version $\widetilde{\mathbf{u}}$ of $\mathbf{u}=+1$ can be obtained based on the proportion of $\mathbf{y}$ to $\mathbf{x}$ in a 1-dimensional case.
Note that the values of $\mathbf{x}$ and $\mathbf{y}$ are not normalized in the 1-dimensional case.
As for the 2-dimensional case of Fig.~\ref{fig:2dimensional_representation},  the normalized vectors $\mathbf{y}^\prime$ and $\mathbf{x}^\prime$ are on the unit-circle, and we have $\mathbf{u}=\left[\frac{1}{\sqrt2},\frac{1}{\sqrt2}\right]^T$. Firstly, Bob calculates the mapping function between $\mathbf{y^\prime}$ and $\mathbf{x^\prime}$, corresponding to $\alpha$, which physically represents the phase rotation operation.
After Alice receives the mapping function, she uses it to get the noisy version $\widetilde{\mathbf{u}}$ of $\mathbf{u}$ by rotating $\mathbf{x}^\prime$ with the same angle $\alpha$.
Similarly, for the 3-dimensional case seen in Fig.~\ref{fig:3dimensional_representation}, the mapping function can be calculated based on $\mathbf{y}^\prime$ and $\mathbf{u}$ on the surface of the unit-radius sphere. Then the noisy version of $\mathbf{u}$, namely $\widetilde{\mathbf{u}}$ can be obtained by applying the same mapping function to $\mathbf{x}^\prime$.
As for how strong the noise is, it depends on the quality of the QuC, which is modelled by a virtual equivalent binary-input AWGN (BI-AWGN) ClC characterized in Fig. \ref{fig:virutal_quantum_channel}. This is  reminiscent of classical modulation and transmission through the {AWGN} channel \cite{leverrier2008multidimensional}. 
After that, FEC decoding can be applied and finally the reconciled key is generated.

To elaborate a little further, the 2-dimensional reconciliation of a segment is exemplified to illustrate this process. Firstly, after Alice and Bob finish their quantum-domain transmission and detection, sifting and parameter estimation, as well as normalization, they have two sequences, which are $\mathbf{x}_1^\prime=[0.8865, -0.4626]^T$, $\mathbf{y}_1^\prime=[0.9748, -0.229]^T$. Let us assume that the random bit stream after interleaving at Bob's side is $\mathbf{b}_1^\prime=[0, 0]^T$ along with the corresponding $\mathbf{u}_1=[0.7071, 0.7071]^T$. Then the resultant mapping matrix can be calculated as $	\mathbf{M_1}\left(\mathbf{y}_1^\prime,\mathbf{u}_1\right)=\left[\begin{array}{ll}
	-0.7618& -0.6479\\
	0.6479& -0.7618 \\
\end{array}\right]$,
where $\mathbf{M}_{1}\left(\mathbf{y}_1^\prime,\mathbf{u}_1\right)=\sum\nolimits_{d=1}^2 {\alpha}_1^d \mathbf{A}_{2}^d$.
Note that the pair of orthogonal matrices used in this 2-dimensional scheme are $\mathbf{A}_2^1=\begin{bmatrix}
	1 &0  \\
	0 &1 \\
\end{bmatrix}$, $\mathbf{A}_2^2=\begin{bmatrix}
	0 &-1  \\
	1 &0\\
\end{bmatrix}$. Furthermore, ${\alpha}_1^d$ is the specific element of
$\bm{\alpha}_1\left(\mathbf{y}_1^\prime,\mathbf{u}_1\right)=\left(\mathbf{A}_2^1\mathbf{y}_{1}^\prime,\mathbf{A}_2^2\mathbf{y}_{1}^\prime\right)^T\cdot\mathbf{u}_1$, which is the coordinate of the vector $\mathbf{u}_1$ under the orthonormal basis $\left(\mathbf{A}_2^1\mathbf{y}_{1}^\prime,\mathbf{A}_2^2\mathbf{y}_{1}^\prime\right)$ \cite{leverrier2008multidimensional}. 
Based on this, the sequence $\widetilde{\mathbf{u}}_1$ at Alice's side after data mapping becomes $\widetilde{\mathbf{u}}_1=\mathbf{M}\left(\mathbf{y}_1^\prime,\mathbf{u}_1\right)\mathbf{x}_1^\prime=[0.8632, 0.5049]^T$, which is a noisy version of $\mathbf{u}_1$. Furthermore, the noise in $\widetilde{\mathbf{u}}_1$ is capable of reflecting the noise level of the quantum transmission between Alice and Bob.
Therefore, in our ensuing discourse, the QuC is modelled by an equivalent BI-AWGN ClC.
\subsection{LDPC code}
{LDPC} codes constitute a class of linear block codes defined by a sparse {parity-check matrix (PCM)} $\mathbf{H}$ of size $(N-K) \times N, K\leq N$, where $N$ is the number of columns in $\mathbf{H}$ and it is also known as the block length, while $(N-K)$ is the rank of $\mathbf{H}$. Hence, the code rate is $R=K/N$.
A $[N, K]$-regular {LDPC} code is defined as the null space of a sparse {PCM}, where each row of $\mathbf{H}$ contains exactly $d_c$ ones, which is also called the degree $d_c$ of check nodes (CNs) . Each column of  $\mathbf{H}$ contains exactly $d_v$ ones, which is also called the degree $d_v$ of variable nodes (VNs) . Both the degrees of CNs and VNs are small compared to the number of rows in $\mathbf{H}$.
An {LDPC} code is classified as being irregular if the row weight $d_c$ and column weight $d_v$ are not constant.
It is often helpful to use the so-called Tanner graph to represent the {PCM} $\mathbf{H}$ \cite{Tanner1981}.
In the Tanner graph representation, there are two types of nodes, which are the {VNs} (or code-bit nodes) and {CNs} (or constraint nodes), respectively.
If an element of $\mathbf{H}_{i,j}$ is equal to one, then {CN} $i$ denoted as $c_i$ is connected by an edge to {VN} $j$ denoted as $v_j$ in the Tanner graph. Otherwise, there is no  connection between them. The notion of degree distribution is used for characterizing the check and variable node degrees \cite{Luby2001}.
\begin{figure}[tbp]
	\centering	\includegraphics[width=\linewidth]{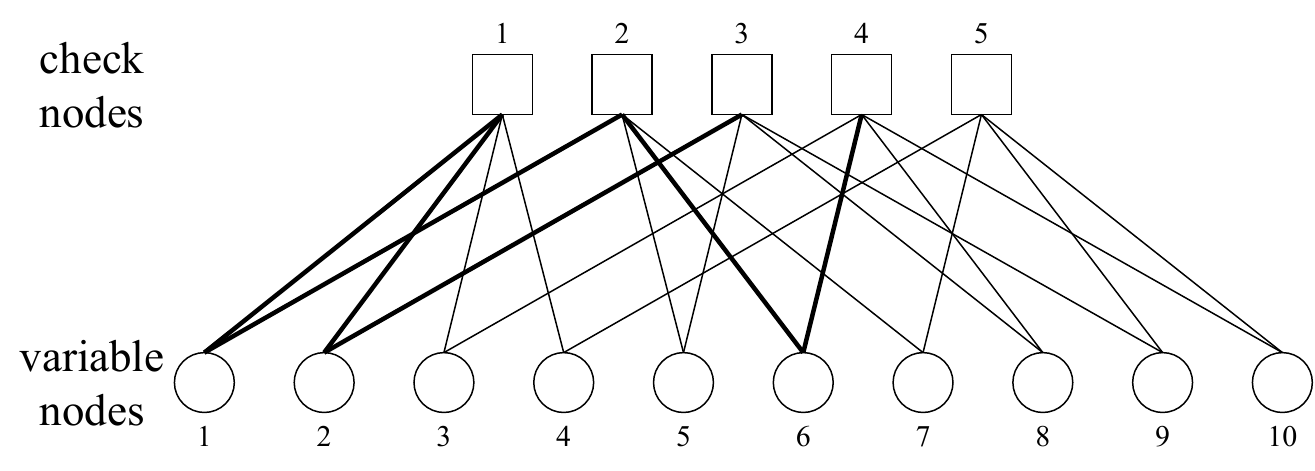}
	\caption{The Tanner graph for the code given in Eq. (\ref{exapmle_Tanner_graph})}.
	\captionsetup{format=hang}
	\label{model:Tanner_graph}
\end{figure}
For example, as shown in Fig.~\ref{model:Tanner_graph}, for the first {VN}, there are two edge connections seen in bold lines with the first and second {CN}.
In a similar fashion, the second {VN} is connected with the first and third {CN}.
The corresponding {PCM} $\mathbf{H}$ is formulated as
\begin{equation}\label{exapmle_Tanner_graph}
	\mathbf{H}=\left[\begin{array}{llllllllll}
		1 & 1 & 1 & 1 & 0 & 0 & 0 & 0 & 0 & 0 \\
		1 & 0 & 0 & 0 & 1 & 1 & 1 & 0 & 0 & 0 \\
		0 & 1 & 0 & 0 & 1 & 0 & 0 & 1 & 1 & 0 \\
		0 & 0 & 1 & 0 & 0 & 1 & 0 & 1 & 0 & 1 \\
		0 & 0 & 0 & 1 & 0 & 0 & 1 & 0 & 1 & 1
	\end{array}\right]_{10 \times 5},
\end{equation}
which is a $[10,5]$-regular {LDPC} code having the code length of $N=10$ and code rate of $R=0.5$, the row weight is $d_c=4$ and the column weight is $d_v=2$.
The notation of $[N, K]$-regular {LDPC} code used from now on to represent regular {LDPC} codes having a code length of $N$, and information length of $K$.

{LDPC} decoding is popularly performed using the BP algorithm \cite{ryan2009channel}, which is an iterative message-passing algorithm commonly used for inference based on graphical models such  as factor graphs \cite{Kschischang2001FactorGA}.
In the context of {LDPC} codes, the decoding procedure attempts to find  a valid codeword by iteratively exchanging the probabilistic information represented by the log-likelihood ratio (LLR) between the {CN} and {VN} along the edges of the Tanner graph until the parity-check condition is satisfied or  the maximum affordable number of iterations is reached.
More explicitly, we modify the classic Gallager SPA \cite{ryan2009channel} for QKD systems, as seen in Algorithm \ref{alg:alg1}, where both the codeword transmitted through the QuC and the side information constituted by the syndrome transmitted through the authenticated ClC are the  inputs of the modified SPA.
%There are various kinds of message-passing decoding algorithm, hence, we mainly introduce the most classical Gallager {SPA} as below \cite{ryan2009channel}, which is the most common variant of belief propagation algorithm.
%\begin{algorithm}
%	\caption{An algorithm with caption}\label{alg:cap}
%	\begin{algorithmic}
	%		\Require $n \geq 0$
	%		\Ensure $y = x^n$
	%		\State $y \gets 1$
	%		\State $X \gets x$
	%		\State $N \gets n$
	%		\While{$N \neq 0$}
	%		\If{$N$ is even}
	%		\State $X \gets X \times X$
	%		\State $N \gets \frac{N}{2}$  \Comment{This is a comment}
	%		\ElsIf{$N$ is odd}
	%		\State $y \gets y \times X$
	%		\State $N \gets N - 1$
	%		\EndIf
	%		\EndWhile
	%	\end{algorithmic}
%\end{algorithm}
\begin{algorithm}
	\caption{The  Modified  Sum-Product Algorithm of Gallager}\label{alg:alg1}
	\begin{algorithmic}[1] 
		\STATE \textbf{Initialization:} Initialize {LLR} at each VN, $v=1,2,..., n$ for the appropriate channel model. Then, for all $i, j$ for which
		$h_{i,j}=1$, set $L_{v\rightarrow c}^l=L_{v\rightarrow c}^0$.
		\STATE \textbf{CN update:} Compute outgoing CN messages $L_{c \rightarrow v}$ for each CN using \\
		%		\begin{equation*}
			%		\mu_{c v}^l=(-1)^{S_c} \cdot 2 \tanh ^{-1}\left(\prod_{v^{\prime} \in V_c \backslash v} \tanh \left(\frac{\lambda_{v^{\prime} c}^{l-1}}{2}\right)\right),
			%		\end{equation*}\\
		\begin{center}
			$	L_{c \rightarrow v}^t=(-1)^{\mathbf{s}_B(c)} \cdot 2 \tanh ^{-1}\left(\prod_{v^{\prime} \in V_c \backslash v} \tanh \left(\frac{L_{v^{\prime} \rightarrow  c}^{t-1}}{2}\right)\right)$,
		\end{center}
		and then transmit to the {VN}.
		\STATE \hspace{0.0cm}\textbf{VN update:} Compute outgoing {VN} messages $L_{ v \rightarrow c }$  for each {VN} using\\
		\begin{center}
			$
			L_{v \rightarrow c}^t= \begin{cases}L_{v \rightarrow c}^0 \\
				L_{v \rightarrow c}^0+\sum_{c^{\prime} \in C_v \backslash c} L_{c^{\prime} \rightarrow v}^t & \text { if } t \geq 1\end{cases} $,
		\end{center}
		and then transmit to the {CN}.
		\STATE \hspace{0.0cm}\textbf{LLR total:} For $v=1,2,...,n$ compute
		\begin{center}
			$L_{v}^{total}=L_{v \rightarrow c}^0+\sum_{c^{\prime} \in C_v} L_{c^{\prime} \rightarrow v}^t$.
		\end{center}
		\STATE \textbf{Stopping criterion:} Hard decision and early termination check:
		\begin{center}
			$
			\hat{C}_v^{(t)} = \begin{cases}0, & L_v^{total} \geq 0 \\ 1, & \text { otherwise }\end{cases}.
			$
		\end{center}
		If $\hat{\mathbf{C}} \mathbf{H}^T={\mathbf{s}_B} $ or the number of affordable iterations reaches the maximum limit, stop; else, go to step 2.
	\end{algorithmic}
	\label{alg1}
\end{algorithm}

In Algorithm \ref{alg:alg1}, Step 1 prepares the {LLR} input values at each {VN}. All {VN}-to-{CN} messages arriving from {VN} $v$ to {CN} $c$ are initialized to the received {LLR}, denoted as $L_{v\rightarrow c}^0$ at the output of the channel before the first message-passing iteration. Then, Step 2 to Step 5 illustrate the process of finding the most likely codeword by iterative soft information exchange between {CN} and {VN}, until the syndrome defined by $\hat{\mathbf{C}} \mathbf{H}^T$ becomes zero, or the maximum affordable number of decoding iterations is reached.
To elaborate further, in Step 2, $L_{c \rightarrow v}^t$ is the message arriving from {CN} to {VN} in iteration $t$, and $C_v \backslash c$ denotes all the {CNs} connected to {VN} $v$, except for {CN} $c$.
In Step 3, $L_{v \rightarrow c}^t$ is the message coming from {VN} to {CN} in iteration $t$, and $V_c \backslash v$ is the set of  {VNs} connected to  {CN} $c$, except for {VN} $v$.

In contrast to the conventional {SPA} decoding algorithm, both the contaminated codeword received from the QuC and the side information received from the ClC are entered into the modified SPA of Algorithm \ref{alg:alg1}.
Normally, the side information refers to the  syndrome calculated by Bob in the context of {LDPC} codes.
Hence, the {SPA} decoding algorithm has to be modified.
Specifically, we have to change the {CN} update operation, which is Step 2 in Algorithm \ref{alg1}, based on the syndrome $\mathbf{s}_B$ from Bob  received by Alice.
The modified {CN} update operation can be formulated as \cite{mani2021error}
\begin{equation}\label{CN_with_syndrome}
	L_{c \rightarrow v}^t=(-1)^{\mathbf{s}_B(c)} \cdot 2 \tanh ^{-1}\left(\prod_{v^{\prime} \in V_c \backslash v} \tanh \left(\frac{L_{v^{\prime} \rightarrow  c}^{t-1}}{2}\right)\right),
\end{equation}
where $\mathbf{s}_B(c) \in \{0,1\}$ represents the parity value at index $c$.
It is plausible that if the syndrome is  $\mathbf{s}_B(c)=0$, the {CN} update operation remains the same as that of the conventional SPA. Otherwise, for $\mathbf{s}_B(c)=1$, the {CN} update operation would flip the sign of the outgoing messages.

\subsection{The relationship between FEC codes and segment vectors $\widetilde{\mathbf{u}}_i$ after mapping}
\begin{figure*}[!htbp]
	\centering	\includegraphics[width=0.8\linewidth]{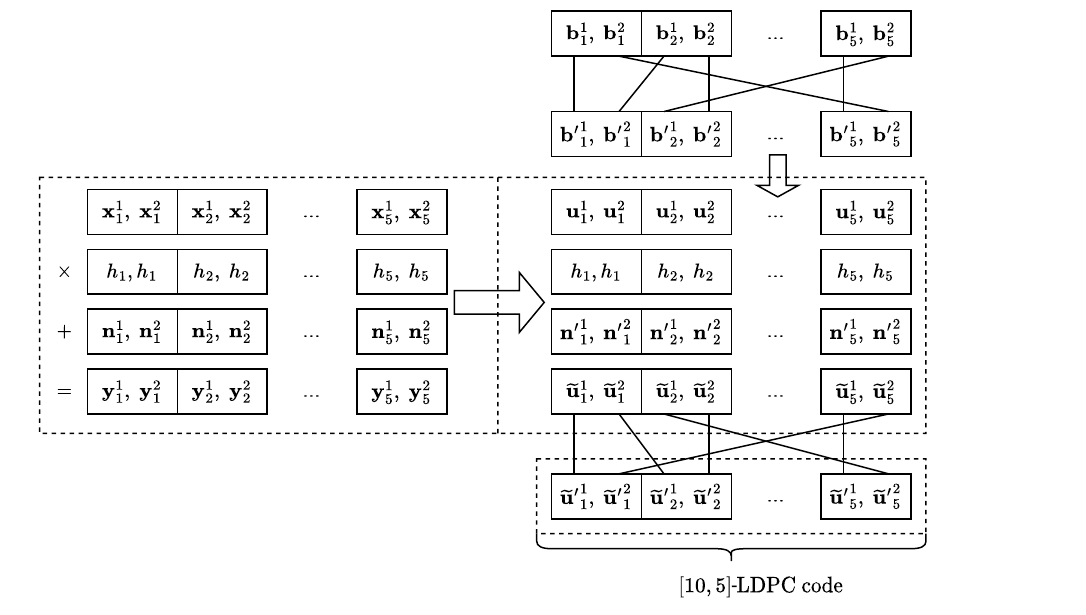}
	\caption{The relationship between FEC codes and segmented vectors. Note that a [10,5] LDPC code is applied and 2-dimensional reconciliation is adopted. Correspondingly, $N=10$, $D=2$, and $I=N/D=5$.}
	\captionsetup{format=hang}
	\label{model:mapping_process}
\end{figure*}
Once an equivalent ClC has been setup for the QuC, an FEC scheme is needed to proceed. Therefore, in this section, we aim for clarifying how to connect the segment vectors $\widetilde{\mathbf{u}}_i$ after mapping with FEC codes.
%\footnote{Note that, even though the QuC is only considered as an AWGN channel in our paper, we try to clarify this relationship for a QuC with fading as well.}

In Fig.~\ref{model:mapping_process}, we consider 2-dimensional reconciliation and a [10,5] LDPC code. The PCM of Eq.~(\ref{exapmle_Tanner_graph}), is used for illustrating the relationship between the FEC codes and segmented vectors $\mathbf{u}_i$.
In Fig.~\ref{model:mapping_process}, the dashed box at the left represents the relationship between the pair of Gaussian sequences, namely $\mathbf{y}$ and $\mathbf{x}$.
Since we consider a [10,5] LDPC code and a 2-dimensional reconciliation scheme ($D=2$), each of the pair of Gaussian sequences of length $N=10$, is divided into $I=N/D=5$ segments, yielding $\mathbf{x}=[\mathbf{x}_1;\mathbf{x}_2;\mathbf{x}_3;\mathbf{x}_4;\mathbf{x}_5]$ and  $\mathbf{y}=[\mathbf{y}_1;\mathbf{y}_2;\mathbf{y}_3;\mathbf{y}_4;\mathbf{y}_5]$, each of which contains 2 Gaussian elements, i.e. $\mathbf{x}_i=[\mathbf{x}_i^1,\mathbf{x}_i^2]$ for $i=1,2,...,5$ and $\mathbf{y}_i=[\mathbf{y}_i^1,\mathbf{y}_i^2]$ for $i=1,2,...,5$.
Furthermore, it is assumed that within each segment the channel's fading coefficients remain constant. For example, we have  $\mathbf{h}_1=[\mathbf{h}_1^1;\mathbf{h}_1^2]=[h_1,h_1]$.
On the other hand, the bit stream $\mathbf{b}$ generated by Bob's QRNG seen in Fig.~\ref{reconciliation_protocol_in_QKD_whole}(b) is correspondingly divided into 5 segments, i.e. $\mathbf{b}=[\mathbf{b}_1;\mathbf{b}_2;\mathbf{b}_3;\mathbf{b}_4;\mathbf{b}_5]$, each of which contains two elements, i.e. $\mathbf{b}_i=[\mathbf{b}_i^1,\mathbf{b}_i^2]$ for $i=1,2,...,5$.
After interleaving, the new bit stream $\mathbf{b}^\prime$ is obtained, which is also partitioned into 5 segments, i.e. $\mathbf{b}^\prime=[\mathbf{b}_1^\prime;\mathbf{b}_2^\prime;\mathbf{b}_3^\prime;\mathbf{b}_4^\prime;\mathbf{b}_5^\prime]$, where $\mathbf{b}_i^\prime=[{\mathbf{b}_i^\prime}^1,{\mathbf{b}_i^\prime}^2]$ for $i=1,2,...,5$.
In light of this, the affect of the channel's fading coefficient $\mathbf{h}=[\mathbf{h}_1;\mathbf{h}_2;\mathbf{h}_3;\mathbf{h}_4;\mathbf{h}_5]$ and noise $\mathbf{n}=[\mathbf{n}_1;\mathbf{n}_2;\mathbf{n}_3;\mathbf{n}_4;\mathbf{n}_5]$ in the QuC are used for representing the relationship between the modulated sequences $\mathbf{u}=[\mathbf{u}_1;\mathbf{u}_2;\mathbf{u}_3;\mathbf{u}_4;\mathbf{u}_5]$ based on $\mathbf{b^\prime}$ and $\widetilde{\mathbf{u}}=[\widetilde{\mathbf{u}}_1;\widetilde{\mathbf{u}}_2;\widetilde{\mathbf{u}}_3;\widetilde{\mathbf{u}}_4;\widetilde{\mathbf{u}}_5]$.
Note that it is assumed that the fading coefficients are known at both sides, and the noise variances of $\mathbf{n}=[\mathbf{n}_1;\mathbf{n}_2;\mathbf{n}_3;\mathbf{n}_4;\mathbf{n}_5]$ and $\mathbf{n}^\prime=[\mathbf{n}_1^\prime;\mathbf{n}_2^\prime;\mathbf{n}_3^\prime;\mathbf{n}_4^\prime;\mathbf{n}_5^\prime]$ are the same even thought the exact value of noise $\mathbf{n}^\prime$ is not the same as $\mathbf{n}$ in the QuC.
After deinterleaving, a reordered sequence $\widetilde{\mathbf{u}}^\prime=[\widetilde{\mathbf{u}}^\prime_1;\widetilde{\mathbf{u}}^\prime_2;\widetilde{\mathbf{u}}^\prime_3;\widetilde{\mathbf{u}}^\prime_4;\widetilde{\mathbf{u}}^\prime_5]$ of $\widetilde{\mathbf{u}}$ is derived, which represents the corrupted sequences of $\mathbf{b}$.
Therefore, the sequence $\widetilde{\mathbf{u}}^\prime$ of length 10 will be fed into the [10,5] LDPC decoder.

\section{System descriptions and comparisons}\label{system comparison}
In this section, our new codeword-based reconciliation system will be proposed, following the critical appraisal of the state-of-the-art. 
More explicitly, four reconciliation systems will be presented in this section. In a nutshell,
\begin{enumerate}
\item \textbf{System A} represents the conventional LDPC-coded reconciliation scheme relying on the idealistic simplifying assumption that the ClC used for syndrome transmission is error-free.
	\item \textbf{System B} takes into account the fading and noise effects of the ClC, where a separate LDPC code is required for both the QuC and the ClC. Note that System B is a practical version of System A. 
	{\color{black}{\item \textbf{System C} is proposed to demonstrate that the bit-difference vector-based side information can play the same role as the syndrome of Systems A and B. Hence System C has the same performance as System A and System B.}}
	\item \textbf{System D} represents the proposed codeword-based reconciliation scheme suitable for any arbitrary FEC code. Hence the family of powerful IRCCs can also be applied to achieve a near-capacity performance for both the QuC and ClC.
\end{enumerate}
Note that the following reconciliation systems mainly focus on the details of the reconciliation step within the QKD protocol.
More specifically, the BI-AWGN equivalent QuC of Fig.~\ref{fig:virutal_quantum_channel} is  used here for the description of the reconciliation post-processing step.

\begin{figure*}[htbp] 
	\begin{center}
		\includegraphics[width=0.7\linewidth]{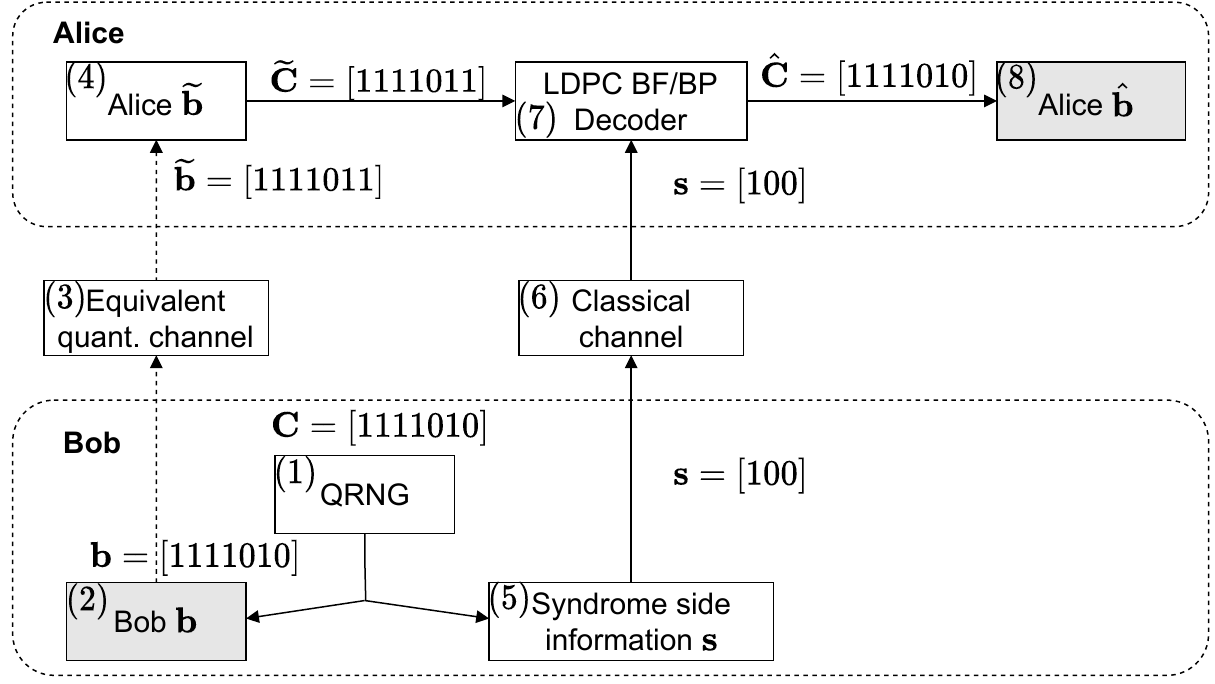}
		\caption{\textit{System} $A$ - \textbf{the ideal LDPC-coded syndrome-based reconciliation scheme} in the CV-QKD system relying on the  BF/BP decoding algorithm. Note that, the dashed arrow represents the
			bit stream sent from Bob to Alice through the equivalent QuCs as illustrated in Section~\ref{system descriptions}-\ref{equivlanet_setup}.}
		\label{fig:exame_LDPC_reconciliation_system_BF}
	\end{center}
\end{figure*}

\subsection{System A: the ideal syndrome-based LDPC-coded scheme}
\textbf{\textit{System A}}: The first reconciliation system shown in Figure~\ref{fig:exame_LDPC_reconciliation_system_BF} is the BF/BP decoding algorithm based  LDPC-coded CV-QKD reconciliation scheme, where the ClC used for syndrome transmission is assumed to be error-free. The algorithmic steps are described as follows.
\begin{enumerate}
	\item[(a)] 
	Bob randomly generates a bit stream $\mathbf{C}$ using a QRNG, and we view this as the initial raw key $\mathbf{b}$ at his side.
	Note that, the QRNG generates classical random numbers.
	The length of this is determined by the codeword length of the predefined PCM $\mathbf{H}$.
	The PCM is known at both sides. Note that, the bit stream $\mathbf{b}$ at Bob's side does not have to be a legitimate codeword, because the final objective is to obtain a reconciled key.
	More specifically, in the reverse reconciliation scheme, the bit stream generated at Bob's side is the reference key, and Alice has to acquire this as the final key.
	Let us consider the single-error correcting [7,4,1]  Bose-Chaudhuri-Hocquenghem (BCH) code as our rudimentary example, and assume that the bit stream generated by the QRNG in block (1) of Fig.~\ref{fig:exame_LDPC_reconciliation_system_BF} is $\mathbf{C}=[1111010]$. Then Bob treats this random bit stream as the initial key $\mathbf{b}=[1111010]$ in block (2) of Fig.~\ref{fig:exame_LDPC_reconciliation_system_BF}.
	
	\item[(b)]  Bob transmits this bit stream $\mathbf{b}=[1111010]$ through a QuC to Alice, which is modelled by the equivalent ClC constructed in Fig.~\ref{fig:virutal_quantum_channel} and represented by block (3) of Fig.~\ref{fig:exame_LDPC_reconciliation_system_BF}.
	The channel-contaminated sequence received by Alice is denoted by $\widetilde{\mathbf{b}}=[1111011]$ in block (4), which is corrupted in the last bit position.
	\item[(c)]  Meanwhile, based on the QRNG output Bob calculates the syndrome, say $\mathbf{s}=[100]$ in block (5) and transmits it as side information to Alice through the authenticated ClC of block (6), which is assumed to be perfectly \textit{noiseless and error-free}.
	\item[(d)]  Alice takes the bit stream $\widetilde{\mathbf{b}}$ inferred at the output of the QuC, which may or may not be a legitimate codeword, and forwards it as namely $\widetilde{\mathbf{C}}=[1111011]$ to the decoder. Decoding is carried out by the corresponding FEC decoder  with the aid of the syndrome bits she received  through the ClC (6) and gets the decoded result of $\hat{\mathbf{C}}=[1111010]$ at the output of block (7). Based on this, Alice gets the decoded codeword as the final reconciled key, which is $\mathbf{\hat{b}}=[1111010]$ shown in block (8). Observe that this is the same as Bob's bit stream $\mathbf{b}$, provided that there are no decoding errors.
	This is the case, if the QuC inflicts no more than a single error, since the [7,4,1] code can only correct a single error. It is important to mention here that if the classical syndrome-transmission channel inflicts errors, this would results in catastrophic corruption of the QuCs' output. This issue will be addressed by System B.
	%Note that System A is an LDPC-coded CV-QKD reconciliation scheme, and the [7,4,1] BCH code is just an example to introduce this scheme. 	
\end{enumerate}

\begin{figure*}[htbp] 
	\begin{center}
		\includegraphics[width=0.7\linewidth]{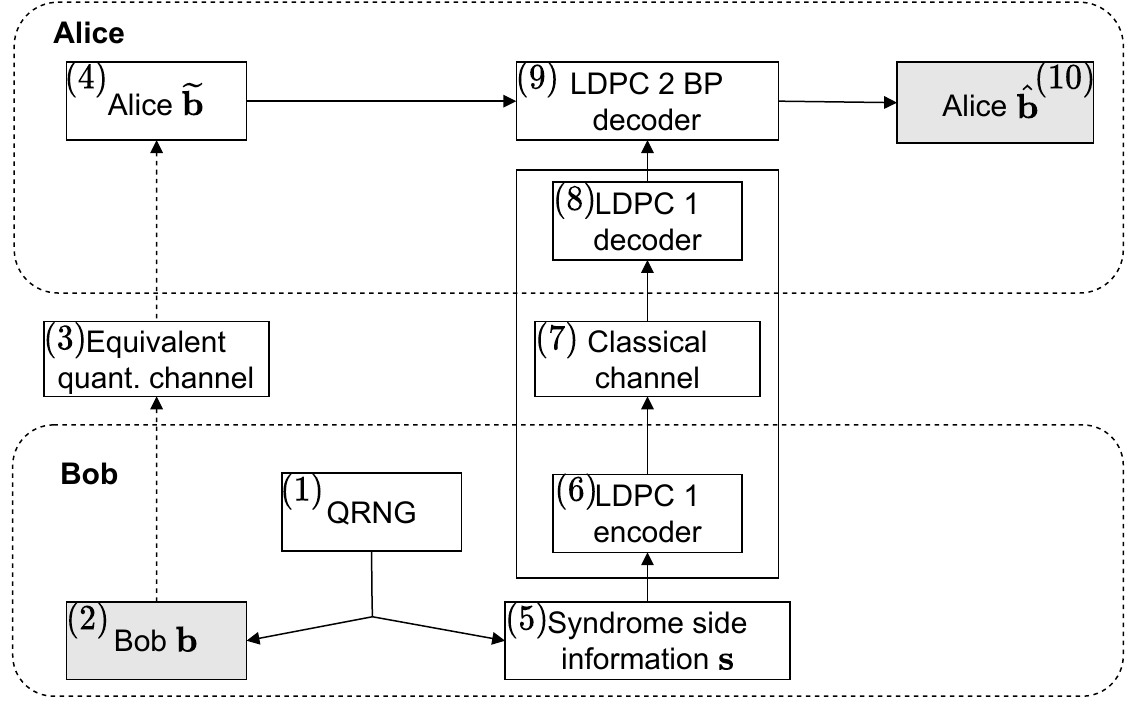}
		\caption{\textit{System} $B$ - \textbf{the practical LDPC-coded syndrome-based reconciliation scheme} in the CV-QKD system with BP decoding algorithm.
			Compared to System A, System B no longer assumes that the ClC is error-free, where both ClC and QuC require data protection by error correction.}
		\label{fig:exame_LDPC_reconciliation_system_BP}
	\end{center}
\end{figure*}

\subsection{System B: the practical syndrome-based LDPC-coded scheme}
\textbf{\textit{System B}}: Following the above rudimentary BCH-coded example to introduce how System A works, let us now detail a prctical LDPC code based scheme. System B of Fig.~\ref{fig:exame_LDPC_reconciliation_system_BP} represents the BP decoding algorithm based CV-QKD reconciliation scheme. %It is shown that most steps are similar to System A.
%Hence, similar steps are briefly described whereas some different blocks will be highlighted as follows.
In contrast to System A, System B no longer assumes that the ClC is error-free. Hence both the classical and the QuC require error correction. Let us consider a [1024,512] LDPC code as our example to introduce System B.
More explicitly, the operational steps of System B are
\begin{enumerate}
	\item[(a)]   Bob randomly generates a 1024-bit stream using the QRNG of Fig.~\ref{fig:exame_LDPC_reconciliation_system_BP}, and views this as the initial key $\mathbf{b}$ at his side. Note that, the QRNG generates classical random numbers. Again, the length of this is determined by the codeword length of the predefined LDPC PCM $\mathbf{H}$, which is known to both sides.
	\item[(b)]   Bob transmits this bit stream $\mathbf{b}$ through the equivalent QuC to Alice, who receives the bit stream as $\widetilde{\mathbf{b}}$.
	\item[(c)]  Meanwhile, Bob calculates the syndrome based on the QRNG output - namely $\mathbf{b}$ - as the side information $\mathbf{s}$ and transmits it to Alice through the authenticated ClC protected by the LDPC encoder in block (6) of Fig.~\ref{fig:exame_LDPC_reconciliation_system_BP}.
	\textit{Note that}, the rectangular frame shown in Fig.~\ref{fig:exame_LDPC_reconciliation_system_BP} that encompasses blocks (6)-(8), constitutes a separate FEC-aided data protection for the ClC, which relies on the LDPC 1 code.
	%To elaborate further, let us consider a [1024, 512] LDPC code as our example. 
	The dimensionality of the PCM of such LDPC codes in our example is $512\times1024$, and hence the syndrome $\mathbf{s}=\mathbf{H}\cdot\mathbf{b}$ calculated from Bob has the length of 512 bits.
	After the FEC scheme applied to the syndrome protection, which is protected by another [1024, 512] LDPC code, the encoded syndrome has the length of 1024 bits.
	Then, after being decoded at Alice's side by the LDPC 1 decoder (8), the syndrome $\mathbf{s}$ having 512 bits is recovered.
	In the literature \cite{leverrier2008multidimensional,mani2021error,Mani21,Zhang2020,guo2020free,shirvanimoghaddam2016design,milicevic2017key,milicevic2018quasi,ai2018quantum,Zhou2021,ai2022optimised,bloch2005efficient}, the ClC is assumed to be \textit{noiseless and error-free}, but a realistic ClC tends to inflict both fading and noise. 
	Hence the ClC's LDPC 1 scheme of Fig.~\ref{fig:exame_LDPC_reconciliation_system_BP} may not be able to eliminate all errors imposed on the syndrome. Therefore, the performance of practical FEC schemes in the classical syndrome-transmission channel is taken into account in System B. 
	\item[(d)]  Then Alice carries out \textit{BP} decoding of the information received over the QuC with the aid of the syndrome bits to get $\mathbf{\hat{b}}$, as seen in block (9).
	%We should find a better way to distinguish them so as to get the legends consistent....(modify it later....)
\end{enumerate}
Note that, the syndrome-based scheme is limited to FEC codes that rely on syndromes, whereas other codes such as polar codes and CCs cannot be applied. Therefore, the bit-difference vector-based scheme (System C) is proposed to tackle this issue, which is described as follows. 

%\begin{figure*}[t] 
%	\begin{center}
%		\includegraphics[width=.8\linewidth]{figures/redesigned_reconciliation_system4_4.pdf}
%		\caption{\textit{System} $C^{-}$ - \textbf{the new LDPC-coded codeword-based reconciliation scheme} designed for CV-QKD systems and using the BP decoding algorithm.
%			%Note that, the basic (7, 4) Hamming code is utilized as an example to illustrate this scheme.
%		}
%		\label{fig:redesigned_LDPC_reconciliation_system}
%	\end{center}
%\end{figure*}
\begin{figure*}[t] 
	\begin{center}
		\includegraphics[width=.8\linewidth]{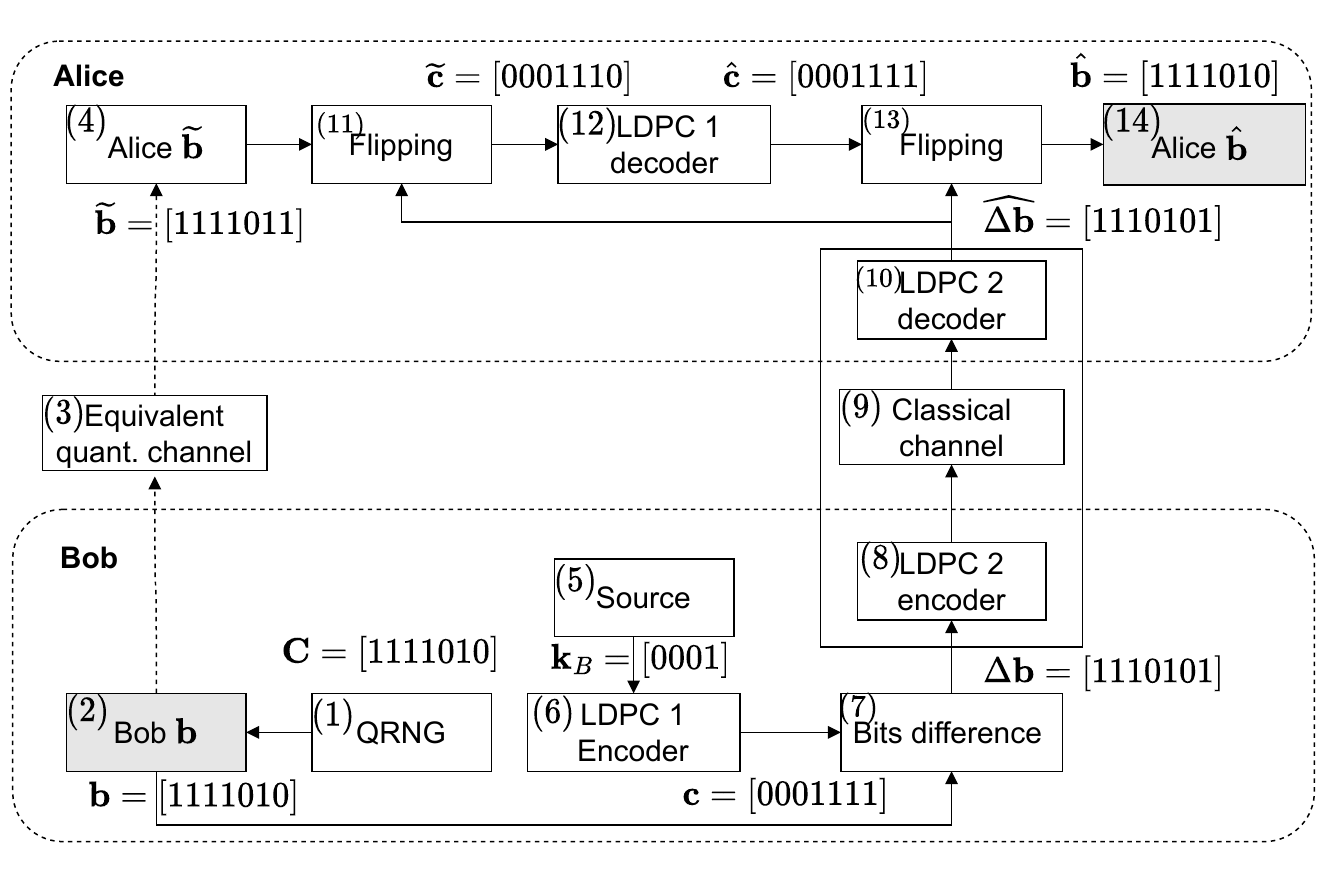}
		\caption{{\color{black}{\textit{System C} - \textbf{the LDPC-coded bit-difference vector-based reconciliation scheme} designed for CV-QKD systems and using the BP decoding algorithm.}}
			%Note that, the basic (7, 4) Hamming code is utilized as an example to illustrate this scheme.
		}
		\label{fig:redesigned_LDPC_reconciliation_systemX}
	\end{center}
\end{figure*}
\subsection{System C: the proposed bit-difference vector-based LDPC-coded scheme}
\textbf{\textit{System C}} of Figure~10, is our proposed scheme, where the final key generated by the QRNG is transmitted through the QuC and the syndromes of System B are replaced by the bit-difference vector. %\footnote{Note that, there are two different LDPC codes utilized in QuC and classical, respectively, which is the same case for System B. The one used in QuC is based on the NR LDPC codes with some fixed PCM $\mathbf{H}$ patterns, whereas the other one used in the ClC is based on randomly constructed PCM $\mathbf{H}$.}  
For convenience, both the QuC and the ClC may adopt the same kind of FEC codes, albeit they may have different length.
The corresponding steps are described as follows.
\begin{enumerate}
	\item[(a)] The functions of block (1) to (4) in Fig.~10 are the same as described in System A. Here, again a simple [7,4,1] BCH code is used as our rudimentary example. Specifically, the bit stream $\mathbf{b}=[1111010]$ may be obtained from the QRNG, which generates a bit stream C=[1111010], and it is transmitted from Bob to Alice through the QuC, resulting the corrupted bit stream $\widetilde{\mathbf{b}}=[1111011]$ at Alice's side. This has a single error in the last position.
	\item[(b)] In contrast to the way of calculating the syndrome in Systems A and B, a legitimate codeword $\mathbf{c}=[0001111]$ is required for deriving the bit-difference vector $\Delta\mathbf{b}=[1110101]$ based on blocks (5) to (7) in Fig.~10, where $\mathbf{k}_B=[0001]$ represents the corresponding random information bits used to obtain $\mathbf{c}$. 
	\item[(c)] Based on the received and protected bit-difference vector $\widehat{\Delta\mathbf{b}}=[1110101]$ at the output of block (10) in Fig.~10, Alice flips the bits of $\widetilde{\mathbf{b}}$ in those specific positions, where a logical 1 occurs in $\widehat{\Delta\mathbf{b}}$ at the output of block (3) in Fig.~10 to arrive at $\widetilde{\mathbf{c}}=[0001110]$  at the output of (11) before decoding.
	\item[(d)] Alice then decodes the  bit stream $\widetilde{\mathbf{c}}=[0001110]$ to arrive at $\hat{\mathbf{c}}=[0001111]$  after (12) to get the key $\hat{\mathbf{b}}=[1111010]$ at the output of block (13), which is ideally the same as $\mathbf{b}$ at Bob's side.
\end{enumerate}
Observe in Fig.~9 (System B) and Fig.~10 (System C) that there are two LDPC decoders at Alice's side. By contrast, there is merely a single LDPC encoder and a low-complexity syndrome calculation scheme at Bob's side in System B, while two LDPC encoders are required at Bob's side in System C. Since Alice has to perform computationally demanding LDPC decoding twice in order to infer the final key, this is not a balanced-complexity system\footnote{{\color{black}{Even though System C is not a balanced-complexity system, there are practical scenarios, where having a balanced complexity is not imperative, such as in ground station to unmanned aerial vehicle (UVA) quantum communication \cite{9663283,9656915}, etc.}}}. 
%On the other hand, System B is limited to the FEC scheme applied to the QuC transmission, since not every kind of FEC codes possess the syndrome property, for example the polar codes.
%Hence, it is not fair enough to compare system performances among different FEC schemes using the same reconciliation system.
{\color{black}{In light of these considerations, the new codeword-based reconciliation System D was proposed for arriving at a solution having a balanced-complexity, where Alice and Bob have a similar complexity, as required in device-to-device (D2D) systems for example \cite{wang2017uaka,7345413}, which is described as follows.}}
\subsection{System D: the proposed practically generic scheme}
\begin{figure*}[t] 
	\begin{center}
		\includegraphics[width=.8\linewidth]{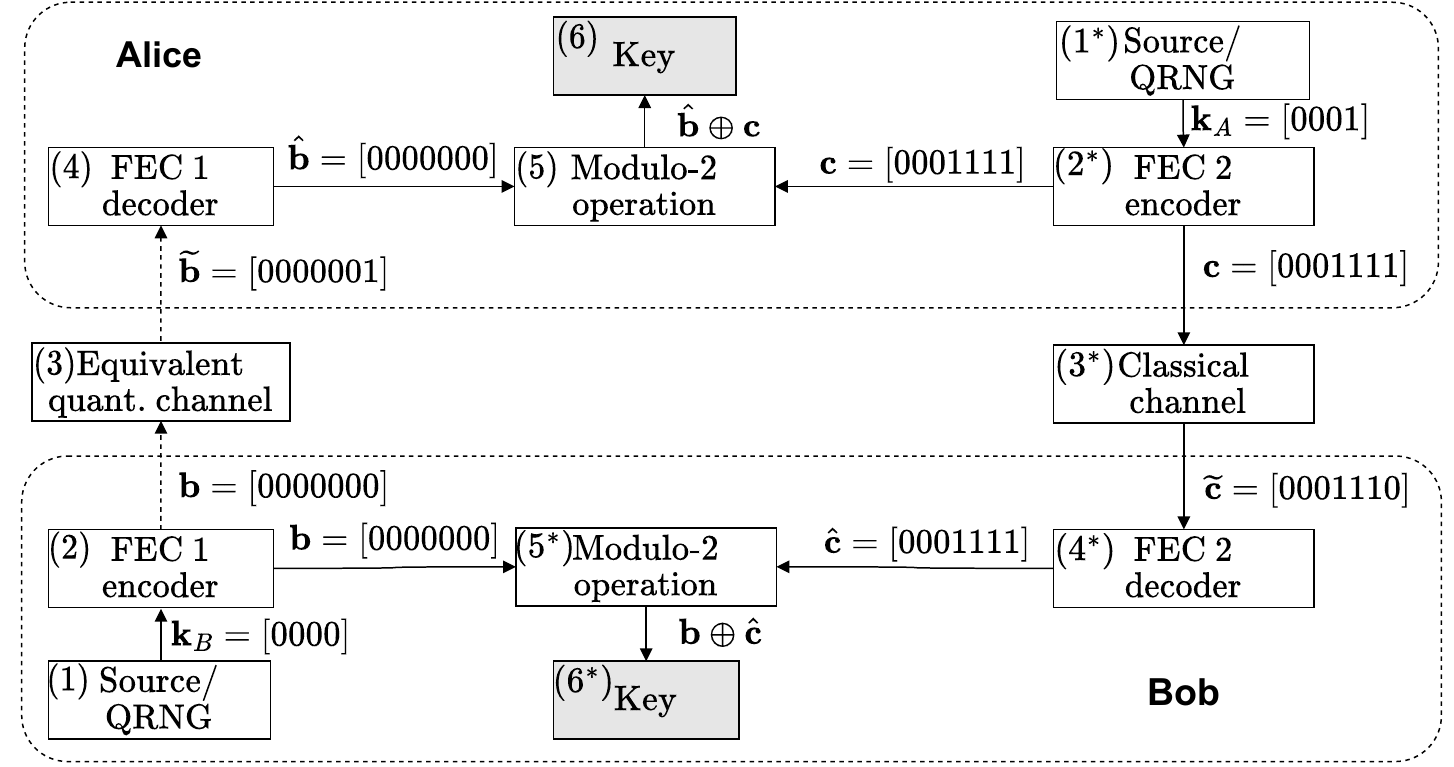}
		\caption{\textit{System D} - \textbf{the proposed practically generic reconciliation scheme} designed for CV-QKD systems.
			%Note that, the basic (7, 4) Hamming code is utilized as an example to illustrate this scheme.
		}
		\label{fig:redesigned_LDPC_reconciliation_system_generic}
	\end{center}
\end{figure*}

\textbf{\textit{System D}} of Figure~\ref{fig:redesigned_LDPC_reconciliation_system_generic}, is our proposed scheme that utilizes a pair of FEC codes %\footnote{Note that, there are two different LDPC codes utilized in QuC and classical, respectively, which is the same case for System B. The one used in QuC is based on the NR LDPC codes with some fixed PCM $\mathbf{H}$ patterns, whereas the other one used in the ClC is based on randomly constructed PCM $\mathbf{H}$.} 
to protect both the QuC and the classical authenticated channel. 
For convenience, both the QuC and the ClC may adopt the same kind of FEC codes.
The corresponding steps are described as follows.
\begin{enumerate}
	\item[(a)]  Both Bob and Alice generate a legitimate codeword based on a pair of predefined PCMs $\mathbf{H}_1$ and $\mathbf{H}_2$, which are $\mathbf{b}$ and $\mathbf{c}$. Consider again a simple [7,4,1] BCH code as our rudimentary example, where the pair of legitimate codewords are $\mathbf{b}=[0000000]$ and $\mathbf{c}=[0001111]$, respectively, as indicated in Fig.~\ref{fig:redesigned_LDPC_reconciliation_system_generic}. The corresponding uncoded information bits are for example $\mathbf{k}_B=[0000]$ and $\mathbf{k}_A=[0001]$, respectively.
	\item[(b)]  Bob transmits his legitimate codeword $\mathbf{b}$ through the QuC, which is modelled again by the equivalent ClC of Fig.~\ref{fig:virutal_quantum_channel}. On the other hand, Alice transmits her legitimate codeword $\mathbf{c}$ through the ClC, which may inflict errors. Note that, the 2 LDPC codes in Fig.~\ref{fig:redesigned_LDPC_reconciliation_system_generic} do not have to be exactly the same code, whose PCMs are the same. 
	However, for convenience, in our study, it is assumed that both QuC and ClC may adopt the same kind of FEC codes, which have exactly the same PCM.
	The codewords transmission over both the QuC and the ClC is independent.
	More specifically, the codeword $\mathbf{c}$ is transmitted the same as that in conventional wireless communication, whilst the codeword $\mathbf{b}$ is transmitted with the aid of the equivalent QuC of Fig.~\ref{fig:virutal_quantum_channel}, where the relationship between the Gaussian signals transmitted over the QuC and the random bit stream generated by QRNG is leveraged as can be seen in Fig.~\ref{reconciliation_protocol_in_QKD_whole}(b).
	\item[(c)]  Both Alice and Bob carry out LDPC BP decoding to get $\mathbf{\hat{b}}=[0000000]$ and $\mathbf{\hat{c}}=[0001111]$, respectively\footnote{{\color{black}{As for handling decoding failures, it is assumed to be identical to that in the conventional LDPC-based reconciliation scheme of \cite{milicevic2017key}, which is based on the classic cyclic redundancy check. Specifically, the system opts for discarding the sifted keys, if decoding failure occurs. Yet, a slight difference is that our codeword-based reconciliation needs two separate steps to check whether decoding is successful or not. We can only proceed to the next step when both parts are correct.}}}. 
	\item[(d)]  Furthermore, Modulo-2 operation is carried out at both sides to obtain the final key for both Alice and Bob, which are  $\mathbf{\hat{b}}\oplus\mathbf{c}$ and $\mathbf{b}\oplus\mathbf{\hat{c}}$, respectively.
\end{enumerate}
%\begin{figure*}[t] 
%	\begin{center}
%		\includegraphics[width=.8\linewidth]{figures/redesigned_reconciliation_system4_general4.pdf}
%		\caption{\textit{System} $C$ - \textbf{the proposed practically generic reconciliation scheme} designed for CV-QKD systems.
%			%Note that, the basic (7, 4) Hamming code is utilised as an example to illustrate this scheme.
%		}
%		\label{fig:redesigned_LDPC_reconciliation_system_generic}
%	\end{center}
%\end{figure*}

%As a further advance, a generically applicable System D is illustrated in Fig.~\ref{fig:redesigned_LDPC_reconciliation_system_generic}.
%The difference between System D and System C lies in the FEC schemes. In System C, the FEC schemes applied before transmission over the QuC are LDPC codes, since we aim for making comparisons between System B and System C using the same coding scheme. Then, a more generic reconciliation scheme is concerned for System D, which has no coding scheme limitation. Hence, the new reconciliation System D is more flexible than that of System C.
%For convenience, both the QuC and the ClC may adopt the same kind of FEC codes.
The proposed System D is summarized in Algorithm \ref{alg:description_system_D}.
\begin{algorithm}
	\caption{Description of \textit{System} \textit{D}}\label{alg:description_system_D}
	\begin{algorithmic}[1] 
		\STATE \textbf{Codeword generation:} \\Both Alice and Bob generate a legitimate codeword, which are $\mathbf{c}$ and $\mathbf{b}$.
		\STATE \textbf{Codeword transmission:} \\Bob transmits his legitimate codeword $\mathbf{b}$ through the equivalent QuC, which is the same process as in the System A, B and C. Meanwhile, Alice transmits her legitimate codeword $\mathbf{c}$ through the ClC.
		\STATE \hspace{0.0cm}\textbf{Decoding:} \\Both Alice and Bob carry out FEC decoding.
		\STATE \textbf{Modulo-2 operation:} \\Modulo-2 operation is implemented at both sides to obtain the final key for both Alice and Bob, which are  $\mathbf{\hat{b}}\oplus\mathbf{c}$ and $\mathbf{b}\oplus\mathbf{\hat{c}}$, respectively.
	\end{algorithmic}
\end{algorithm}
As a benefit of this design, first of all, the proposed codeword-based - rather than syndrome-based - QKD reconciliation scheme protects both the QuC and ClC. Secondly, the system has a similar complexity for both Alice and Bob, each of whom has a FEC encoder and a FEC decoder. Thirdly, System D makes QKD reconciliation compatible with a wide range of  FEC, including polar codes and the family of CCs. We will demonstrate in Section \ref{simulation} that this design allows us to achieve a near-capacity performance for both the QuC and for the ClC.

\subsection{Systems comparison}
{\color{black}{
%In summary, the comparison of these four different systems, i.e. System A, B, C and D, is as follows, which is summarized in Table \ref{table:systems_comparison}.
%More specifically, all these systems use the same equivalent QuC, but System A assumes the ClC is error-free while System B, C and D consider noises and fading in the ClC.
%Secondly, in the System A, B and C the final key is the key that is transmitted over the QuC from Bob to Alice, which can be defined as the quantum key (QK), while the final key in System D is constituted of two parts, namely one is the same QK as in System A, B and C, and the other part of the key is transmitted over the ClC from Alice to Bob, which is defined as the classical key (CK). 
%Thirdly, for the side information, syndromes are transmitted from Bob to Alice through the ClC in both System A and System B, while instead a bits-difference is a replacement of syndrome in System C. However, there is no such side information in System D since the CK is instead transmitted through the ClC and is a part of the final key.
%Lastly, only LDPC codes can be applied to both System A and System B, while any kinds of FEC codes can be applied to System C and D, especially IRCC can be used to achieve near-capacity performance but at the cost of higher complexity compared to System B. What's more, System D has a better complexity balance compared to System C.
%The pros and cons are summarized in Table \ref{table:systems_comparison} for the practical System B, C and D except for the ideal one System A.

In summary, the comparisons between System A (ideal syndrome-based CV-QKD), System B (practical syndrome-based CV-QKD), System C (practical bit-difference vector based CV-QKD) and System D (codeword-based CV-QKD) are summarized in Table~\ref{table:systems_comparison}.
More specifically, all four systems use the same equivalent QuC, but in System A we assume that the ClC is error-free, while in Systems B, C and D we consider realistic noise and fading in the ClC.
Secondly, as for the side information, syndromes are transmitted from Bob to Alice through the ClC in both System A and System B. By contrast, instead of using the syndrome, System C transmits te bit-difference vector from Bob to Alice through the ClC, making the system compatible with any FEC. Furthermore, System D transmits the CK from Alice to Bob through the ClC, making the FEC decoding complexity balanced between both sides.
Lastly, only LDPC codes can be applied to both System A and System B, while any kinds of FEC codes can be applied to System C and D. We opted for powerful IRCCs to achieve near-capacity performance.}}
\begin{table*}[!htb]
	\centering
	\caption{{\color{black}{Comparisons between four different systems.}}}
	{ 	\begin{tabular}{|p{3cm}|>{\centering\arraybackslash}m{3cm}|>{\centering\arraybackslash}m{3cm}|>{\centering\arraybackslash}m{3cm}|>{\centering\arraybackslash}m{3cm}|}
			%\toprule
			\hline
			\quad & System A & System B & System C&System D\\
			\hline
			Equivalent QuC&\multicolumn{4}{c|}{BI-AWGN channel}\\\hline
			ClC&Error-free&\multicolumn{3}{c|}{Noise and fading}\\\hline
			QK  &Bob\textrightarrow Alice&Bob\textrightarrow Alice&Bob\textrightarrow Alice&Bob\textrightarrow Alice\\\hline
			%CK  &-&-&-&Alice\textrightarrow Bob\\\hline
			\multirow{2}{*}{Side information (CK)} &\multicolumn{2}{c|}{Syndromes} &Bit-difference &CK \\
			&\multicolumn{2}{c|}{(Bob\textrightarrow Alice)}  &(Bob\textrightarrow Alice)& (Alice\textrightarrow Bob)\\\hline
			FEC types &Only LDPC& Only LDPC&Any&Any\\\hline\hline
			\multirow{3}{3cm}{Improvements over syndrome-based CV-QKD \cite{Laudenbach2018}}& &&Near-capacity, &Near-capacity, \\
			& -&- &compatible to any FEC & balanced complexity,\\
			& && & compatible to any FEC\\\hline
			
			%Cons &- &Unbalanced complexity, and limited to only LDPC &Unbalanced complexity and higher complexity&Higher complexity\\
			%\bottomrule
			%\hline
	\end{tabular}}
	\label{table:systems_comparison}
\end{table*}

%\begin{figure*}[htbp] 
%	\begin{center}
	%		\includegraphics[width=.6\linewidth]{figures/general_model_of_reconciliation_scheme_LDPC_BF.pdf}
	%		\caption{System $B$ - a general model of reconciliation scheme in the QKD system where BP decoding algorithm is utilised here. Moreover, the dash array represents the
		%			quantum transmission that not only contains the transmission of
		%			quantum states through the channel but also contains the process
		%			of converting the classical data $y_A$ to quantum sates at Alice's
		%			end and that of converting the received quantum states to classical
		%			data $y_B$ at Bob's end.}
	%		\label{fig:exame_LDPC_reconciliation_system_BP}
	%	\end{center}
%\end{figure*}

%\begin{figure*}[t] 
%	\begin{center}
	%		\includegraphics[width=0.8\linewidth]{figures/redesigned_reconciliation_system4_general.pdf}
	%		\caption{System $C$ - toy exame of the redesigned LDPC-coded reconciliation system. Note that, the exame utilised here is based on (7, 4) Hamming code, but is not restricted to this kind of code.}
	%		\label{fig:BER_redesigned_LDPC_reconciliation_system}
	%	\end{center}
%\end{figure*}
\section{Secret key rate analysis}\label{SKR}
{\color{black}{Note that the security level of the proposed System C and D is the same as that of System A and B, since the difference between them only lies in the side information.
		More specifically, we can only proceed with the reconciliation steps of Fig. 3, when the QK is securely received through the QuC, which obeys the Heisenberg's uncertainty principle.
		Therefore, even if Even steals the side information from the ClC, the final key still cannot be recovered. This is true for Systems A-D.
		Nonetheless, there are three distinct advantages for the proposed System D.
		Firstly, it is compatible with any FEC code, rather than being limited to LDPC codes. Secondly, it has a balanced complexity for Alice and Bob, which is particularly favourable in wireless device-to-device scenarios.
		Lastly, it exhibits near-capacity performance, where the SKR is close to the PLOB. This is achieved by using IRCCs for protecting both the QuC and ClC, making the SNRs required for error-free quantum and classical transmissions near-optimal.
%Note that the security analysis of the proposed System C is the same as that in System A and B, namely the security analysis in the literature since the difference between them only lies in the side information.
%As for System D, its security is also the same as that in System A and System B. More specifically, only when the QuC is detected safe enough, can we proceed our following reconciliation step, which contains both the QK and the additional CK generation process. Hence, the security is guaranteed by the QuC albeit Eve can obtain the additional CK. This is like the mechanism in the current syndrome-based QKD system, the security is guaranteed by the QuC even though Eve obtains the syndrome information transmitted through the ClC.
%Nonetheless, there are three advantageous points in our proposed System D.
%Firstly, it is compatible to any kinds of FEC codes not limited to LDPC codes. Secondly, it has balanced complexity.
%Lastly, it has near-capacity property, namely the SKR can be achieved closer to the PLOB bound when using IRCC in the proposed System C. What's more, the performance of ClC is also near capacity, indicating that the SNR required for the protection of classical transmission is relatively lower.
}}

The SKR is defined as \cite{milicevic2018quasi}
\begin{equation}\label{eq_SKR}
	K_{f}=\gamma \left(1-P_B\right)\left[\beta I_{AB}-\chi_{BE}-\triangle\left(N_{\text{privacy}}\right)\right],
\end{equation}
where $\gamma$ denotes the proportion of the key extractions in the total number of data exchanged by Alice and Bob, while $P_B$ represents the BLER in the reconciliation step.
Furthermore, $I_{AB}$ is the classical MI between Alice and Bob based on their shared correlated data, and $\chi_{BE}$ represents the Holevo information \cite{Laudenbach2018} that Eve can extract from the information of Bob. Finally,  $\triangle\left(N_{\text{privacy}}\right)$ represents the finite-size offset factor with the  finite-size $N_{\text{privacy}}$\footnote{Note that the finite-size offset can be viewed as a penalty term imposed by the imperfect parameter estimation step as shown in Fig.~\ref{reconciliation_protocol_in_QKD_whole} when using finite length  data. The value of $N_{\text{privacy}}$ set in our analysis is $10^{12}$, which is a value chosen in most of the literature.}. It was proven in \cite{leverrier2010finite} that when $N_{\text{privacy}}>10^4$, this factor can be simplified as 
\begin{equation}
	\Delta\left(N_{\text {privacy }}\right) \approx 7 \sqrt{\frac{\log _2(2 / \epsilon)}{N_{\text {privacy }}}},
\end{equation}
where $\epsilon$ represents the security parameter\footnote{This security parameter corresponds to the failure probability of the whole protocol, implying that the protocol is assured to perform as requested except for a probability of at most $\epsilon$. The value of $\epsilon$ is chosen to be $10^{-10}$ in our following analysis, which is widely used in the literature.} for the protocol.  
As for $\beta\in[0, 1]$, it represents the reconciliation efficiency, which is defined as \cite{Zhou2021,Laudenbach2018}
\begin{equation}\label{reconciliation_coefficient}
	\begin{aligned}
		\beta&=\frac{R}{C}=\frac{R}{0.5\log_2(1+\text{SNR})},
	\end{aligned}
\end{equation}
where $R$ represents the transmission rate, and $C$ is referred to as the one-dimensional Shannon capacity \cite{ryan2009channel,richardson2008modern}, which is given by the MI as follows \cite{milicevic2018quasi}: 
\begin{equation}\label{I_AB}
	C=I_{AB}=\frac{1}{2} \log _{2}(1+\mathrm{SNR})=\frac{1}{2} \log _{2}\left(\frac{V+\xi_{\mathrm{total}}}{1+\xi_{\mathrm{total}}}\right),
\end{equation}
where $V_A=V_s+1$ and $V_s$ is Alice's modulation variance\footnote{The modulation variance here represents the variance of Gaussian signals used in the modulator of CV-QKD.}, while
$\xi_{\mathrm{total}}$ is the total amount of noise between Alice and Bob,
which can be expressed as
\begin{equation}\label{definition_of_total_noise}
	\xi_{\text {total }}=\xi_{\text {line }}+\frac{\xi_{\text {hom }}}{T},
\end{equation}
where $\xi_{\mathrm{hom}}=\frac{1+v_{e l}}{\eta}-1$ is the homodyne detector's noise, and $v_{e l}$ stands for the electric noise, while $\eta$ represents the detection efficiency. Furthermore, $\xi_{\text {line }}=\left(\frac{1}{T}-1\right)+\xi_{\mathrm{ch}}$ represents the channel noise from the sender Alice, where ${T}$ represents the path loss and $\xi_{\mathrm{ch}}$ 
%$\xi_{\mathrm{ch}}=(W-1)(1-T)/T$ 
is the excess noise \cite{Weedbrook2012b} (i.e. imperfect modulation noise, Raman noise, phase-recovery noise, etc.). Assuming a single-mode fiber having an attenuation of $\alpha=0.2\;\mathrm{dB/km}$, the distance-dependent path loss of such a channel is $T=10^{-\alpha L/10}$, where $L$ denotes the distance between the two parties.
%Hence the MI can be calculated in a general way, which is
%\begin{equation}
%	\begin{aligned}
	%	I_{AB}&= 0.5\log\left(1+SNR\right)\\
	%	&=0.5\log\left(1+\frac{V_{mod}}{1+\xi_{total}}\right)\\
	%	&= 0.5\log\left(1+\frac{TV_{mod}}{1+\xi}\right)
	%	\end{aligned}
%\end{equation}
%Next, we can derive the Holevo information based on the paper \cite{Laudenbach2018},
%\begin{equation}
%	\chi_{B,E}=G\left(\frac{\lambda_{1}-1}{2}\right)+G\left(\frac{\lambda_{2}-1}{2}\right)-G\left(\frac{\lambda_{3}-1}{2}\right),
%\end{equation}
%where $\lambda_1$, $\lambda_2$ and $\lambda_3$ are symplectic eigenvalues, and details about them can be seen in Section \ref{Security_analysis}.

%As for the Holevo information between Bob and Eve, based on the fact that Eve is capable of purifying the whole system \cite{Laudenbach2018}, 
The Holevo information between Bob and Eve can be calculated as follows \cite{Laudenbach2018}
\begin{equation}
	%\begin{aligned}
	\chi_{E B}  =S\left(\rho_E\right)-S{\left(\rho_{E \mid B}\right)} =S{\left(\rho_{A B}\right)}-S{\left(\rho_{A\mid B}\right)},
	%\end{aligned}
\end{equation}
where $S(\cdot)$ is the von Neumann entropy defined in \cite{Laudenbach2018}. The von Neumann entropy of a Gaussian state $\rho$ containing $M$ modes can be written in terms of its symplectic eigenvalues \cite{holevo1999capacity}
\begin{equation}
	S\left( \rho  \right) = \sum\limits_{m= 1}^M {G\left( {{\upsilon _m}} \right)},
\end{equation}
where
\begin{equation}\label{g_function}
	G\left( \upsilon  \right) = \left( {\frac{{\upsilon  + 1}}{2}} \right){\log _2}\left( {\frac{{\upsilon  + 1}}{2}} \right) - \left( {\frac{{\upsilon  - 1}}{2}} \right){\log _2}\left( {\frac{{\upsilon  - 1}}{2}} \right).
\end{equation}
To elaborate on Eq.~(\ref{g_function}), generally these symplectic eigenvalues can be calculated based on the covariance matrix (CM) $\bf{V}$ of the Gaussian state using the formula \cite{Weedbrook2010}
\begin{equation}\label{symplectic_general_calculation}
	\upsilon  = \left| {i{\boldsymbol{\Omega} \bf{V}}} \right|,\quad \upsilon  \ge  1,
\end{equation}
where $\boldsymbol{\Omega}$ defines the symplectic form given by
\begin{equation}
	\boldsymbol{\Omega}:=\bigoplus_{m=1}^M \boldsymbol{\omega}=\left(\begin{array}{ccc}
		\boldsymbol{\omega} & & \\
		& \ddots & \\
		%& & \\
		& & \boldsymbol{\omega}
	\end{array}\right), \boldsymbol{\omega}=\left(\begin{array}{cc}
		0 & 1 \\
		-1 & 0
	\end{array}\right).
\end{equation}
Here $\bigoplus$ is the direct sum indicating the construction of a block-diagonal matrix $\boldsymbol{\Omega}$ having the same dimensionality as $\mathbf{V}$ by placing $M$ blocks of $\boldsymbol{\omega}$ diagonally.
Eq.~(\ref{symplectic_general_calculation}) indicates that first we have to find the eigenvalue of the matrix $i{\boldsymbol{\Omega}\bf{V}}$ and then take the absolute values.
However, in some circumstances, we can simplify the calculation of the eigenvalues. To elaborate further, firstly we consider a generic two-mode CM in the   form of
\begin{equation}
	{\bf{V}} = \left( {\begin{array}{*{20}{c}}
			{\bf{A}}&{\bf{C}}\\
			{{{\bf{C}}^T}}&{\bf{B}}
	\end{array}} \right).
\end{equation}
Based on \cite{Laudenbach2018}, the symplectic eigenvalues $\upsilon_1$ and $\upsilon_2$ of $\bf{V}$ can be written in the form of \cite{Weedbrook2010}
\begin{equation}\label{symplectic_calculation}
	{\upsilon _{1,2}} = \sqrt {\frac{1}{2}\left( {\Delta  \pm \sqrt {{\Delta ^2} - 4\det {\bf{V}}} } \right)} ,
\end{equation}
where $\det\bf{V}$ represents the determinant of the matrix $\bf{V}$ and we have
\begin{equation}
	\Delta:=\operatorname{det} \mathbf{A}+\operatorname{det} \mathbf{B}+2 \operatorname{det} \mathbf{C}.
\end{equation}

%In particular, let us consider a CM of the form
%\begin{equation}
%	\mathbf{V}=\left(\begin{array}{cc}
	%		a \mathbf{I} & \sqrt{T} c \mathbf{Z} \\
	%		\sqrt{T} c \mathbf{Z} & b \mathbf{I}
	%	\end{array}\right),
%\end{equation}
%where $c \geq 0, T \in[0,1]$ and
%\begin{equation}
%	\mathbf{I}_2=\left(\begin{array}{ll}
	%		1 & 0 \\
	%		0 & 1
	%	\end{array}\right), \quad \mathbf{Z}=\left(\begin{array}{cc}
	%		1 & 0 \\
	%		0 & -1
	%	\end{array}\right)
%\end{equation}
%are the two Pauli matrices.
%We can readily verify that $\operatorname{det} \mathbf{V}=\left(a b-c^{2} T\right)^{2}$ and $\Delta=$ $a^{2}+b^{2}-2 c^{2} T$. As a consequence, the eigenvalues obey the simple expression of
%\begin{equation}
%	\upsilon_{1,2}:=\frac{1}{2}\left[\sqrt{y} \pm(a-b)z\right],
%\end{equation}
%where $y:=(a+b)^{2}-4 c^{2} T \geq 4$.
In light of this, the CM related to the information between Alice and Bob, - namely the mode of $\rho_{{AB}}$ after transmission through the QuC - can be expressed as
%\begin{equation}
%\left( {\begin{array}{*{20}{c}}
		%		{{V_{mod + 1}}{I_2}}&{{V_{mod + 1}}{I_2}}\\
		%		{\sqrt{{V_{mod + 1}}}{I_2}}&{{V_{mod + 1}}{I_2}}
		%\end{array}} \right)
		%\end{equation}
		\begin{equation}\label{CM_after_transmission2}
			\begin{aligned}
				{\mathbf{V} _{{{A}}{{B}}}} &= \left({\begin{array}{*{20}{c}}
						{ {V}_A {\mathbf{I}_2}}&{\sqrt {\eta T\left( V_A^2 -1 \right)} {\mathbf{Z}}}\\
						{\sqrt {\eta T\left( V_A^2 -1 \right)} {\mathbf{Z}}}&{ {\eta T\left(V_A+\xi_{\text{total}}\right)} {\mathbf{I}_2}}
				\end{array}} \right)\\
				&= \left( {\begin{array}{*{20}{c}}
						{a{{\mathbf{I}_2}}}&{c{\mathbf{Z}}}\\
						{c{\mathbf{Z}}}&{b{{\mathbf{I}_2}}}
				\end{array}} \right),
			\end{aligned}
		\end{equation}
		where we have
		\begin{equation}
			\mathbf{I}_2=\left(\begin{array}{ll}
				1 & 0 \\
				0 & 1
			\end{array}\right), \quad \mathbf{Z}=\left(\begin{array}{cc}
				1 & 0 \\
				0 & -1
			\end{array}\right),
		\end{equation}
		which are the two Pauli matrices.
		%		where the variance of Bob $V_B={T{V_{\,\bmod \,}} + 1 + \xi }$ can be also denoted as
		%		\begin{equation}\label{general_represation_of_Vb}
			%			\begin{aligned}
				%				V_{B} &=\eta T_{\mathrm{ch}}\left(V+\xi_{\mathrm{total}}\right)\\
				%				&=T V_{\mathrm{mod}}+1+\xi,
				%			\end{aligned}
			%		\end{equation}
		%		where $I_{AB}$ is the MI between Alice and Bob, $\beta$ is the previously defined reconciliation efficiency, and $\chi_{B,E}$ is the Holevo bound on the information leaked to Eve.
		Therefore, the symplectic eigenvalues of $\rho_{AB}$ required are given by 
		\begin{equation}
			\upsilon_{1,2}^2=\frac{1}{2}\left(\Delta \pm \sqrt{\Delta^2-4 D^2}\right),
		\end{equation}
		where we have:
		\begin{equation}
			\begin{array}{r}
				\Delta=a^2+b^2-2 c^2, \\
				D=a b-c^2 .
			\end{array}
		\end{equation}
		As for the  symplectic eigenvalue of $\rho_{A\mid B}$, it can be shown that:
		\begin{equation}
			\upsilon_{3}=\sqrt{a\left(a-\frac{c^2}{b}\right)}.
		\end{equation}
		Hence, the Holevo information can be formulated as
		\begin{equation}\label{Holevo_three_eigenvalues}
			\chi_{BE}=G\left(\upsilon_{1}\right)+G\left(\upsilon_{2}\right)-G\left(\upsilon_{3}\right),
		\end{equation}
		where $\upsilon_1$, $\upsilon_2$ and $\upsilon_3$ are symplectic eigenvalues. Upon substituting Eq.~(\ref{I_AB}) and  Eq.~(\ref{Holevo_three_eigenvalues}) into Eq.~(\ref{eq_SKR}), the corresponding SKR can be obtained.
		
		In summary, SKR versus distance $L$ performance metric, used in our following analysis are as follows.
		\begin{itemize}
			\item Once the BLER versus SNR performance is obtained, a fixed BLER corresponds to a fixed SNR. 
			\item The noise term $\xi_{\mathrm{total}}$ in Eq.~(\ref{I_AB}) is a function of $L$. Hence, the value of $V_A$ is adjusted for each $L$ to satisfy the fixed SNR based on Eq.~(\ref{I_AB}).
			\item Once $V_A$ is adjusted for each $L$, $\chi_{BE}$ can be obtained, since it is a function of $V_A$. 
			\item Finally, the target SKR versus distance is derived.
		\end{itemize}
		
		\section{Performance analysis}\label{simulation}
		In this section, our BLER performance comparisons will be presented for different reconciliation schemes.
		Moreover, the SKR versus distance performance indicator will be analyzed.
		The common simulation parameters\footnote{Note that the code length and code rate used in both the QuC and ClC are the same.}, which are used in our LDPC based reconciliation scheme are summarized in Table~\ref{table:param_LDPC}.
		\begin{table}[htbp]
			\centering
			\caption{Simulation parameters.}
			\begin{tabular}{|l|r|}
				\hline
				Parameter                   &  Value \\\hline \hline
				{\color{black}{Coding rate (fixed)}}                  &  0.5 \\\hline
				Code length          &   1024  \\\hline
				Decoding algorithm           & BF/BP \\\hline
				Maximum  number of iterations                     & 50 \\\hline
				Quantum equivalent channel type                 & {BI-AWGN}   \\\hline
				ClC type                 & {AWGN/Rayleigh}   \\\hline
				QuC quality & SNR\\\hline
				ClC quality &$SNR^C$\\\hline	
				%	Column weight   &3 \\\hline
			\end{tabular}
			\label{table:param_LDPC}
			%\vspace{-0.5cm}
		\end{table}
		
		\subsection{Performance comparison between BF and BP decoding in System A }
\begin{figure}[t]
			\centering
			\includegraphics[width=3.3in]{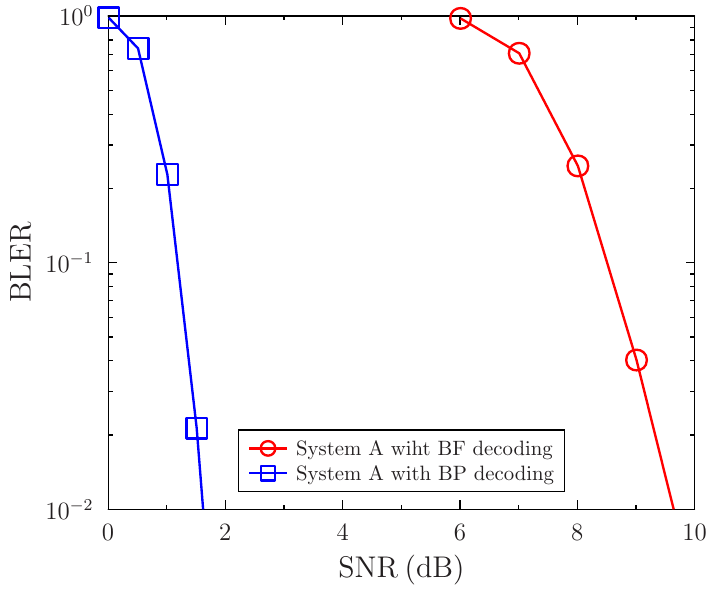}%
			\label{fig:LDPC_QKD_BLER_system12}
			%\hfil 
			%	\subfloat[]{\includegraphics[width=3.3in]{Figures/systemA_B_comparison_BER2.eps}%
				%		\label{fig:LDPC_QKD_BER_system12}}
			\caption{Performance comparison between System A of Fig.~\ref{fig:exame_LDPC_reconciliation_system_BF} and System B of Fig.~\ref{fig:exame_LDPC_reconciliation_system_BP}. The code length and code rate of the LDPC code are 1024 and 0.5, respectively.  BF decoding  is used in System A and BP decoding is utilized in System B. The classical authenticated channel is assumed to be error-free.}
			\label{fig:LDPC_QKD_system12}
		\end{figure}

		Firstly, the performance comparison between the BF and BP decoding in System A is presented by Fig.~\ref{fig:LDPC_QKD_system12}, where the classical authenticated channel is assumed to be error-free.
		Observe from Fig.~\ref{fig:LDPC_QKD_system12} that as expected, BP decoding outperforms BF decoding.
		Since the BLER performance is a key performance factor in the  SKR of Eq.~(\ref{eq_SKR}), the BP decoding algorithm will be adopted in the rest of performance analysis.
		%\begin{figure}[!htbp]
		%	\begin{subfigure}{\textwidth}
			%		\centering
			%		\includegraphics[width=0.5\linewidth]{Figures/comparison_BLER.eps}
			%		\caption{BLER}
			%		\label{fig:LDPC_QKD_BLER_system1}
			%	\end{subfigure}
		%	%\hfill
		%	%\newline
		%	\begin{subfigure}{\textwidth}
			%		\centering
			%		\includegraphics[width=.5\linewidth]{Figures/comparison_BER.eps}
			%		\caption{BER}
			%		\label{fig:LDPC_QKD_BER_system1}
			%	\end{subfigure}
		%	
		%	\caption{Performance comparisons between System 1 and System 2. The code length of 0.5 code rate LDPC code is 1024. BF decoding algorithm is utilised in System 1 and BP decoding is utilised in System 2. Note that the classical authenticate channel is assumed to be error-free.}
		%	\label{fig:LDPC_QKD_system12}
		%\end{figure}
		
		%\begin{figure}[!htbp]
		%	\centering
		%	\subfloat[1]{
			%		\label{fig:1}
			%		\begin{minipage}[c]{0.45\textwidth}
				%			\centering
				%			\includegraphics[width=\textwidth]{Figures/comparison_BER.eps}\\
				%		\end{minipage}
			%	}\\
		%	\subfloat[2]{
			%		\label{fig:2}
			%		\begin{minipage}[c]{0.45\textwidth}
				%			\centering
				%			\includegraphics[width=\textwidth]{Figures/comparison_BER.eps}
				%		\end{minipage}
			%	}
		%	\caption{xxx}
		%	\label{fig:2figs}
		%\end{figure}

		\subsection{Performance comparison among System B, System C and System D}
\begin{figure}[!t]
			\centering
			\includegraphics[width=3.3in]{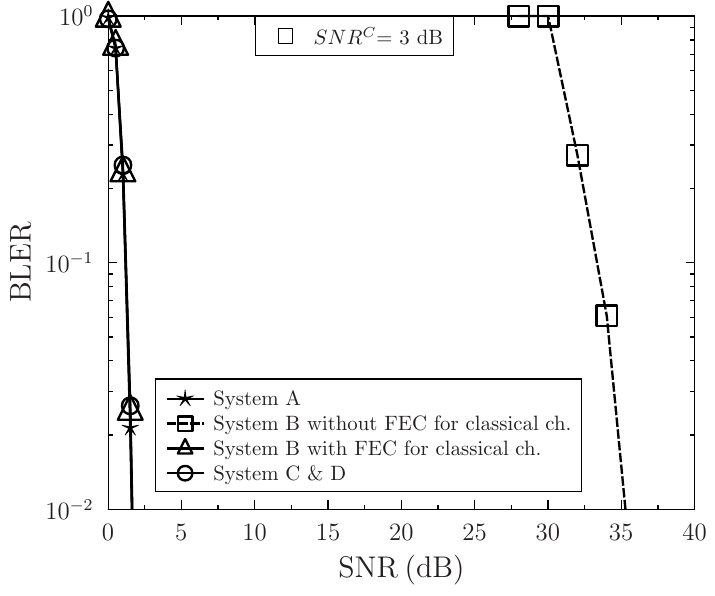}%
			\label{fig:LDPC_QKD_BLER_system23_AWGN}
			%\hfil 
			%	\subfloat[]{\includegraphics[width=3.3in]{Figures/comparsion_for_system_B_C_AWGN_BER2.eps}%
				%		\label{fig:LDPC_QKD_BER_system23_AWGN}}
			\caption{Performance comparison among System B, System C and System D. The code length and code rate of the LDPC code are 1024 and 0.5, respectively. BP decoding algorithm is used in System B, System C and System D, as well as System A. The  authenticated ClC is assumed to be an \textbf{AWGN} channel and the corresponding $SNR^C$ is 3 dB.}
			\label{fig:LDPC_QKD_AWGN_system23}
		\end{figure}

\begin{figure}[t]
			\centering
			\includegraphics[width=3.3in]{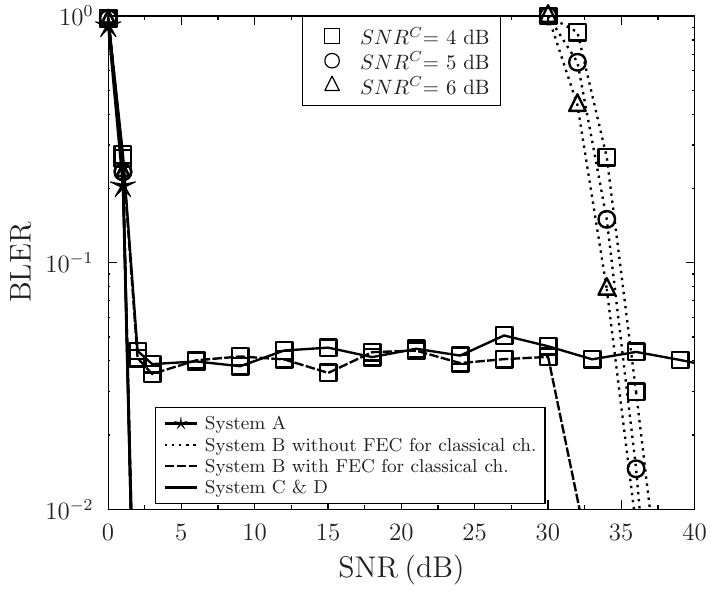}%
			%\label{fig:LDPC_QKD_BLER_system23_rayleigh}
			\caption{Performance comparison among System B, System C and System D. The code length and code rate of the LDPC code are 1024 and 0.5, respectively. BP decoding algorithm is used in System B, System C and System D, as well as System A.The  authenticated ClC is assumed to be a \textbf{Rayleigh} fading channel.}
			\label{fig:LDPC_QKD_system23_rayleigh}
		\end{figure}

		Let us now compare System B of Fig.~\ref{fig:exame_LDPC_reconciliation_system_BP} and System C of Fig.~\ref{fig:redesigned_LDPC_reconciliation_systemX} as well as System D of Fig.~\ref{fig:redesigned_LDPC_reconciliation_system_generic}, given that the authenticated channel is no longer error-free.
		Instead, an AWGN channel and an uncorrelated Rayleigh fading channel as well as perfect channel estimation are assumed for the classical side-information in  Fig.~\ref{fig:LDPC_QKD_AWGN_system23} and Fig.~\ref{fig:LDPC_QKD_system23_rayleigh}, respectively.
		
		%Let's commence with the analysis of BLER and BER performance with LDPC coding in the classical wireless communication under AWGN and Rayleigh channel as shown in Fig.~\ref{fig:comparison_AWGN_Rayleigh}. The code length and code rate of the LDPC code are 1024 and 0.5, respectively.
		%The PCM $\mathbf{H}$ is randomly constructed given the fixed column weight of 3.
		%It can be seen in Fig.~\ref{fig:LDPC_QKD_BLER_system23_AWGN} that the corresponding performance under AWGN channel outperforms that under Rayleigh channel by a SNR gap that around 2.3 dB at BLER=$10^{-3}$.
		%We note that the point-to-point communication performance of Fig.~\ref{fig:comparison_AWGN_Rayleigh} is a reference for choosing SNR for the ClC in Fig.~\ref{fig:LDPC_QKD_AWGN_system23} and Fig.\ref{fig:LDPC_QKD_system23_rayleigh}.
		
		%Fig.\ref{fig:LDPC_QKD_AWGN_system23} presents the performance comparison  between System B and System C where the ClC is modelled as an AWGN channel, and FEC codes are applied to protect the ClC.
		%Take the performance with System B where the ClC is error-free as a benchmark, it can be observed that the performance in System C as well as in System B with FEC codes for ClC can achieve almost the same performance as the benchmark.
		Fig.~\ref{fig:LDPC_QKD_AWGN_system23} demonstrates that the performance of the uncoded System B is severely degraded, when the ClC is contaminated by AWGN and hence it is no longer error-free, which confirms that error correction is required for both the ClC and QuC. By contrast, it can be seen in Fig.~\ref{fig:LDPC_QKD_AWGN_system23} that when System B, System C and System D employed FEC to protect their ClC, they no longer suffer from performance loss compared to the scenario of the idealistic assumption of having an error-free ClC. The ClC of $SNR^C=3$dB is sufficient for supporting  System B, System C and System D for approaching  the performance of the error-free ClC, which is shown by the solid line associated with stars, representing the BP-based performance of System A.
		%It can be seen in Fig.~\ref{fig:LDPC_QKD_AWGN_system23} that System C with FEC for ClC does not impose performance loss to System B with the idealistic assumption of error-free ClC.
		%The ClC quality is chosen to be $SNR^C=3$ dB.
		%based on Fig.~\ref{fig:comparison_AWGN_Rayleigh}.
		%On the other hand, the performance extremely degrades in System B that the ClC contains errors but without FEC for it, compared with the benchmark, which in a sense can reflect the importance of FEC codes given a realistic imperfect channel.
		%Furthermore, Fig.~\ref{fig:LDPC_QKD_AWGN_system23} demonstrates that the performance of System B is severely degraded, when the ClC has AWGN and hence it is no longer error-free, which confirms that error correction is required for both ClC and QuC.
		
		Similarly, Fig.~\ref{fig:LDPC_QKD_system23_rayleigh} provides our performance comparison, when the ClC is modelled by an uncorrelated Rayleigh fading channel having $SNR^C=4, 5, 6$dB.
		It can be seen in Fig.\ref{fig:LDPC_QKD_system23_rayleigh} that  System B operating without error protection for the ClC performs worst, requiring excessive SNR.
		By contrast, Fig.\ref{fig:LDPC_QKD_system23_rayleigh} shows that when FEC is applied to the ClC, at say $SNR^C=5,6$dB, System B, System C and System D approach the idealistic scenario of an error-free ClC.
		%This is due to that the corresponding performance in the ClC given such SNR is high enough to present a extremely low error floor.
		By contrast, an error floor is encountered by both System B , System C and System D at $SNR^C=4$dB, which is too low to mitigate the errors imposed by the Rayleigh faded ClC.
		% as shown in Fig.~\ref{fig:comparison_AWGN_Rayleigh}.
		%Additionally, as for System C the solid square line, it always has a lot of errors with $SNR^C=4$ dB in the ClC, hence the overall performance contains an error floor and it can not be removed away. Whilst in System B the dash square line, the error floor appears in a range of different qualities of QuC and the performance curve continue to drop after the channel quality continuously increases. This is due to that the performance with coding cannot be worse than that without coding.
		
\begin{figure}[!htbp]
			\centering
			\vspace{-0.5cm}
			\includegraphics[width=3.3in]{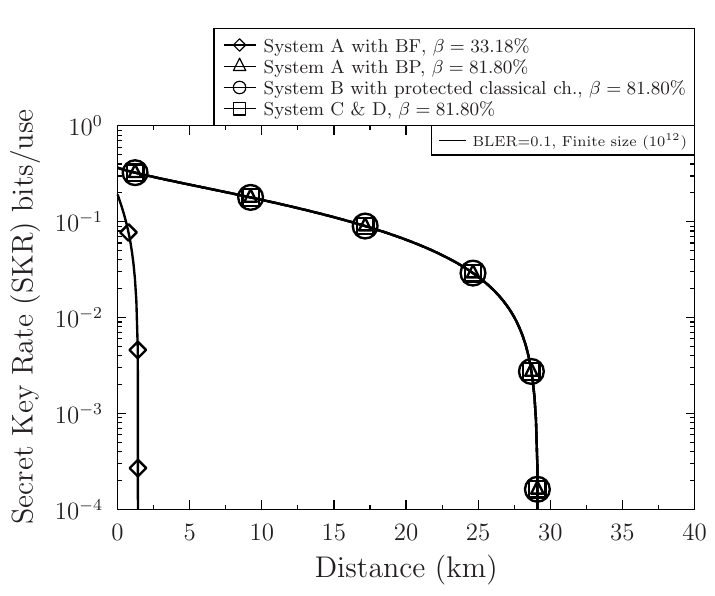}
			%\vspace{-0.5cm}
			\caption{The secret key rate analysis versus distance. The values of different reconciliation efficiency are calculated based on the corresponding SNR at the threshold of {BLER} equals to 0.1.}
			\vspace{-0.0cm}
			\label{fig:SKR_comparison_system_A_B_C_LDPC}
		\end{figure}

		Based on the BLER performances shown above, the corresponding SKR versus distance comparison is portrayed in Fig. \ref{fig:SKR_comparison_system_A_B_C_LDPC}. 
		%As can be seen that, the SKR will decrease with the increase of distance.
		The parameters are as follows: the modulation variance is adjusted to get a certain target SNR, which is related to the BLER threshold of 0.1 utilized for comparison; the excess noise is $\xi_{\text{ch}}=0.002$; the efficiency of the homodyne detector is $\eta=0.98$; the attenuation of a single-mode optical fibre is $\alpha=0.2 $dB/km, and the electric noise is $v_{el}=0.01$.
		More explicitly, Fig.~\ref{fig:SKR_comparison_system_A_B_C_LDPC} demonstrates that the maximum secure distance of System A using BF decoding is limited at around 1km, while that of System A using BP decoding is about 30 km.
		A similar performance as that of System A using BP decoding is attained for System B for a protected ClC at $SNR^C=3$dB, which is a sufficiently high $SNR^C$. 
		System C and System D also achieve a similar SKR performance, as evidenced by Fig.~\ref{fig:SKR_comparison_system_A_B_C_LDPC}.
		Note that the SKR versus distance  performance of System B without protecting the ClC is not shown here, because it is extremely low at such low reconciliation efficiency. 
		%\begin{figure*}[t]
		%	\centering
		%	\subfloat[]{\includegraphics[width=3.3in]{Figures/system_comparison_BLER2_modified2.eps}%
			%		\label{fig:LDPC_QKD_BLER_system23_rayleigh}}
		%	\hfil 
		%	\subfloat[]{\includegraphics[width=3.3in]{Figures/system_comparison_BER2_modified2.eps}%
			%		\label{fig:LDPC_QKD_BER_system23_rayleigh}}
		%	\caption{Performance comparison between System B and System C. The code length and code rate of the LDPC code are 1024 and 0.5, respectively. BP decoding algorithm is used in both System B and System C. Note that the classical authenticate channel is assumed to be a Rayleigh fading channel.}
		%	\label{fig:LDPC_QKD_system23_rayleigh}
		%\end{figure*}
		\subsection{Performance comparison among different FEC codes in System D}
		In this section, comparisons have been made among three different types of FEC codes, which are LDPC codes, CC and IRCC, respectively. {\color{black}{The number of LDPC decoding iteration is 50 and that of IRCC decoding is 30.}}
		%Note that the EXIT chart such as in Fig.~\ref{fig:EXIT_chart_for_IRCC_code} can be applied to design the IRCC code so as it can offer a better performance to the system.
		%More explicitly, we mainly focus on the comparison among different FEC codes in System D the codeword based reconciliation system in AWGN channel, since it  is demonstrated that the performance of System C can achieve the same performance of System B considering an imperfect channel with FEC codes given high enough $SNR^C$.
		
		\begin{figure}[ht]
			\centering
			\includegraphics[width=3.3in]{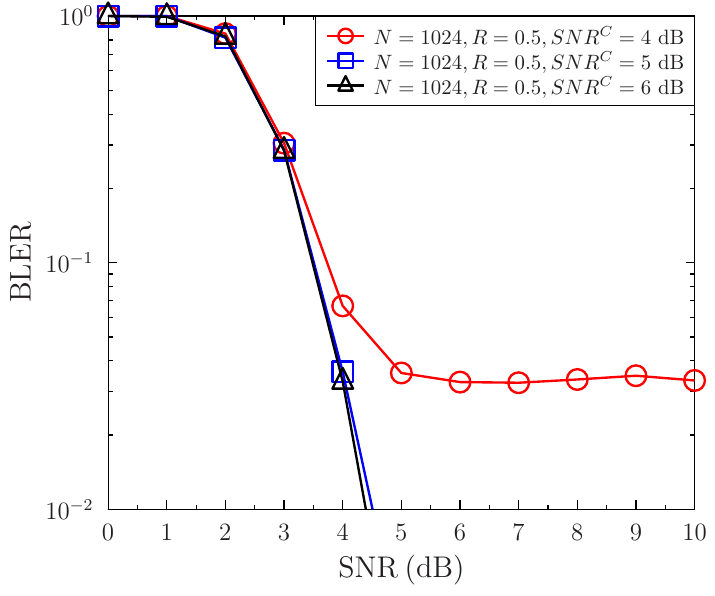}%
			\label{fig:CC_QKD_BLER_system23_AWGN}
			%\hfil 
			%	\subfloat[]{\includegraphics[width=3.3in]{Figures/comparison_AWGN_BER_CC2.eps}%
				%		\label{fig:CC_QKD_BER_system23_AWGN}}
			\caption{Performance comparison in System D with CC. The code length and code rate of the CC code are 1024 and 0.5, respectively. The  authenticated ClC is assumed to be a \textbf{AWGN}  channel.}
			\label{fig:CC_QKD_system23_AWGN}
		\end{figure}
		\begin{figure}[ht]
			\centering
			\includegraphics[width=3.3in]{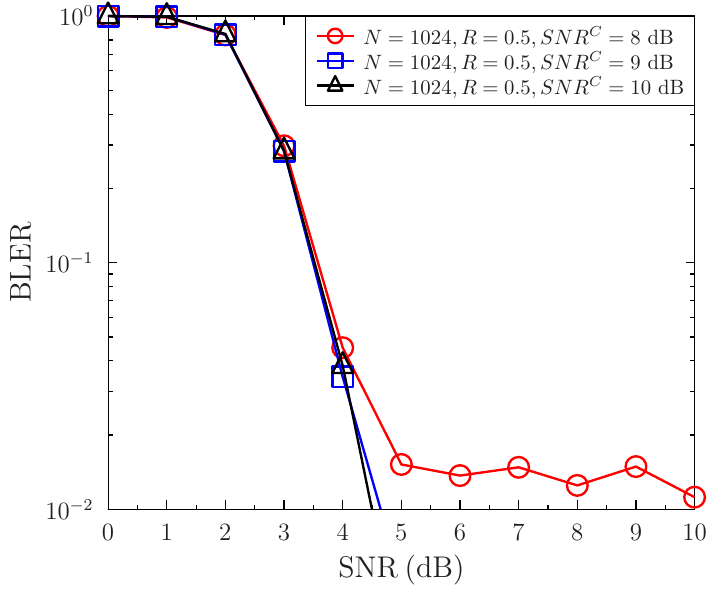}%
			\label{fig:CC_QKD_BLER_system23_rayleigh}
			%\hfil 
			%	\subfloat[]{\includegraphics[width=3.3in]{Figures/comparison_AWGN_Rayleigh_BER_CC2.eps}%
				%		\label{fig:CC_QKD_BER_system23_rayleigh}}
			\caption{Performance comparison in System D with CC. The code length and code rate of the CC code are 1024 and 0.5, respectively. The  authenticated ClC is assumed to be a \textbf{Rayleigh} fading  channel.}
			\label{fig:CC_QKD_system23_rayleigh}
		\end{figure} 
		Fig.\ref{fig:CC_QKD_system23_AWGN} and Fig.\ref{fig:CC_QKD_system23_rayleigh} characterize the performance of our codeword-based reconciliation scheme using a 1/2-rate CC of constraint-length 7 under AWGN and Rayleigh channels, respectively.
		The same trend can be observed in Fig.~\ref{fig:CC_QKD_system23_AWGN} and Fig.~\ref{fig:CC_QKD_system23_rayleigh}, where a higher $SNR^C$ of the ClC leads to reduced error floor.
		%What's more, a relatively higher $SNR^C$ is needed to have low enough error floor in Rayleigh channel to achieve a better performance for the wholist system as compared with that in AWGN channel.
		We note that as expected, compared to the AWGN scenario of Fig.~\ref{fig:CC_QKD_system23_AWGN}, the Rayleigh scenario of Fig.~\ref{fig:CC_QKD_system23_rayleigh} requires a higher $SNR^C$ for achieving a low BLER.

		\begin{figure}[htbp]
			\centering
			\subfloat[Diagram of interative double EXIT chart matching~\cite{MauderIrCC2009}]{\includegraphics[width=3.3in]{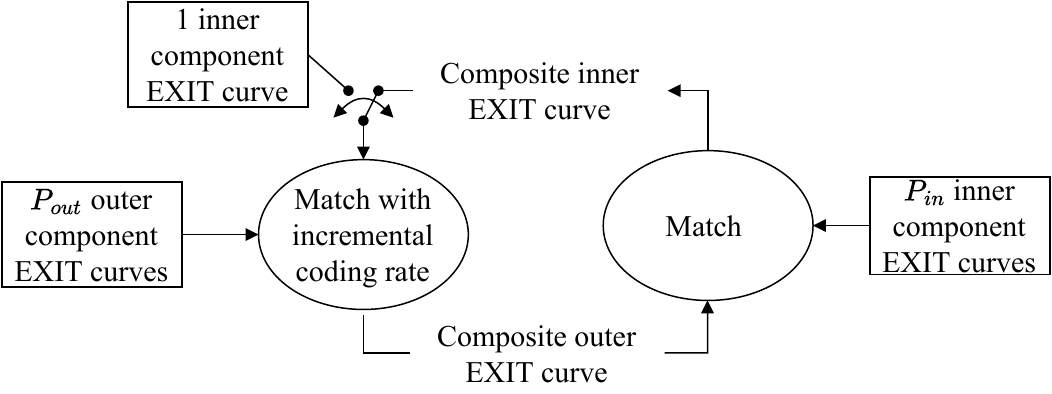}%
				\label{fig:Diagram_of_interative_double_EXIT_chart_matching}
				%\caption{Performance comparison with different FEC codes in System C. The code length and code rate of different codes are 1024 and 0.5, respectively.  Note that the classical authenticate channel is assumed to be a AWGN channel.}
			}
			%\hfil
			\vspace{0.2cm} 
			\\
			\subfloat[Schematic of the IRCC encoding and decoding~\cite{WuIrCC2009}]{\includegraphics[width=3.3in]{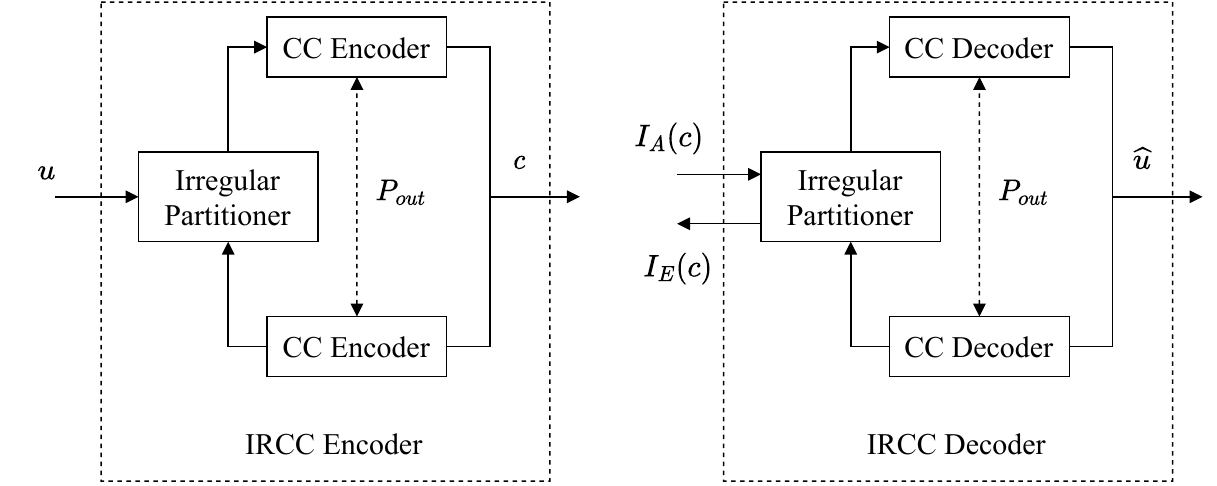}%
				\label{fig:Schematic_of_the_IRCC_encoder_decoder}
				%\caption{Performance comparison with different FEC codes in System C. The code length and code rate of different codes are 1024 and 0.5, respectively.  Note that the classical authenticate channel is assumed to be a AWGN channel.}
			}
			\caption{Schematic of IRCC codes.}
			\label{fig:Schematic_of_IRCC_codes}
		\end{figure}
		\begin{figure}[htbp]
			\centering
			\vspace{-0.5cm}
			\includegraphics[width=3.3in]{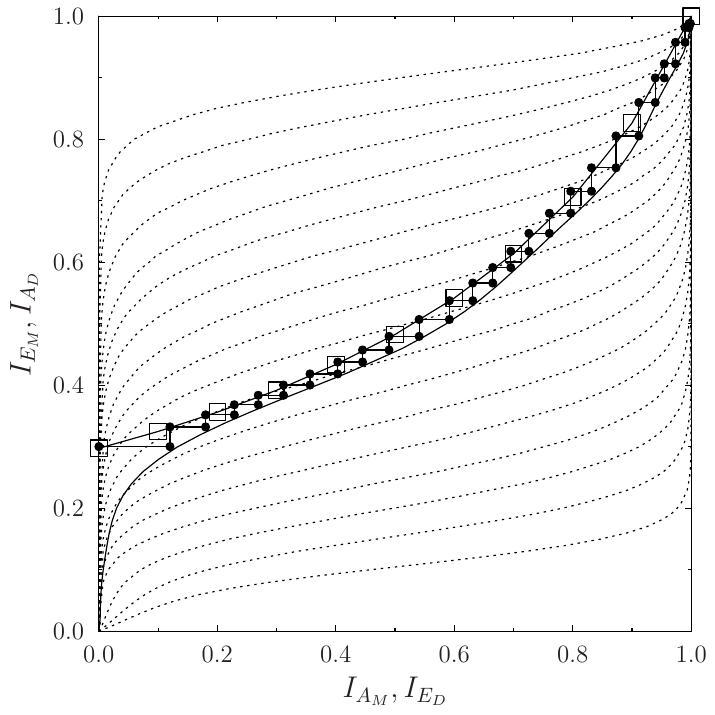}
			%\vspace{-0.5cm}
			\caption{EXIT chart and a decoding trajectory of IRCC and URC coded BPSK having a block-length of $10^5$, communicating over a classical AWGN  channel.}
			\label{fig:EXIT_chart_for_IRCC_code}
			\vspace{-0.0cm}
		\end{figure}
		Let us now consider the most sophisticated FEC scheme of this study, namely the IRCC used, which was discussed in great detail in \cite{MauderIrCC2009,WuIrCC2009} and shown in Fig.~\ref{fig:Schematic_of_IRCC_codes}, where $P_{out}$ and $P_{in}$ represents the number of irregular coding components used.
		In Fig.~\ref{fig:Diagram_of_interative_double_EXIT_chart_matching}, the extrinsic information transfer (EXIT) chart matching process detailed in \cite{Hanzo2013exit} is briefly illustrated, and the process of IRCC encoding and decoding is shown in Fig.~\ref{fig:Schematic_of_the_IRCC_encoder_decoder}. The EXIT charts \cite{Brink2001,hanzo2009near,Xutwodecades2017} and the iterative decoding trajectory of IRCC and Unity Rate Code (URC) coded BPSK modulation  communicating over classical AWGN channel are portrayed in Fig.~\ref{fig:EXIT_chart_for_IRCC_code}. More explicitly, the dotted EXIT curves seen in Fig.~\ref{fig:EXIT_chart_for_IRCC_code} correspond to  17 component CCs having coding rates ranging from 0.1 to 0.9 with a step size of 0.05. The IRCC design assigns different-length segments to different-rate component codes, so that a narrow tunnel is formed between the inner URC-BPSK coding component's EXIT curve and that of the outer IRCC decoder, as seen in Fig.~\ref{fig:EXIT_chart_for_IRCC_code}.
		It was shown in \cite{Hanzo2013exit} that the open tunnel area is proportional to the distance from capacity. More explicitly, as this area tends to zero, the scheme tends to approach the capacity.
		Hence, the presence of the narrow but open decoding tunnel of Fig.~\ref{fig:EXIT_chart_for_IRCC_code} indicates decoding convergence at a low SNR that approaches the capacity limit. The IRCC fractions of the component codes are found to be [0.0120603 0 0 0 0 0 0.605992 0.0780007 0 0 0 0.0672488 0.177274 0 0 0 0.0594503] for the 17 subcodes used in Fig.~\ref{fig:EXIT_chart_for_IRCC_code}. To elaborate briefly, for a 1000-bit IRCC the cod-rate of 0.05 is used for 0.0120603$\cdot$1000$\approx$12 bits. Then the code-rates of 0.1, 0.15, 0.2, 0.25 and 0.3 have 0 weight, so they are unused.
		The code-rate of 0.35 has a weight of 0.605992, hence it is used for 0.605992$\cdot$1000$\approx$606 bits and this process is applied to the remaining code-rates as well.

\begin{figure}[ht]
			\centering
			\includegraphics[width=3.3in]{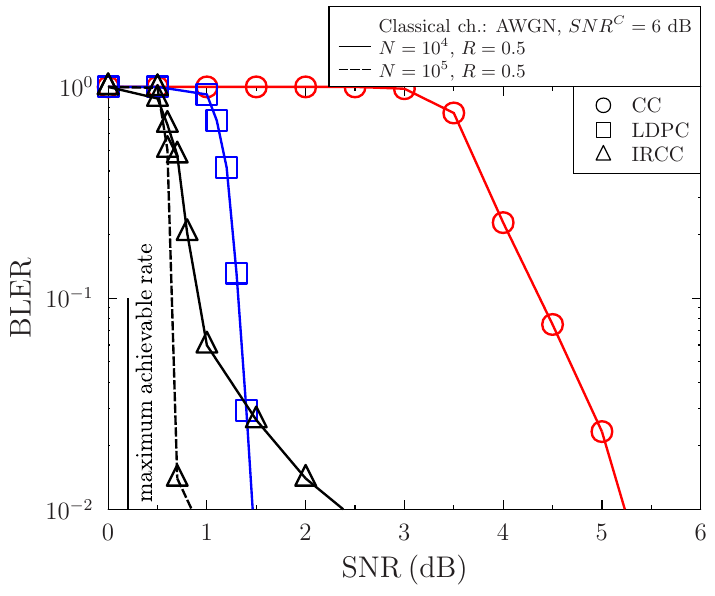}%
			\label{fig:QKD_BLER_system23_AWGN_CC_LDPC_IRCC_FL_1e4}
			%\hfil 
			%	\subfloat[]{\includegraphics[width=3.3in]{Figures/comparison_AWGN_BER_LDPC_CC_IRCC2_FL_1e4_AWGN.eps}%
				%		\label{fig:QKD_BER_system23_AWGN_CC_LDPC_IRCC_FL_1e4}}
			\caption{Performance comparison of different FEC codes in System D of Fig.~\ref{fig:redesigned_LDPC_reconciliation_system_generic}. The code length and code rate of different codes are $10^4$ and 0.5, respectively. The authenticated ClC is an \textbf{AWGN} channel.}
			\label{fig:QKD_system23_AWGN_CC_LDPC_IRCC_FL_1e4}
		\end{figure}

\begin{figure}[!htbp]
			\centering\includegraphics[width=3.3in]{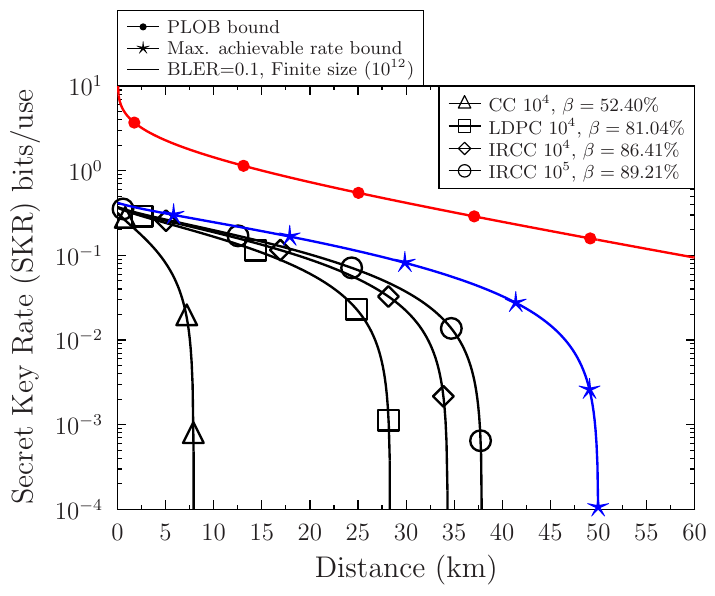}
			%\vspace{-0.5cm}
			%\caption{update for Fig. \ref{fig:SKR_d__target_SNR_mod_comparison_SKRmethod2_LDPC_CC_IRCC2}. Add the SKR for IRCC and polar code with N=1024 as well as the Pirandola-Laurenza-Ottaviani-Banchi (PLOB) \cite{Pirandola2017} bound and the maximum achievable rate bound.}
			%\vspace{-0.0cm}
			%\hfil 
			%	\subfloat[]{\includegraphics[width=3.3in]{figures/SKR3_comparison_N_1024_LDPC_CC_IRCC_FL_1e4.eps}
				%	%\vspace{-0.5cm}
				%	%\caption{update for Fig. \ref{fig:SKR_d__target_SNR_mod_comparison_SKRmethod2_LDPC_CC_IRCC2}. Add the SKR for IRCC and polar code with N=1024 as well as the Pirandola-Laurenza-Ottaviani-Banchi (PLOB) \cite{Pirandola2017} bound and the maximum achievable rate bound.}
				%	\vspace{-0.0cm}
				%	\label{fig:SKR_d__target_SNR_mod_comparison_SKRmethod2_LDPC_CC_IRCC2_update}}
			\caption{The secret key rate versus distance. The values of different reconciliation efficiencies are shown in Table \ref{table:reconciliation_BLER_SNR_2} at the {BLER} threshold of 0.1. The other parameters are as follows: the modulation variance is adjusted to get a target SNR, the excess noise is $\xi_{ch}=0.002$, the efficiency of the homodyne detector is $\eta=0.98$, the attenuation of a single-mode optical fibre is $\alpha=0.2$dB/km, and the electric noise is $v_{el}=0.01$. The corresponding PLOB \cite{Pirandola2017} bound and the maximum achievable rate bound are shown as well.}
			\label{fig:SKR2_comparison_N_1024_LDPC_CC_IRCC_FL_1e4}
		\end{figure}

\begin{figure}[ht]
			\centering
			\includegraphics[width=3.3in]{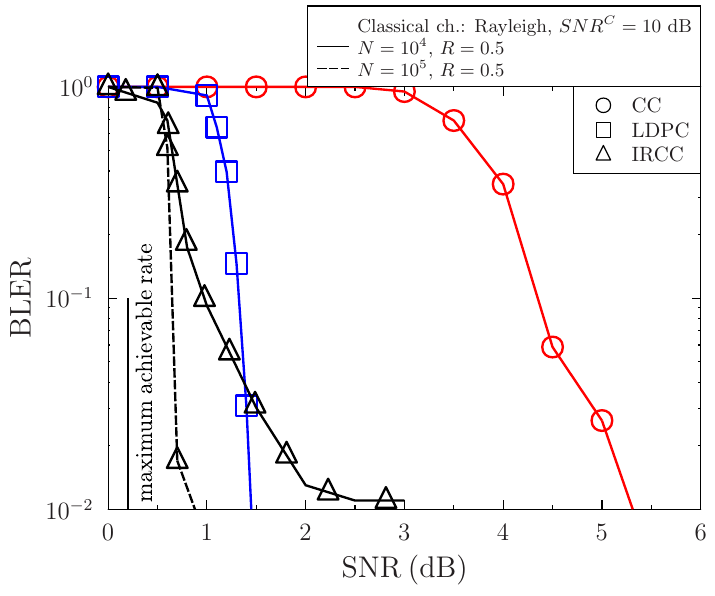}%
			\label{fig:QKD_BLER_system23_Rayleigh_CC_LDPC_IRCC_Fl_1e4}
			%\hfil 
			%	\subfloat[]{\includegraphics[width=3.3in]{Figures/comparison_AWGN_BER_LDPC_CC_IRCC2_FL_1e4_Rayleigh.eps}%
				%		\label{fig:QKD_BER_system23_Rayleigh_CC_LDPC_IRCC_Fl_1e4}}
			\caption{Performance comparison of different FEC codes in System D of Fig.~\ref{fig:redesigned_LDPC_reconciliation_system_generic}. The code length and code rate of different codes are $10^4$ and 0.5, respectively. The authenticated ClC is a \textbf{Rayleigh} channel.}
			\label{fig:QKD_system23_Rayleigh_CC_LDPC_IRCC_FL_1e4}
		\end{figure}

		The BLER of the codeword based reconciliation scheme (System D) of a variety of FEC codes is shown in Fig.~\ref{fig:QKD_system23_AWGN_CC_LDPC_IRCC_FL_1e4}. The corresponding $(\text{BLER},\beta)$ pair can be obtained as tabulated in Table~\ref{table:reconciliation_BLER_SNR_2}.
		In light of the BLER performance comparison among different FEC codes,
		%\begin{table} [!htbp]
		%	\centering
		%	\caption{Reconciliation efficiency of different FEC codes at different BLER thresholds, that is $\beta_1$at BLER equals to 0.1 and $\beta_2$ at BLER equals to 0.2, followed by the corresponding SNRs. The code length and code rate of them are set the same, which are $N=1024$, $R=0.5$.}
		%	\begin{tabular}{c|c|c|c|c}
			%		\hline
			%		 &\multicolumn{2}{c|}{BLER=0.1} & \multicolumn{2}{c}{BLER=0.5}\\\hline\hline%\cline{1-5}
			%		Code type& $\beta_1(\%)$& $SNR_1$ & $\beta_2(\%)$ & $SNR_2$ \\\hline
			%		LDPC code  & 69&72&1.73&1.62 \\
			%		CC  &71 &74&1.65&1.55\\
			%		polar code & 73& 76&1.58&1.49\\
			%		IRCC  & 76& 78&1.49&1.43\\\hline
			%	\end{tabular}
		%	\label{table:reconciliation_BLER_SNR}
		%\end{table}
		%\begin{table} [tbp]
		%	\centering
		%	\caption{Reconciliation efficiency of different FEC codes at the BLER threshold that equals to 0.1, together with the corresponding SNRs. The code length and code rate of them are set the same, which are $N=1024$, $R=0.5$, the classical authenticated channels are AWGN  and Rayleigh channel, respectively.}
		%	\begin{tabular}{ccccc}
			%		\hline
			%		&\multicolumn{2}{c}{AWGN} & \multicolumn{2}{c}{Rayleigh}\\\cline{2-5}
			%		Code type& $SNR_1$(dB) &$\beta_1(\%)$ &$SNR_2$(dB)&  $\beta_2(\%)$  \\\hline
			%		CC& 3.5 &58.98&3.5&58.98 \\
			%		LDPC code  &1.25 &81.80&1.25& 81.80\\
			%		IRCC  & 2.0& 72.99&2.0&72.99\\\hline
			%	\end{tabular}
		%	\label{table:reconciliation_BLER_SNR}
		%\end{table}
		the corresponding SKR versus distance performances of different FEC code based reconciliation schemes can be obtained with the aid of the reconciliation efficiencies as shown in Fig.~\ref{fig:SKR2_comparison_N_1024_LDPC_CC_IRCC_FL_1e4}.
		For the same BLER, for example BLER=0.1, given the same block length of  $10^4$ bits, the reconciliation performances associated with IRCC, LDPC and CC exhibit different reconciliation efficiencies, which are $86.41\%$, $81.04\%$ and $52.40\%$, respectively. Therefore, the SKR performance of the IRCC scheme is the best. More explicitly, the maximum secure distance associated with the IRCC code (the diamond solid line) is longer than that of LDPC (the square solid line) and of the CC (the square dash line) code. Furthermore, the SKR at each specific secure distance associated with IRCC code is higher than that of the LDPC or CC codes. To elaborate further, the maximum secure distance of the IRCC code with BLER=0.1, $\beta=86.41\%$ is around 35km, whereas the corresponding  maximum secure distance of the LDPC (BLER=0.1, $\beta=81.04\%$) and CC (BLER=0.1, $\beta=52.40\%$) codes are around 28km and 8km, respectively.
		The same conclusion can be drawn for the comparison between LDPC and CC codes at BLER=0.01.
		Moreover, Fig.~\ref{fig:SKR2_comparison_N_1024_LDPC_CC_IRCC_FL_1e4} demonstrates that the SKR performance of a longer block length of $N=10^5$ is superior to that of $N=10^4$, since a longer block length can offer near-capacity performance, hence leading to a longer secure transmission distance of around 37km. 
		Note that the vertical line shown in Fig.~\ref{fig:QKD_system23_AWGN_CC_LDPC_IRCC_FL_1e4} represents the minimum SNR required to achieve near-error-free transmission. It is obtained based on \cite{Xutwodecades2017,hanzo2009near,NgmimoCapacity}
		\begin{equation}
			C^{DCMC}(SNR)=1-\frac{1}{2}\sum_{i=0}^{1}\mathbb{E}\left\{\log_2\left[\sum_{\bar{i}=0}^{1}\exp\left(\Psi_{i,\bar{i}}\right)\right]\right\},
		\end{equation}
		where we have $\Psi_{i,\bar{i}}=\frac{-\|s^i-s^{\bar{i}}+n\|^2+\|n\|^2}{N_0}$, $s^i$ represents the BPSK symbols, while $n$ is the noise, whose distribution obeys $n \sim\mathcal{CN}(0,N_0)$.
		The corresponding SNR can be obtained by solving 	$C^{DCMC}(SNR)=0.5$, since we consider BPSK and $R=0.5$, FEC codes. The same capacity line is also drawn in Fig.~\ref{fig:QKD_system23_Rayleigh_CC_LDPC_IRCC_FL_1e4}.
		\begin{table} [tbp]
			\centering
			\caption{Reconciliation efficiency of different FEC codes calculated from Eq.~(\ref{reconciliation_coefficient}) at the BLER threshold  that equals to 0.1, together with the corresponding SNRs. The code length and code rate of them are the same for all of them, which are $N=10^4$, $R=0.5$. The  authenticated ClCs are AWGN and uncorrelated Rayleigh channel.}
			\begin{tabular}{lcccc}
				\hline
				&\multicolumn{2}{c}{AWGN} & \multicolumn{2}{c}{Rayleigh}\\\cline{2-5}
				Code type& SNR(dB) &$\beta(\%)$ &SNR(dB)&  $\beta(\%)$  \\\hline
				CC& 4.4 &52.40&4.4&52.40 \\
				LDPC code  &1.31 &81.04&1.31& 81.40\\
				IRCC  & 0.9& 86.41&1.0&85.06\\
				IRCC ($10^5$)   &0.7 &89.21&0.7&89.21\\ \hline
			\end{tabular}
			\label{table:reconciliation_BLER_SNR_2}
		\end{table}
		%In addition, the SKR comparison among different kinds of codes are illustrated in Fig.~\ref{fig:SKR2_comparison_N_1024_LDPC_CC_IRCC_FL_1e4}.

		%Another set of comparison where the ClC is Rayleigh fading channel, are discussed below. We can decide to choose one of them or keep both sets of them in the paper. 
		
		%\begin{figure*}[ht]
		%	\centering
		%	\subfloat[]{\includegraphics[width=3.3in]{Figures/comparison_AWGN_BLER_LDPC_CC_IRCC2_FL_1024_Rayleigh.eps}%
			%		\label{fig:QKD_BLER_system23_Rayleigh_CC_LDPC_IRCCFL_1024}}
		%	\hfil 
		%	\subfloat[]{\includegraphics[width=3.3in]{Figures/comparison_AWGN_BER_LDPC_CC_IRCC2_FL_1024_Rayleigh.eps}%
			%		\label{fig:QKD_BER_system23_Rayleigh_CC_LDPC_IRCC_FL_1024}}
		%	\caption{Performance comparison with different FEC codes in System C. The code length and code rate of different codes are 1024 and 0.5, respectively.  Note that the classical authenticate channel is assumed to be a Rayleigh channel.}
		%	\label{fig:QKD_system23_Rayleigh_CC_LDPC_IRCC_FL_1024}
		%\end{figure*}
		
		%\begin{table} [tbp]
		%	\centering
		%	\caption{Reconciliation efficiency of different FEC codes at different BLER thresholds, that is $\beta_1$ at BLER equals to 0.1 and $\beta_2$ at BLER equals to 0.01, together with the corresponding SNRs. The code length and code rate of them are set the same, which are $N=1024$, $R=0.5$, the classical authenticated channel is Rayleigh fading channel.}
		%	\begin{tabular}{ccccc}
			%		\hline
			%		&\multicolumn{2}{c}{BLER=0.1} & \multicolumn{2}{c}{BLER=0.01}\\\cline{2-5}
			%		Code type& $SNR_1$(dB) &$\beta_1(\%)$ &$SNR_2$(dB)&  $\beta_2(\%)$  \\\hline
			%		
			%		CC& 3.5 &58.98&2.5&67.83 \\
			%		LDPC code  &1.25 &81.80&0.67& 89.69\\
			%		IRCC  & 2.0& 72.99&1.0&85.06\\\hline
			%	\end{tabular}
		%	\label{table:reconciliation_BLER_SNR_Rayleigh1}
		%\end{table}

		%\begin{table} [tbp]
		%	\centering
		%	\caption{Reconciliation efficiency of different FEC codes at different BLER thresholds, that is $\beta_1$ at BLER equals to 0.1 and $\beta_2$ at BLER equals to 0.01, together with the corresponding SNRs. The code length and code rate of them are set the same, which are $N=10^4$, $R=0.5$, the classical authenticated channel is Rayleigh fading channel.}
		%	\begin{tabular}{ccccc}
			%		\hline
			%		&\multicolumn{2}{c}{BLER=0.1} & \multicolumn{2}{c}{BLER=0.01}\\\cline{2-5}
			%		Code type& $SNR_1$(dB) &$\beta_1(\%)$ &$SNR_2$(dB)&  $\beta_2(\%)$  \\\hline
			%		CC& 4.4 &52.40&3.70&57.42 \\
			%		LDPC code  &1.31 &81.04&1.15& 83.08\\
			%		IRCC  & 1.0& 85.06&0.65&89.93\\
			%		IRCC ($10^5$)   &0.7 &89.21&0.6&90.65\\ \hline
			%	\end{tabular}
		%	\label{table:reconciliation_BLER_SNR_Rayleigh2}
		%\end{table}
		On the other hand, based on the reconciliation efficiencies seen in
		%Table \ref{table:reconciliation_BLER_SNR} and 
		Table \ref{table:reconciliation_BLER_SNR_2} and inferred from Fig.~\ref{fig:QKD_system23_AWGN_CC_LDPC_IRCC_FL_1e4} as well as Fig.~\ref{fig:QKD_system23_Rayleigh_CC_LDPC_IRCC_FL_1e4}, the reconciliation efficiencies are similar for the Rayleigh-faded and for the AWGN ClC.
		This is because the reconciliation efficiencies are mainly determined by the QuC quality characterized by the equivalent channel SNR, provided that the ClC quality is high enough for ensuring that the errors from the classical transmission do not unduly erode the overall system performance, as demonstrated in Fig.~\ref{fig:QKD_system23_Rayleigh_CC_LDPC_IRCC_FL_1e4}.
		Intuitively, a higher $SNR^C$ is required in Rayleigh faded ClCs compared to the $SNR^C$ in an AWGN based ClC to achieve nearly the same system performance. Therefore, given that the $\beta$s are nearly the same, the SKR of a Rayleigh faded ClC is similar to that in Fig.~\ref{fig:SKR2_comparison_N_1024_LDPC_CC_IRCC_FL_1e4}.
		%But, we can see if it is necessary to keep both parts of analysis.
		
		\section{Conclusions}\label{conclusion}
		The codeword based reconciliation concept was proposed as a general reconciliation scheme that can be applied in conjunction with diverse FEC codes.
		This is a significant improvement because the popular syndrome-based LDPC-coded reconciliation scheme can only be applied for FEC codes that possess syndromes.
		Furthermore, in contrast to the general assumption that the classical authenticated channel is error-free and noiseless, a realistic ClC has been considered, which may contain errors. We investigated the performance of our QKD systems when the classical authenticated channel is modelled as an AWGN channel or a Rayleigh channel.
		We demonstrated that when the ClC quality is sufficiently high, the QKD system will have a relatively low BLER. An error floor is exhibited by the system, when the ClC has errors due to employing a weak channel code or when the ClC quality is too low.
		More specifically, we have investigated LDPC codes, CC and IRCCs assisted CV-QKD schemes. It was demonstrated that the IRCC associated system performs the best among them, followed by the LDPC codes, whilst the CC code performs the  worst.
		In light of this, the SKR versus distance performance of different FEC codes using optical fibre as the QuC has been compared.
		It was demonstrated that near-capacity FEC codes such as IRCC can provide higher reconciliation efficiency, hence they can offer a longer secure transmission distance.
		\appendix[Mapping function of multidimensional reconciliation]\label{Mapping}
		\textbf{Mapping function calculation:} Bob calculates the mapping function $\mathbf{M}_{i}\left(\mathbf{y}_i^\prime,\mathbf{u}_i\right)$ for each 8-element vector, which meets $\mathbf{M}_{i}\left(\mathbf{y}_i^\prime,\mathbf{u}_i\right)\mathbf{y}_{i}^\prime=\mathbf{u}_{i}$, using the following formula:
		\begin{equation}
			\mathbf{M}_{i}\left(\mathbf{y}_i^\prime,\mathbf{u}_i\right)=\sum\nolimits_{d=1}^8 {\alpha}_i^d \mathbf{A}_{8}^d,
		\end{equation}
		where ${\alpha}_i^d$ is the $d$-th element of $\bm{\alpha}_i\left(\mathbf{y}_i^\prime,\mathbf{u}_i\right)=\left(\alpha_i^1,\alpha_i^2,...,\alpha_i^8\right)^T$, which is the coordinate of the vector $\mathbf{u}_i$ under the orthonormal basis $\left(\mathbf{A}_8^1\mathbf{y}_{i}^\prime,\mathbf{A}_8^2\mathbf{y}_{i}^\prime,...,\mathbf{A}_8^8\mathbf{y}_{i}^\prime\right)$ and it can be expressed as $\bm{\alpha}_i\left(\mathbf{y}_i^\prime,\mathbf{u}_i\right)=\left(\mathbf{A}_8^1\mathbf{y}_{i}^\prime,\mathbf{A}_8^2\mathbf{y}_{i}^\prime,...,\mathbf{A}_8^8\mathbf{y}_{i}^\prime\right)^T\mathbf{u}_i$. Note that $\mathbf{A}_8^d, d=1,2,...,8$ is the orthogonal matrix of size $8\times 8$ provided in the Appendix of \cite{leverrier2008multidimensional}.
		Note that the 8 orthogonal  matrices used in our scheme are listed in Eq.~(\ref{orthogonal_matrices}).
		\begin{figure*}
			\begin{equation}\label{orthogonal_matrices}
				%\small
				\begin{aligned}
					&\mathbf{A}_8^1=\begin{bmatrix}
						1 &0 & 0 &0 &0 &0 &0 &0 \\
						0 &1 & 0 &0 &0 &0 &0 &0 \\
						0 &0 & 1 &0 &0 &0 &0 &0 \\
						0 &0 & 0 &1 &0 &0 &0 &0 \\
						0 &0 & 0 &0 &1 &0 &0 &0 \\
						0 &0 & 0 &0 &0 &1 &0 &0 \\
						0 &0 & 0 &0 &0 &0 &1 &0 \\
						0 &0 & 0 &0 &0 &0 &0 &1 \\
					\end{bmatrix},
					~~~~~~~~~\mathbf{A}_8^2=\begin{bmatrix}
						0 &-1 & 0 &0 &0 &0 &0 &0 \\
						1 &0 & 0 &0 &0 &0 &0 &0 \\
						0 &0 & 0 &1 &0 &0 &0 &0 \\
						0 &0 & -1&0 &0 &0 &0 &0 \\
						0 &0 & 0 &0 &0 &1 &0 &0 \\
						0 &0 & 0 &0 &-1 &0 &0 &0 \\
						0 &0 & 0 &0 &0 &0 &0 &-1 \\
						0 &0 & 0 &0 &0 &0 &1 &0 \\
					\end{bmatrix},\\
					&\mathbf{A}_8^3=\begin{bmatrix}
						0 &0 & -1 &0 &0 &0 &0 &0 \\
						0 &0 & 0 &-1 &0 &0 &0 &0 \\
						1 &0 & 0 &0 &0 &0 &0 &0 \\
						0 &1 & 0&0 &0 &0 &0 &0 \\
						0 &0 & 0 &0 &0 &0 &1 &0 \\
						0 &0 & 0 &0 &0 &0 &0 &1 \\
						0 &0 & 0 &0 &-1 &0 &0 &0 \\
						0 &0 & 0 &0 &0 &-1 &0 &0 \\
					\end{bmatrix},
					\mathbf{A}_8^4=\begin{bmatrix}
						0 &0 & 0 &-1 &0 &0 &0 &0 \\
						0 &0 & 1 &0 &0 &0 &0 &0 \\
						0 &-1 & 0 &0 &0 &0 &0 &0 \\
						1 &0 & 0&0 &0 &0 &0 &0 \\
						0 &0 & 0 &0 &0 &0 &0 &1 \\
						0 &0 & 0 &0 &0 &0 &-1 &0 \\
						0 &0 & 0 &0 &0 &1 &0 &0 \\
						0 &0 & 0 &0 &-1 &0 &0 &0 \\
					\end{bmatrix},\\
					&\mathbf{A}_8^5=\begin{bmatrix}
						0 &0 & 0 &0 &-1 &0 &0 &0 \\
						0 &0 & 0 &0 &0 &-1 &0 &0 \\
						0 &0 & 0 &0 &0 &0 &-1 &0 \\
						0 &0 & 0&0 &0 &0 &0 &-1 \\
						1 &0 & 0 &0 &0 &0 &0 &0 \\
						0 &1 & 0 &0 &0 &0 &0 &0 \\
						0 &0 & 1 &0 &0 &0 &0 &0 \\
						0 &0 & 0 &1 &0 &0 &0 &0 \\
					\end{bmatrix},
					\mathbf{A}_8^6=\begin{bmatrix}
						0 &0 & 0 &0 &0 &-1 &0 &0 \\
						0 &0 & 0 &0 &1 &0 &0 &0 \\
						0 &0 & 0 &0 &0 &0 &0 &-1 \\
						0 &0 & 0&0 &0 &0 &1 &0 \\
						0 &-1 & 0 &0 &0 &0 &0 &0 \\
						1 &0 & 0 &0 &0 &0 &0 &0 \\
						0 &0 & 0 &-1 &0 &0 &0 &0 \\
						0 &0 & 1 &0 &0 &0 &0 &0 \\
					\end{bmatrix},\\
					&\mathbf{A}_8^7=\begin{bmatrix}
						0 &0 & 0 &0 &0 &0 &-1 &0 \\
						0 &0 & 0 &0 &0 &0 &0 &1 \\
						0 &0 & 0 &0 &1 &0 &0 &0 \\
						0 &0 & 0&0 &0 &-1 &0 &0 \\
						0 &0 & -1 &0 &0 &0 &0 &0 \\
						0 &0 & 0 &1 &0 &0 &0 &0 \\
						1 &0 & 0 &0 &0 &0 &0 &0 \\
						0 &-1 & 0 &0 &0 &0 &0 &0 \\
					\end{bmatrix},
					\mathbf{A}_8^8=\begin{bmatrix}
						0 &0 & 0 &0 &0 &0 &0 &-1 \\
						0 &0 & 0 &0 &0 &0 &-1 &0 \\
						0 &0 & 0 &0 &0 &1 &0 &0 \\
						0 &0 & 0&0 &1 &0 &0 &0 \\
						0 &0 & 0 &-1 &0 &0 &0 &0 \\
						0 &0 & -1 &0 &0 &0 &0 &0 \\
						0 &1 & 0 &0 &0 &0 &0 &0 \\
						1 &0 & 0 &0 &0 &0 &0 &0 \\
					\end{bmatrix}.
				\end{aligned}
			\end{equation}
			{\noindent} \rule[-10pt]{18cm}{0.05em}
		\end{figure*}
		\bibliographystyle{IEEEtran}
		\bibliography{Myreference2}
\end{document}